\documentclass[pra,aps,floatfix,floats,superscriptaddress,notitlepage]{revtex4-1}

\pdfoutput=1
\usepackage[colorlinks=true,linkcolor=blue, citecolor=blue, urlcolor=blue, bookmarks]{hyperref}
\usepackage{epsfig}
\usepackage{soul}
\usepackage{enumerate}
\usepackage{amsmath,amssymb,mathtools,tensor,braket,empheq, tikz}
\usepackage{color,tikz,theorem}

\newcommand{\be}{\begin{equation}}
\newcommand{\ee}{\end{equation}}
\newcommand{\bea}{\begin{eqnarray}}
\newcommand{\eea}{\end{eqnarray}}

\definecolor{violet}{rgb}{0.62,0,1}
\definecolor{lightblue}{rgb}{0.12,0.56,1}
\definecolor{green}{rgb}{0.13,0.55,0.13}
\definecolor{myred}{RGB}{202,52,51}
\definecolor{myblue}{RGB}{101,147,245}

\definecolor{violet}{rgb}{0.62,0,1}
\definecolor{lightblue}{rgb}{0.62,0,1}
\definecolor{green}{rgb}{0.13,0.55,0.13}

\newcommand{\1}{\hat{ 1}}

\newcommand{\limth}{{\textstyle \lim_{th}}}
\newcommand*\circled[1]{\tikz[baseline=(char.base)]{
            \node[shape=circle,draw,inner sep=.5pt] (char) {#1};}}
\newcommand{\limsc}{{\textstyle \lim_{sc,\zeta}}}

\def\eqref#1{(\ref{#1})}

\usepackage{tcolorbox}

\begin{document}

\title{Generalized-Hydrodynamic approach to Inhomogeneous Quenches:\\
	Correlations, Entanglement and Quantum Effects}

\author{Vincenzo Alba}
\affiliation{Institute for Theoretical Physics, Universiteit van Amsterdam, Science Park 904, Postbus 94485, 1098 XH Amsterdam, The Netherland}
\author{Bruno Bertini}
\affiliation{Rudolf Peierls Centre for Theoretical Physics, Clarendon Laboratory, Oxford University, Parks Road, Oxford OX1 3PU, United Kingdom}
\author{Maurizio Fagotti}
\affiliation{Universit\'e Paris-Saclay, CNRS, LPTMS, 91405, Orsay, France}
\author{Lorenzo Piroli}
\affiliation{Max-Planck-Institut f\"ur Quantenoptik, Hans-Kopfermann-Str.~1, 85748 Garching, Germany}
\author{Paola Ruggiero}
\affiliation{Department  of  Quantum  Matter  Physics,  University  of  Geneva,24  Quai  Ernest-Ansermet,  CH-1211  Geneva,  Switzerland}

\begin{abstract}
We give a pedagogical introduction to the Generalized Hydrodynamic approach to inhomogeneous quenches in integrable many-body quantum systems. We review recent applications of the theory, focusing in particular on two classes of problems: bipartitioning protocols and trap quenches, which represent two prototypical examples of broken translational symmetry in either the system initial state or post-quench Hamiltonian. We report on exact results that have been obtained for generic time-dependent correlation functions and entanglement evolution, and discuss in detail the range of applicability of the theory. Finally, we present some open questions and suggest perspectives on possible future directions.

\end{abstract}

\maketitle

\tableofcontents

\section{Introduction}
\label{sec:introduction}

If an isolated many-body quantum system is prepared in a pure state and left to evolve unitarily, its state remains pure at all times, eventually experiencing quantum recurrence and revivals. This is, however, no longer the case if one looks at the density matrix reduced to finite regions of space. The latter can be driven to a stationary state by the global unitary dynamics, provided that the whole system is taken to be infinitely large. This simple observation has led, over the past two decades, to the development of a rich literature aiming at understanding concepts of local relaxation and thermalisation in isolated systems, based on first principle investigations~\cite{polkovnikov2011colloquium,polkovnikov2011colloquium,eisert2015quantum,gogolin2016equilibration,dalessio2016quantum}. This effort has been triggered in no small part by recent revolutionary experiments on trapped ultracold atomic gases~\cite{bloch2008many,cazalilla2011one,greiner2002collapse,kinoshita2006quantum,hofferberth2007non,haller2009realization,trotzky2012probing,cheneau2012light,gring2012relaxation,schneider2012fermionic,fukuhara2013microscopic,pagano2014one,langen2015experimental,jepsen2020spin,schweigler2021decay, ronzheimer2013expansion, xia2015quantum, choi2016exploring}, where an unprecedented degree of isolation and control can be achieved.

An early but central realization of these studies has been that the constrains imposed by conservation laws with local spatial density bring about qualitative differences in the relaxation process. Indeed, while in the generic case one observes local \emph{thermalisation}~\cite{deutschQuantum1991,srednicki1994chaos,rigol2008thermalization,cazalilla2010focus}, in \emph{integrable} systems --- characterised by an extensive number of such ``local'' conservation laws --- the stationary state is described by a non-thermal statistical ensemble, the so-called Generalized Gibbs Ensemble (GGE)~\cite{rigol2007relaxation}. The exceptional non-equilibrium features of integrable systems can be traced back to the fact that their \emph{entire spectrum} can be described in terms of \emph{stable} quasiparticles. This represents an invaluable aide for the theoretical description and lead for instance to the exact determination of spectrum and scattering matrix in such systems~\cite{korepin1997quantum, takahashi2005thermodynamics}.  

Given the complexity of non-equilibrium many-body phenomena, theoretical research initially focused on simplified protocols where the essential physics can be revealed. The most famous example in the recent literature is certainly the \emph{quantum quench}~\cite{calabrese2006time,calabrese2007quantum}:\ a system is prepared in a homogeneous initial state and left to evolve unitarily under a local Hamiltonian. Even in this idealised setting, it was not obvious whether integrability could lead to exact results in the presence of interactions. Indeed, although their spectrum is known, the structure of the eigenstates of interacting integrable systems is notoriously complicated~\cite{korepin1997quantum}, placing most of the quantities of interest out of the reach of direct approaches. In fact, the development of analytic methods to provide non-trivial predictions on the post-quench dynamics has been an important achievement of recent research~\cite{calabrese2016introduction}. 

An immediate obstacle is that GGEs are very hard to handle in the presence of interactions, essentially because they incorporate an extensive number of constraints. This issue, which at first seriously jeopardised the very utility of GGEs, has been resolved by introducing more practical alternative representations for these statistical ensembles~\cite{fagotti2013reduced, cassidy2011generalized, caux2013time}. In particular, a convenient one is the \emph{generalized microcanonical} description~\cite{cassidy2011generalized, caux2013time}. This approach is based on the principle of equivalence of statistical ensembles and consists in representing the GGE using a \emph{single}, appropriately chosen, eigenstate of the Hamiltonian, which is commonly referred to as the \emph{representative eigenstate}. This allows one to compute all expectation values in terms of the quasi-momentum distribution function of the stable quasiparticles in the representative eigenstate. The emerging physical picture is thus reminiscent of the standard statistical-mechanical description of non-interacting quantum gases at thermal equilibrium, whose thermodynamic information is fully encoded in the momentum distribution --- or occupation numbers --- of the free modes. In translational invariant systems this approach can be used to determine exactly the stationary values of relevant observables even in the presence of non-trivial interactions~\cite{caux2013time, caux2016quench, ilievski2016string}. 

Translational invariance, however, is more than a mere technical assumption and lies at the very heart of the aforementioned theoretical framework:\ the very qualitative fact that local subsystems relax to stationary states hinges on the presence of translational symmetry. On the other hand, in many interesting physical settings these hypotheses are necessarily violated. For example this happens in cold-atom experiments, due to finite number of particles and the presence of confining potentials, or in general in all settings exhibiting a non-trivial transport of conserved charges. The theory of Generalized Hydrodynamics (GHD) was originally introduced precisely to address the latter issue, namely to extend the study of quenches in integrable systems to inhomogeneous settings~\cite{bertini2016transport,castro-alvaredo2016emergent}. GHD is based on the realisation that when translational symmetry is broken, we can still achieve an efficient late-time description of local subsystems in terms of space-time-dependent quasi-stationary states. These states can once again be characterised microcanonically in terms of their quasiparticle quasi-momentum distribution functions. In its basic formulation the GHD approach is valid at the hydrodynamic scale, but it can be extended to include genuine ``quantum'' effects, such as the spreading of entanglement and quantum correlations. Furthermore, differently from Luttinger-liquid hydrodynamics, it is not restricted to low energies.

Since its introduction, GHD has experienced a rapid development, already partially covered in a review~\cite{bertini2020finitetemperature} and a set of lecture notes~\cite{doyon2020lectures}, proving to be a very versatile and powerful tool with many applications (most of which are discussed in other contributions to this volume), and being eventually experimentally verified~\cite{schemmer2019generalized,malvania2020generalized} (see in particular the review by Bouchoule and Dubail in this volume). The aim of this manuscript, which sets it apart from the other aforementioned surveys, is to provide a self-contained  pedagogical presentation of the theory, with a special focus on the study of inhomogeneous quantum quenches. In particular, we will consider two prototypical quantum quenches where translational symmetry is broken by either the initial state (``bipartitioning protocols'') or the post-quench Hamiltonian (``trap quenches''), and review recent results in these two important types of problems. We point out that particular types of these quenches are also treatable by other means (see, e.g., Refs.~\cite{vidmar2017emergent, vidmar2017emergent2, zhang2021emergent}). Moreover, there exist inhomogeneous quantum quenches after which observables exhibit behaviours qualitatively similar to the ones considered here but that cannot be addressed using generalised hydrodynamics~\cite{eisler2020front,gruber2021entanglement,zauner2015time,eisler2016universal,eisler2018hydrodynamical}. Both these aspects will not be covered in this review.

The rest of this article is organised as follows. We begin in Sec.~\ref{sec:GHD_intro} by introducing the general formalism of GHD. There we present a pedagogical discussion focusing on a simple non-interacting model, which we use to highlight the main conceptual and formal points. We first review the case of a homogeneous quench (Sec.~\ref{sec:homogeneous_quench}) and show how one can build on its solution to treat the simplest case of inhomogeneous quench dynamics: the bipartitioning protocol, i.e., the sudden junction of different homogeneous states (Sec.~\ref{sec:bipartitioning}). More general inhomogeneous quenches are discussed in Sec.~\ref{sec:inhomogeneous}, while in Sec.~\ref{sec:interactions} we discuss the changes occurring in the presence of interactions. We continue with Sec.~\ref{sec:localphysics}, which contains a review of several works where GHD has been applied to the study of bipartitioning protocols, focusing in particular on the main physical implications of the results. Next, Sec.~\ref{sec:v} is devoted to the dynamics of quantum entanglement, while Sec.~\ref{sec:vPR} focuses on recent developments to capture quantum fluctuations around a inhomogeneous background (namely, to compute generic time-dependent correlation functions for a particular class of states). Sec.~\ref{sec:MF} presents a discussion about genuinely quantum effects in inhomogeneous quenches and how GHD can be extended to account for them. Finally, Sec.~\ref{sec:conclusions} contains our conclusions and a brief discussion of some of the open questions.

\section{GHD approach to Inhomogeneous Quenches}
\label{sec:GHD_intro}

In this section we provide a self-contained introduction to the main concepts of GHD, seen as a method to characterise the late-time non-equilibrium dynamics of inhomogeneous integrable systems. Since the latter relies heavily upon the formalism developed to treat the homogeneous case, we begin by considering homogeneous quenches and briefly review some of the aspects that are directly relevant for the inhomogeneous generalisation (for a comprehensive review we refer the reader to  Ref.~\cite{essler2016quench}). Before doing that, however, we introduce a simple system of non-interacting fermions which we will use as a paradigm. In Secs.~\ref{sec:homogeneous_quench}--\ref{sec:inhomogeneous} we use this system to exemplify the main features of GHD. The appropriate modifications of the treatment to account for interacting integrable interactions are discussed in Sec.~\ref{sec:interactions}. 

\subsection{A simple integrable model}
\label{sec:toy-model}

Let us consider a system of spinless fermions on the lattice described by the following pairing Hamiltonian, with coupling $J$ and chemical potential $h$ 
\be
\label{eq:FP_hamiltonian}
H(h)=-J \sum_{x=-L/2+1}^{L/2} (c^\dag_x c^{\phantom{\dag}}_{x+1}+c^\dag_{x+1} c^{\phantom{\dag}}_{x}+c^\dag_x c^{{\dag}}_{x+1}+c_{x+1} c_{x}+ 2 h c^\dag_{x} c^{\phantom{\dag}}_{x})+2 J h L \,.
\ee
Here we set the lattice spacing $a$ to one, we assumed the number of sites $L$ to be even, we denoted by $c^{\dag}_x, c^{\phantom{\dag}}_x$ a set of fermionic creation and annihilation operators fulfilling the canonical anti-commutation relations
\be
\label{eq:CAR}
\{c^\dag_x,c^\dag_y\}=0=\{c^{\phantom{\dag}}_x,c^{\phantom{\dag}}_y\},\qquad\qquad \{c^\dag_x,c_y\}=\delta_{x,y},
\ee
and we assumed periodic boundary conditions {$(c^\dag_{L+x}=c^\dag_{x})$. Note that in the above equations we set $\hbar=1$: we will abide by this convention throughout the entire review except for Sec.~\ref{sec:MF}, where  the $\hbar$ dependence will be restored for pedagogical reasons.

The Hamiltonian \eqref{eq:FP_hamiltonian} is mapped onto a particular spin-1/2 chain (the celebrated Transverse Field Ising Model) through a Jordan-Wigner transformation~\cite{sachdev2007quantum}. The mapping from fermions to spins, however, is non-local and forces one to distinguish between two sectors that differ in the boundary conditions. Specifically, periodic boundary conditions for the fermions are mapped into periodic or anti-periodic boundary conditions for the spins and vice versa. This requires a slightly more refined analysis but does not change the main results that we are going to present, which are independent of the boundary conditions. To avoid this inessential complication we stick to the fermionic form \eqref{eq:FP_hamiltonian}. 

Since the Hamiltonian \eqref{eq:FP_hamiltonian} is a \emph{quadratic form} of fermionic operators and in addition is \emph{translational invariant}, it is directly diagonalised by a combination of Fourier and Bogoliubov transformations. Namely, it can be written as 
\be
H(h)=2 J \sum_{k} \varepsilon(k) \left(b_{k}^{\dagger} b_{k}^{\phantom{\dag}}-\frac{1}{2}\right)\,.
\label{eq:diagonal_form}
\ee
Here the sum runs over a discrete, or ``quantized", set of rational multiples of $\pi$ 
\be
\sum_{k} f(k) = \sum_{j=1}^{L} f\Bigl(\frac{2 \pi}{L} j -\pi\Bigr),
\ee
and we introduced the ``dispersion relation" 
\be\label{eq:disp_rel}
\varepsilon(k)=\sqrt{1+h^{2}-2 h \cos k},
\ee
and the ``Bogoliubov fermions''
\be
b_0 = \frac{1}{\sqrt L}\sum_{x=1}^L c^\dag_x,\qquad\qquad 
b_{k\neq0}  =
\frac{1}{2 \sqrt L}\sum_{x=1}^L e^{i k x} e^{i \Theta_h(k)/2}  (c_x-c_x^\dag)+ e^{i k x} e^{- i \Theta_h(k)/2}  (c_x+c_x^\dag).
\ee
Here 
\be
e^{i \Theta_h(k)}=\frac{h-e^{ik}}{\varepsilon(k)}\,,
\label{eq:bogoliubov}
\ee
is commonly referred to as the Bogoliubov angle. Note that $b^{\dag}_k, b^{\phantom{\dag}}_p$ fulfil the canonical commutation relations~\eqref{eq:CAR} with $x$ and $y$ replaced by $k$ and $p$.  We also stress that the quantization conditions for the quasi-momenta appearing in the sum \eqref{eq:diagonal_form}, i.e.,   
\be\label{eq:free_quantization}
e^{i L k_j}=1
\ee
do not couple different momenta.

From the representation \eqref{eq:diagonal_form} we see that the Hamiltonian is \emph{non-interacting}: it is written as an independent sum of \emph{mode operators}
\be
\label{eq:modes}
n_k = b_{k}^{\dagger} b_{k}^{\phantom{\dag}},\qquad\qquad [n_k,n_p]=0\,,
\ee
describing the occupation of a single Bogoliubov mode (a mode $k$ has energy $\varepsilon(k)$). This means that the eigenstates of the Hamiltonian are readily written as 
\be
\ket{\{k_j\}_{j=1}^N}=b^\dagger_{k_1}\cdots b^\dagger_{k_N}\ket{\Omega}\,,
\ee
where $\ket{\Omega}$ is the vacuum state such that $b_{k}\ket{\Omega}=0$ for all $k$.  

Let us now proceed to illustrate three key concepts that will play an important role in the upcoming discussion.

\subsubsection{Conserved charges}
The Hamiltonian~\eqref{eq:FP_hamiltonian} commutes with the following set of operators 
\be
Q_{n}= \sum_{k} q_{n}(k) \left(n_k-\frac{1}{2}\right)\qquad n=0,1,2,\ldots,L-1,
\label{eq:Charges}
\ee
where $q_{n}(k)$ are called ``bare charges'' or ``single-particle eigenvalues'' and are defined as
\be
q_{2n}(k)=2J \varepsilon(k)\cos(n k),\qquad\qquad q_{2n+1}(k)=2J  \sin( (n+1) k)\,, \qquad\qquad n=0,1,2,\ldots.
\ee
One can directly see that $Q_{n}$ commute with the Hamiltonian because they are  linear combinations of the mode operators (note in particular that $Q_{0}=H$). These conserved operators, or ``charges", have the following three key properties: 
\begin{itemize}
\item[(i)] They can all be written as sums of densities that are \emph{local in space}, i.e., act non-trivially only on a finite number of neighbouring sites. Specifically, we have~\cite{prosen2000exact, fagotti2013reduced}
\be
Q_{n} = \sum_{x=1}^L q_{n,x} +\text{const}\qquad n=0,1,2,\ldots,L-1,
\label{eq:Charges2}
\ee
with~\cite{Note1, Note3}\footnotetext[1]{Note that there is some freedom in the definition of $q_{n,x}$, as it is always possible to add to them an operator that can be chosen as density of the trivial charge $0$ (e.g., $\sigma_n^x-\sigma_{n-1}^x$).} \footnotetext[3]{Note that we chose the value of the constant in \eqref{eq:densityp} such that $q_{2n,x}$ has expectation value zero on the vacuum $\ket{\Omega}$.}   
\begin{align}
&\!\!\!\!\!\!\!\!\!q_{2m,x} = - \frac{J}{2} [(c^\dag_x-c_x) (c^\dag_{x+ m+1}+c^{\phantom{\dag}}_{x+ m+1}+c^\dag_{x- m+1}+c^{\phantom{\dag}}_{x-m+1})- h (c^\dag_x-c_x) (c^\dag_{x+ m}+c^{\phantom{\dag}}_{x+ m}+c^\dag_{x- m}+c^{\phantom{\dag}}_{x-m})],\label{eq:densityp}\\
&\!\!\!\!\!\!\!\!\!q_{2m+1,x}= - J i  (c^\dag_x c^{\phantom{\dag}}_{x+m+1} - c^\dag_{x+m+1} c^{\phantom{\dag}}_x)\,.\label{eq:densitym}  
\end{align}
Conserved operators with this property are called \emph{local conservation laws} (or \emph{local charges}).
\item[(ii)] Their number is \emph{extensive}, i.e., it is proportional to the volume $L$. 
\item[(iii)] They are linearly independent.
\end{itemize}
These are general features of integrable models. In fact, possessing an extensive set of independent local conserved charges can be considered the defining feature of an integrable system, although, in some cases, the notion of independence can be non-obvious, see, e.g., Ref.~\cite{lychkovskiy2020independence}.

Let us now consider the commutator between $q_{n,x}$ and the Hamiltonian. Since $q_{n,x}$ is local and $H$ has a local density, the commutator is also local. Moreover its sum over the entire chain gives $[ H, Q_{n}]=0$. Therefore, there must exist a \emph{local} operator $j_{n,x}$ such that
\be
[ H, q_{n,x}]= i (j_{n,x}-j_{n,x-1}),
\label{eq:comHdens}
\ee
where we included $i$ in order for $j_{n,x}$ to be Hermitian. The local operator $j_{n,x}$ is called  \emph{current operator} associated with the charge $Q_n$ and is completely specified by \eqref{eq:comHdens} and the condition $\braket{\Omega| j_{n,x}|\Omega}=0$. Note that  \eqref{eq:comHdens} is nothing but a continuity equation and can be brought in the standard form by adopting the Heisenberg picture 
\be
\partial_t q_{n,x}(t)=i[ H, q_{n,x}(t) ]= j_{n,x-1}(t)-j_{n,x}(t)\,,
\label{eq:continuityeq}
\ee
where we used the standard definition of Heisenberg operators $q_{n,x}(t)=e^{iHt}q_{n,x}e^{-iHt}$. As  evident from \eqref{eq:comHdens}, the current does not simply depend on the charge, but on the very definition of its density. What is probably less intuitive is that even the total current $J_{n}=\sum_x j_{n,x}$, which is the sum over the entire chain of the current operator, depends on the definition of the charge density. In particular, our definitions of charge densities~\eqref{eq:densityp}, \eqref{eq:densitym} result in the following total currents~\cite{fagotti2016charges, Note2}\footnotetext[2]{
Note that, while the total currents of the reflection symmetric charges are conserved, the total currents of the odd charges have off-diagonal contributions. Again, this is just a consequence of our conventions: Ref.~\cite{fagotti2020locally} showed indeed that, for non-interacting systems, the densities can be redefined so as to make all total currents conserved.}
\begin{align}
J_{2n} &= \frac{ J h }{2}(Q_{2n+1}-Q_{2n-3}),\label{eq:currentp}\\
J_{2n+1} & =\frac{{ J}^2}{4} \sum_x\Bigl[ (c^{\phantom{\dag}}_x-c^\dag_x)(c^{\phantom{\dag}}_{x+n+2}+c^\dag_{x+n+2})+(c^{\phantom{\dag}}_x+c^\dag_x)(c^{\phantom{\dag}}_{x+n}-c^\dag_{x+n})\Bigr]\,.\label{eq:currentm}  
\end{align}

We stress that the above considerations apply to all integrable models regardless of interactions. In particular, since there is always an extensive number of local conservation laws, we have an extensive number of operatorial continuity equations \eqref{eq:continuityeq}. The main difference encountered in the interacting case is that  one cannot find explicit expressions like \eqref{eq:densityp}--\eqref{eq:densitym} and \eqref{eq:currentp}--\eqref{eq:currentm} for generic charge densities and currents.   
 
\subsubsection{Thermodynamic description of expectation values}
\label{sec:thermo}

Another key property of integrable models is that the expectation values in their eigenstates are expressed in terms of the ``momentum distribution" of the corresponding set of quasiparticles. To understand what this means in our simple example let us look at the following expectation value  
\be
\braket{\{k_j\}_{j=1}^N|c^\dag_{x+\ell} c^{\phantom{\dag}}_{x} |\{k_j\}_{j=1}^N} = \frac{1}{L} \sum_{j=1}^N e^{- i k_j \ell} \cos^2(\Theta_h({k_j})/2)-e^{ i k_j \ell} \sin^2(\Theta_h({k_j})/2) + \frac{1}{L}  \sum_{k} e^{ i k \ell} \sin^2(\Theta_h({k})/2)\,,
\ee
where the sum in the first term of the r.h.s.\ runs over the momenta of the eigenstate $\ket{\{k_j\}_{j=1}^N}$. Considering now the thermodynamic limit
\be
\limth:\qquad L,N\to\infty\qquad \text{with}\qquad d=\frac{N}{L}\quad \text{fixed},
\ee
at fixed $\ell$ we find 
\be
\limth \braket{\{k_j\}_{j=1}^N|c^\dag_{x+\ell} c^{\phantom{\dag}}_{x} |\{k_j\}_{j=1}^N} = \int_{-\pi}^{\pi} \!\!{\rm d}k\, \left\{e^{- i k \ell} \cos^2(\Theta_h({k})/2) \rho(k) + e^{ i k \ell} \sin^2(\Theta_h({k})/2)) (1-\rho(k))\right\} \,,
\label{eq:limEV}
\ee
where $\rho(k)$ is the momentum distribution of the quasiparticles, also known as ``root density" in the literature of integrable models, which appears as the weight of the integrals replacing the sums in the thermodynamic limit
\be
\limth\frac{1}{L}\sum_{j=1}^{N} f(k_j)=\int\mathrm d k\,\,\rho(k)f(k)\,.
\ee
Note that some authors define the root density as the limit of a sequence 
\be
\rho(k) = \limth \frac{1}{L(k_{j+1}-k_{j})}\qquad k\in(k_{j},k_{j+1})\,.
\ee
From the definition of $\rho(k)$, up to $O(L^0)$, we have
\be
2\pi L \rho(k) \Delta k = \#\text{ of momenta in $[k,k+\Delta k]$ present in the state $\ket{\{k_j\}_{j=1}^N}$}.
\ee

Eq.~\eqref{eq:limEV} proves that, in the thermodynamic limit, the expectation value of $c^\dag_{x+\ell} c^{\phantom{\dag}}_{x}$ in the eigenstate $\ket{\{k_j\}_{j=1}^N}$ depends \emph{only} on the state's root density:\ no other property of the eigenstate needs to be retained. A completely analogous reasoning can be applied to the expectation value of all other quadratic combinations of the fermionic operators in $\ket{\{k_j\}_{j=1}^N}$ (provided that their distance is finite in the limit). Moreover, using that the eigenstates fulfil Wick's Theorem for the fermions $c^\dag_{x}, c^{\phantom{\dag}}_{x}$, this statement is extended to all local operators. 

Importantly, the correspondence between eigenstates and root densities is not one-to-one. In fact, exponentially many (in the volume) eigenstates of the Hamiltonian correspond to the same root density. This can be intuitively understood by noting that small changes to the distribution $\{k_j\}_{j=1}^N$ do not change the root density. More precisely by counting all such ineffective changes one finds that the number of eigenstates corresponding to a given $\rho(k)$ is $\approx e^{L S_{YY}[\rho]}$~\cite{yang1969thermodynamics}, where we introduced the so-called Yang-Yang entropy 
\be
\label{eq:S-YY-free}
S_{YY}[\rho]=\int_{-\pi}^{\pi} \!\!{\rm d}k\, \left\{\rho_t(k)\log \rho_t(k)-\rho(k)\log \rho(k)+ \left(\rho_t(k)-\rho(k)\right)\log\left(\rho_t(k)-\rho(k)\right)\right\},
\ee
and  
\be 
\rho_t(k)=\frac{1}{2\pi}
\label{eq:freerhot}
\ee
is the total density, namely the density of ``vacancies" (or slots) that can or not be occupied by the momenta~\cite{takahashi2005thermodynamics, korepin1997quantum}.

\subsubsection{Root density from the charges}

The last key property that we want exemplify here is the so-called ``string-charge duality''~\cite{ilievski2016string}. Namely, the fact that there is a one-to-one correspondence between the root density introduced in the previous section and the expectation values of all charge densities in the thermodynamic limit. 

One direction is straightforward: since the charge densities are local operators, the thermodynamic limit of their expectation values can be expressed in terms of the root density. In particular, we have 
\begin{align}
 \braket{q_{n}}_{\!\rho}\equiv\limth \braket{\{k_j\}_{j=1}^N|q_{n}|\{k_j\}_{j=1}^N} = \int_{-\pi}^{\pi} \!\!{\rm d}k\, q_n(k) \rho(k), \qquad\qquad n=0,1,2,\ldots. \label{eq:chargesrho}
\end{align}
Note that \eqref{eq:chargesrho} has a very simple kinetic theoretical interpretation in terms of the quasiparticles. To find the density of the $n$-th conserved charge one considers the contribution of a particle with momentum $k$ ($q_n(k)$) times the number of particles of momentum in $[k,k+dk]$ divided by $L$ ($\rho(k) {\rm d}k $) and sums over all allowed values of $k$. 

The key point here is that \eqref{eq:chargesrho} can be inverted. Namely, one can use it to find $\rho(k)$ in terms of $\{\braket{q_{n}}_{\!\rho}\}_{n=0,1,\ldots}$. This is a straightforward consequence of the fact that $\{q_n(k)\}_{n=0,1,\ldots}$ is a complete (but not orthogonal) set of functions in $L^2([-\pi,\pi])$. Explicitly we have the following Fourier series expression for the root density  
\be
\rho(k)=  \frac{1}{4\pi J \varepsilon(k)} \braket{q_{0}}_{\!\rho}+\frac{1}{2\pi J \varepsilon(k)}\sum_{n=1}^\infty \braket{q_{2n}}_{\!\rho} \cos(n k)+\frac{1}{2\pi J}\sum_{n=1}^\infty \braket{q_{2n-1}}_{\!\rho} \sin(n k)\,.
\label{eq:rhocharges}
\ee 
Note that the string-charge duality implies that the root density can be written as the expectation of an operator written as an (infinite) linear combination of charge densities. In the non-interacting case one can re-sum the series and obtain 
\be
\rho(k) = \lim_{\mu\rightarrow 0}\limth\braket{\{k_j\}_{j=1}^N|n_\mu(k)|\{k_j\}_{j=1}^N}\,,
\ee
with  
\be\label{eq:rhoop}
n_\mu(k)= \frac{1}{L}\sum_{k'} g_\mu(k-k') n_{k'}\,,
\ee 
and $g_\mu(x)$ is a smooth approximation of the periodic Dirac delta function, i.e. 
\be
\lim_{\mu\to0}g_\mu(x) = \sum_{n=-\infty}^{\infty} \delta(x+2\pi n)\,.
\ee

\subsection{Stationary states after homogeneous quenches}
\label{sec:homogeneous_quench}

Having introduced our toy model \eqref{eq:FP_hamiltonian}, we can now move to consider homogeneous quench problems.  Specifically, let us focus on the following protocol
\begin{itemize}
\item[(i)] Prepare the system in its ground state $\ket{\Psi_0}=\ket{{\rm GS}(h_{0})}$ for a given value $h_0$ of the chemical potential;
\item[(ii)] At time $t=0$ (suddenly) change the chemical potential to $h\neq h_0$;
\end{itemize}
 This means that for $t>0$ the system is no longer in equilibrium and undergoes non-trivial evolution. A crucial point for the following discussion is that both $\ket{\Psi_0}$ and $H(h)$ (and hence also $\ket{\Psi(t)}$) are translationally invariant. 
 
A natural question is whether, after the quench, the system can eventually go back to an equilibrium state, i.e., whether it can \emph{relax}. It is easy to understand that this cannot happen for the system as a whole. Indeed, relaxation is naturally associated with loss of information while the evolution of the system is purely unitary and retains all the information (probabilities are conserved). Nevertheless, relaxation can still be observed considering finite subsystems in the thermodynamic limit because the system can act as an effective bath on its own finite parts. 

In order to make this intuition quantitative, we probe the physics of local subsystems by looking at the expectation values of local observables. More precisely, given a local operator $\mathcal{O}_x$, we consider
\be
\lim_{t\to\infty}\!\limth \langle\Psi_0|\mathcal{O}_x(t)| \Psi_0\rangle=\lim_{t\to\infty}\!\limth\langle\Psi_0|\mathcal{O}_0(t)| \Psi_0\rangle =  \lim_{t\to\infty}\limth \!\sum_{n, m} e^{-i \left(E_n-E_m\right) t}\braket{\Psi_0| m}\braket{n|\Psi_0}\left\langle m|\mathcal{O}_0| n\right\rangle
\label{eq:infinitetime}
\ee
where we denoted by $\ket{n}$ the eigenstates of the Hamiltonian, with associated energy eigenvalues $E_{n}$. If the limit \eqref{eq:infinitetime} exists for any local observable $\mathcal{O}_x$, then we say that the system relaxes locally~\cite{essler2016quench}. In this case one can find a stationary state $\hat \rho_s$ of $H$ (i.e. $[H,\hat{\rho}_s]=0$) such that  
\be
\lim_{t\to\infty}\limth\langle\Psi_0|\mathcal{O}_x(t)| \Psi_0\rangle = \limth  {\rm tr}[\mathcal{O}_0 \hat \rho_{s}],
\label{eq:homoSS}
\ee
for every local operator $\mathcal{O}_x$. In fact, condition \eqref{eq:homoSS} does not specify a state uniquely and can be fulfilled by many different stationary states of $H$. However, all states fulfilling \eqref{eq:homoSS} are equivalent when reduced to finite subsystems in the thermodynamic limit. Therefore they  can be regarded as different \emph{representations} of the same stationary state, very much like the different ensembles of standard statistical mechanics~\cite{essler2016quench}.

For the quench problem under examination local relaxation can be explicitly proven. Indeed, one can find a simple integral representation for $\limth\langle\Psi_0|\mathcal{O}_0(t)| \Psi_0\rangle$ and evaluate the limit of infinite times~\cite{calabrese2012quantuma, calabrese2012quantumb}. This can be done every time that both the Hamiltonian before ($H(h_0)$) and after ($H(h)$) the quench are quadratic in the same variables. Conversely, this is generically  impossible in the presence of interactions. Nevertheless, assuming local relaxation, one can generically find a representation of the stationary state \emph{without solving the dynamics}. This can be done in the following two steps
\begin{itemize}
\item[\circled{H1}] Assume that \eqref{eq:homoSS} holds for some stationary state $\hat \rho_s$. 
\item[\circled{H2}] Fix $\hat \rho_s$ by imposing the conservation of the expectation value of all charge densities. Indeed, since $\ket{\Psi_0}$ is translational invariant, taking the expectation value of \eqref{eq:continuityeq} we have 
\be
\partial_t \limth\braket{\Psi_0|q_{n,0}(t)| \Psi_0} = 0\,,\qquad\qquad n=0,1,2,\ldots, 
\ee
which implies  
\be
\limth \braket{\Psi_0|q_{n,0}(0)| \Psi_0}= \lim_{t\to\infty}\limth\braket{\Psi_0|q_{n,0}(t)| \Psi_0}= \limth   {\rm tr}[q_{n,0} \hat \rho_{s}]\,,\qquad\qquad n=0,1,2,\ldots. 
\label{eq:fixeddensities}
\ee
\end{itemize}

Let us now show that \circled{H1} and \circled{H2} are indeed sufficient to fix the stationary state reached after the quench (i)--(ii). We begin by following Refs.~\cite{cassidy2011generalized, caux2013time} and representing $\hat \rho_s$ microcanonically. Namely we write $\hat \rho_s=\ket{\Psi_s}\!\bra{\Psi_s}$ where  $\ket{\Psi_s}$ is a judiciously-chosen eigenstate of the Hamiltonian. The expectation values in the thermodynamic limit are then fully characterised by a root density $\rho_s(k)$ and from \eqref{eq:fixeddensities} we have 
\be
\int_{-\pi}^{\pi} \!\!{\rm d}k\, q_n(k) \rho_s(k) = \limth \braket{\Psi_0|q_{n,0}| \Psi_0}\,,\qquad\qquad n=0,1,2,\ldots,
\label{eq:rhoscharges}
\ee
which is readily inverted using \eqref{eq:rhocharges}. To find an explicit solution we then only have to 
evaluate the ``initial data" $\limth \braket{\Psi_0|q_{n,0}| \Psi_0}$. In our case this is easily done by expressing $\ket{\Psi_0}$ in terms of the Bogoliubov fermions of the Hamiltonian $H(h)$ (see, e.g., Ref.~\cite{essler2016quench}) and explicitly evaluating the expectation values of \eqref{eq:densityp} and \eqref{eq:densitym}. The result reads 
\be
\limth \braket{\Psi_0|q_{n,0}| \Psi_0} = \int_{-\pi}^{\pi} \!\!{\rm d}k\, q_n(k)  \frac{1-\cos \Delta_{h,h_0}(k)}{4 \pi}\,,
\label{eq:initialdata}
\ee
where (cf.~\eqref{eq:bogoliubov}) 
\be
e^{i \Delta_{h,h_0}(k)} =  e^{i \Theta_h(k)}e^{-i\Theta_{h_0}(k)}\,.
\ee
In fact, in our case we do not need to re-sum \eqref{eq:rhocharges}. Noting that \eqref{eq:initialdata} takes the same form as the l.h.s.\ of \eqref{eq:rhoscharges} and recalling that $\{q_n(k)\}$ is a complete set we immediately conclude 
\be
\label{eq:saddle-point}
\int_{-\pi}^{\pi} \!\!{\rm d}k\, q_n(k)  \left(\rho_s(k)-\frac{1-\cos \Delta_{h,h_0}(k)}{4 \pi}\right)=0 \qquad \forall n \qquad \Rightarrow \qquad\rho_s(k)=\frac{1-\cos \Delta_{h,h_0}(k)}{4 \pi}\,.
\ee
One can explicitly verify that $\rho_s(k)$ agrees with the result found by explicit solution of the dynamics~\cite{calabrese2012quantuma, calabrese2012quantumb}. We point out that, even if we demonstrated it in a particular example, the above procedure to obtain the stationary state from the assumptions \circled{H1} and \circled{H2} relies on the decoupled form of the charges \eqref{eq:Charges}, which are written as sums over independent modes. Therefore, it can be directly applied to all non-interacting systems (its generalisation to interacting integrable systems is described in Sec.~\ref{sec:interactions}).

From a physical point of view, the main message of the above discussion is that, after a given transient (which depends on the details of the initial states, and the Hamiltonian parameters), finite subsystems approach a stationary state that can be determined without solving the dynamics. Note this stationary state is generically non-thermal: see, e.g., the comparison in Fig.~\ref{fig:GGE} between \eqref{eq:saddle-point} and the root density of a thermal state with the same energy density. Importantly, translational symmetry (of post-quench Hamiltonian and initial state) implies that the stationary state is the same for all local subsystems independent of their spatial position. In the next section we will see how this description has to be modified when translational invariance is broken explicitly.

\begin{figure}
\centering
 \includegraphics[width=.5\textwidth]{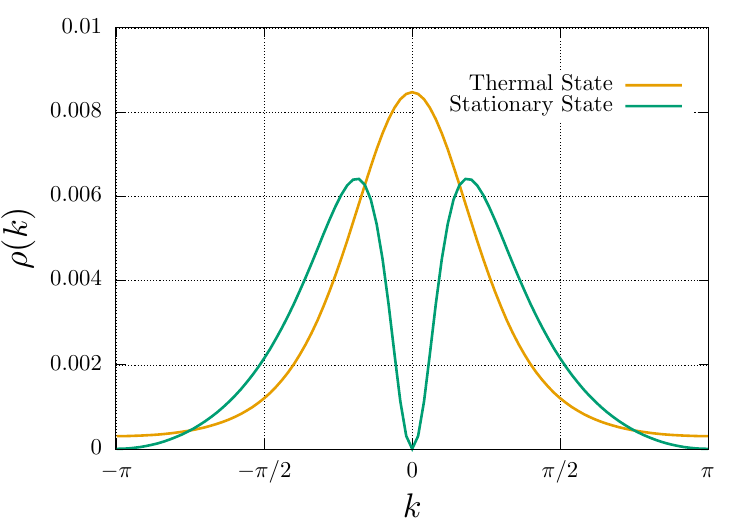}
 \caption{Momentum distribution in the stationary state after the quench $h_0=1.5\mapsto h=2.7$ versus the thermal one at the same energy density.}\label{fig:GGE}
 \end{figure}

\subsection{Stationarity along rays after bipartitioning protocols}
\label{sec:bipartitioning}

Let us now look at a different quench problem. We consider the case in which the system is initially separated in two parts, ``left" and ``right", each prepared in the ground state of the Hamiltonian for different values of the chemical potential, say $h_{\rm L}$ and $h_{\rm R}\neq h_{\rm L}$. Then, at time $t=0$, the two parts are joined together and left to evolve unitarily according to the Hamiltonian $H(h)$, with $h\neq h_{\rm R}\neq h_{\rm L}$ (see Fig.~\ref{fig:bipartitioning}). More precisely, we consider the following initial state 
\be
\ket{\Psi_0}=\ket{{\rm GS}_{\rm L}(h_{\rm L})}\otimes\ket{{\rm GS}_{\rm R}(h_{\rm R})}
\label{eq:bipartitioninginitial}
\ee
where $\ket{{\rm GS}_{\rm L}(h)}$ and $\ket{{\rm GS}_{\rm R}(h)}$ are respectively the ground states of 
\begin{align}
H_{\rm L}(h) =\sum_{x=1-L/2}^{-1} q^{(h)}_{0,x} \quad {\rm with} \quad    c_{-L/2}=0,
\qquad \qquad H_{\rm R}(h) =\sum_{x=1}^{L/2} q^{(h)}_{0,x} \quad {\rm with} \quad   c_{L/2+1}=0,
\end{align}
and $q_{0,x}^{(h)}$ is the energy density operator with field $h$ (cf.~\eqref{eq:densityp}).
Inhomogeneous quench problems of this kind, i.e.\ where the initial state is composed by the junction of two different homogeneous pieces, have been extensively studied in the literature. Of particular interest have been the sudden junction of two half chains prepared at different temperatures~\cite{nozawa2020generalized,ogata2002diffusion, aschbacher2003non, aschbacher2006out,bernard2012energy, deluca2013nonequilibrium, karrasch2013nonequilibrium, eisler2014entanglement, collura2014quantum, collura2014non, deluca2015stationary, bhaseen2015energy, doyon2015non, doyon2015lower, biella2016energy, castro2014thermodynamic, deluca2014energy,vasseur2015expansion, castro-alvaredo2016emergent, bertini2016transport, zotos2016a, kormos2017inhomogeneous, bertini2018low, mazza2018energy, mestyan2018spin,yoshimura2018full, bertini2019transport, karevski2019charge, bertini2018universal} and at different chemical potentials (or fillings)~\cite{misguich2017dynamics, ljubotina2017spin, santos2008transport, santos2009transport, santos2011domain, antal2008logarithmic, antal1998isotropic, antal1999transport, collura2018analytic, calabrese2008time, vidmar2017emergent, eisler2013full, alba2014entanglement, vidmar2015dynamical, sabetta2013nonequilibrium, viti2016inhomogeneous, bertini2016transport, piroli2017transport, eisler2016universal, deluca2017nonequilibrium, hauschild2015sudden, gobert2005real, collura2020domainwall,stephan2017return}, but more general initial states have also been considered~\cite{bertini2016transport,mendl2021highlow,jin2021interplay,torres2014quench,lancaster2010quantum,lancaster2010quenched,moosavi2020emergence,langmann2017time}, particularly in relation to studies on the entanglement growth (see Section~\ref{sec:v} for references).
Here we will refer to inhomogeneous quenches of this kind as ``bipartitioning protocols". 

\begin{figure}
\begin{tikzpicture}[scale = 1.5,radius=3.6cm, delta angle=30, opacity=1]
\def\sqr{1.414213}
\def\subdivisions{60}
\clip (-3.cm,-0.35cm) rectangle (3.4cm,\sqr*2.5cm);

\fill[fill=myred, opacity = 1 
] (-2.5cm,0) -- (0,0) -- (-\sqr*1.5cm,\sqr*1.5cm) -- (-2.5cm,\sqr*1.8cm);
\fill[fill=myblue, opacity = 1] (0,0) -- (2.5cm,0) -- (2.5cm,\sqr*1.8cm) -- (\sqr*1.5cm,\sqr*1.5cm);

\foreach \i[evaluate={\col=\i/(\subdivisions)*100}] in {0,...,\subdivisions}
\draw[myred!\col!myblue, fill=myred!\col!myblue] (0,0) -- (\i*90/\subdivisions+45:3.6cm) arc[start angle=\i*90/\subdivisions+45, end angle=(\i+1)*90/\subdivisions+45] -- cycle;

\shade[top color=white, bottom color = white] (-2.6cm,\sqr*1.8cm) rectangle (2.6cm,\sqr*2.8cm);
\shade[top color=white, bottom color = white] (-3.cm,0) rectangle (-2.4cm,\sqr*2.0cm);
\shade[top color=white, bottom color = white] (3.cm,0) rectangle (2.4cm,\sqr*2.0cm);

\draw [myred,line width = 0.8 mm] (-2.6,0) -- (0,0);
\draw [myblue,line width = 0.8 mm] (0,0) -- (2.6,0);

\node[font=\fontsize{12}{12}\selectfont] (c) at (3.2,2.55) {$t$};

\draw[line width =3, draw=black] (2.9,0) -- (3.06,0);
\draw[-latex, line width =1.5, draw=black] (2.98,0) -- (2.98,2.65);

\node[font=\fontsize{12}{12}\selectfont] (d) at (-1.5,0.2) {$\ket{{\rm GS}_{\rm L}(h_{\rm L})}$};
\node[font=\fontsize{12}{12}\selectfont] (e) at (1.5,0.2) {$\ket{{\rm GS}_{\rm R}(h_{\rm R})}$};

\shade[top color=white, bottom color = white] (-2.6cm,\sqr*1.8cm) rectangle (2.6cm,20cm);
\end{tikzpicture}
\caption{Pictorial representation of the bipartitioning protocol. After the sudden junction of two homogeneous halves a growing region around the junction is affected by the inhomogeneity. This region is contained in a light cone spreading from the junction at the maximal speed attainable in the system, i.e., the speed of the fastest quasiparticles.}\label{fig:bipartitioning}
\end{figure}
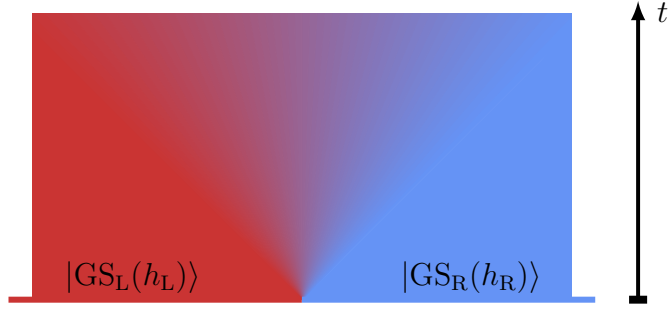

The time-evolving Hamiltonian~\eqref{eq:FP_hamiltonian} is still invariant under translations but, crucially, the initial state is not. This means that the expectation values $\limth \braket{\Psi(t)|\mathcal{O}_x| \Psi(t)}$ maintain a dependence on $x$ and we cannot directly use the strategy of the previous section to find their large-time limit. As we will see, however, the above strategy can be successfully modified at the expense of ``diagonalising" the continuity equations for a system with infinitely many conservation laws.  

Let us first consider point \circled{H1}. Since the expectation values retain a non-trivial dependence on the position we cannot assume \eqref{eq:homoSS}. At the same time it is still reasonable to expect finite subsystems to reach local equilibrium at large enough times. To conciliate these observations, one might think to impose a condition like \eqref{eq:homoSS} with an $x$-dependent stationary state. This idea, however, is too naive: an inhomogeneity in the stationary state inevitably produces dynamics, which makes the infinite-time limit ill defined. In fact, for the specific bipartite geometry of the initial state, true stationarity can still emerge along light cones or``rays", namely for observables moving away from the junction at fixed speed. This can be established by analytic calculations in simple cases (see, e.g.,~\cite{bertini2016determination, kormos2017temperature, viti2016inhomogeneous, sotiriadis2020nonequilibrium}) and by direct numerical calculations. Formally, we have
\be
\limsc\limth\langle\Psi_0|\mathcal{O}_x(t)| \Psi_0\rangle = \limth  {\rm tr}[\mathcal{O}_0 \hat \rho_{s}(\zeta)],
\label{eq:inhomoSSscaling}
\ee
where we introduced the scaling limit 
\be
\limsc:\qquad x,t\to\infty\quad\text{with}\quad\zeta=x/t\quad\text{fixed}.
\label{eq:scalinglimit}
\ee
This ballistic scaling can be intuitively understood by recalling that in integrable models the dynamics is interpreted in terms of moving quasiparticles. In this interpretation the ray dependence comes naturally by noting that observables on a given ray $\zeta$ receive a blend of quasiparticles from the two edges that is fixed by $\zeta$. The ray-dependent stationary state $ \hat \rho_{s}(\zeta)$, known as \emph{locally quasi-stationary state} (LQSS), has been introduced in \cite{bertini2016determination}. 

The modification of point \circled{H2} is more direct. Instead of imposing the conservation of the expectation value of all charge densities, we require them to fulfil the continuity equation 
\be
\partial_t \braket{\Psi_0|q_{n,x}(t)| \Psi_0}+ \braket{\Psi_0|j_{n,x}(t)| \Psi_0}-\braket{\Psi_0|j_{n,x-1}(t)| \Psi_0}=0,
\label{eq:evcontinuity}
\ee
which is just the expectation value of \eqref{eq:continuityeq}. Putting all together, we arrive at the following strategy to predict the late-time properties of the system:
\begin{itemize}
\item[\circled{B1}] Assume that, in the scaling limit $\limsc$, every local subsystem is asymptotically described by $\hat \rho_s(\zeta)$. Namely, assume that \eqref{eq:inhomoSSscaling} holds for every local observable $\mathcal O_x$.  
\item[\circled{B2}] Fix $\hat \rho_s(\zeta)$ by imposing \eqref{eq:evcontinuity}. 
\end{itemize}
To show that this strategy is able to determine the stationary state we proceed as in the homogeneous case. We consider the scaling limit, plug \eqref{eq:inhomoSSscaling} in \eqref{eq:evcontinuity}, and represent the stationary state microcanonically obtaining 
\be
-\zeta \partial_\zeta \braket{q_{n,0}}_{\!\rho_{\zeta}}+ \partial_\zeta \braket{j_{n,0}}_{\!\rho_{\zeta}}\!\!= 0,
\label{eq:evcontinuityscalingmicro}
\ee
where $\braket{\cdot}_{\!\rho}$ denotes the thermodynamic limit of the expectation values in an eigenstate with root density $\rho(k)$, while $\rho_{\zeta}(k)$ is the root density associated with a given ray $\zeta$.

To proceed, we need to express the expectation values in \eqref{eq:evcontinuityscalingmicro} in terms of the root densities. The expression for the charge densities is reported in \eqref{eq:chargesrho} while that for the currents reads~\cite{fagotti2016charges}
\be
 \braket{j_{n,0}}_{\!\rho} =   \int {\rm d} k\,\, q_n(k) \varepsilon'(k) \rho(k).
\label{eq:currentsrho}
\ee
Note that, as for the charge densities \eqref{eq:chargesrho}, also the expectation values of the currents can be interpreted in a kinetic theory fashion. Indeed, viewing the group velocity $\varepsilon'(k)$ as the classical velocity of the quasiparticles with momentum $k$, we see that \eqref{eq:currentsrho} is the expression for the flux of charge $Q_n$ generated by the motion of the quasiparticles.  

\begin{figure}
\centering
 \includegraphics[width=.45\textwidth]{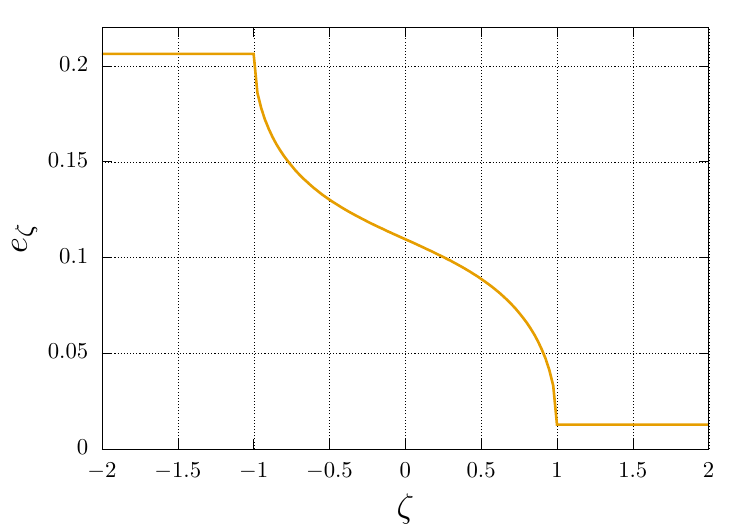}\qquad \includegraphics[width=.45\textwidth]{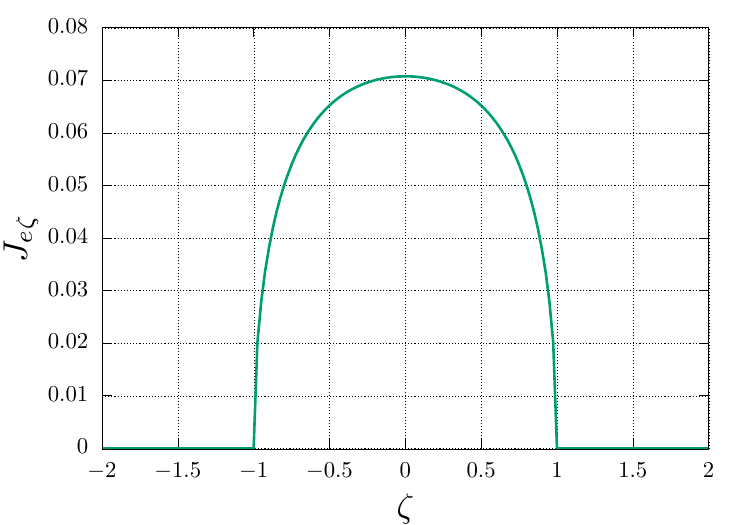}
 \caption{Profiles of energy density (left) and energy current (right) after the bipartitioning protocol starting from the initial state \eqref{eq:bipartitioninginitial} with $h_{\rm L}=0.5$, $h_{\rm R}=2.6$ and $h=1.6$. The maximum and minimum velocities of the quasiparticles are $\zeta_{\rm min}=-1$, $\zeta_{\rm max}=1$, respectively.}\label{fig:profiles}
 \end{figure}

Putting all together and using the completeness of the bare charges $\{q_n(k)\}$ we then find 
\be
-\zeta \partial_\zeta \rho_{\zeta}(k)+\varepsilon'(k) \partial_\zeta \rho_{\zeta}(k) = 0\,.  
\label{eq:continuityscaling}
\ee
The boundary condition for this equation can be found by noting that, since there is a finite speed for the propagation of signals~\cite{lieb1972finite}, observables infinitely far from the junction relax as if the system were homogeneous. Namely  
\begin{align}
\lim_{\zeta\to-\infty}\limsc\limth\langle\Psi_0|\mathcal{O}_x(t)| \Psi_0\rangle &= \limth {\rm tr}[\mathcal{O}_0 \hat \rho_{s, \rm L}],\label{eq:inhomoSSscalingboundaryL}\\
\lim_{\zeta\to\infty}\limsc\limth\langle\Psi_0|\mathcal{O}_x(t)| \Psi_0\rangle &= \limth {\rm tr}[\mathcal{O}_0 \hat \rho_{s, \rm R}],\label{eq:inhomoSSscalingboundaryR}
\end{align}
where $\hat \rho_{s, \rm L}$ and $\hat \rho_{s, \rm R}$ are respectively the stationary states reached after the homogeneous quenches $h_{\rm L}\to h$ and $h_{\rm R}\to h$. Using the result \eqref{eq:saddle-point}, Eqs.~\eqref{eq:continuityscaling}, \eqref{eq:inhomoSSscalingboundaryL}, and \eqref{eq:inhomoSSscalingboundaryR} are solved by  
\be
\rho_{\zeta}(k) = \frac{1}{4\pi} - \frac{\cos\Delta_{h,h_{\rm L}}(k)}{4\pi} \Theta(\varepsilon'(k)-\zeta) - \frac{\cos\Delta_{h,h_{\rm R}}(k)}{4\pi} \Theta(\zeta-\varepsilon'(k))\,,
\label{eq:solutionbipartitionfree}
\ee
where $\Theta(x)$ is the step function. Once again, one can directly verify that $\rho_\zeta(k)$ agrees with the result found via explicit solution of the dynamics.

Eq.~\eqref{eq:solutionbipartitionfree} encodes complete information about the local properties of the system at large times after the quench. In particular, due to the fact that root densities fully specify the expectation values of local observables, it allows one to access their \emph{profiles} throughout the whole light cone $\zeta\in (\zeta_{\rm min}, \zeta_{\rm max})$, where $\zeta_{\rm min}$, $\zeta_{\rm max}$ respectively correspond to the minimum and maximum velocities of the quasiparticles. We note that, despite the discontinuous step function in \eqref{eq:solutionbipartitionfree}, since the position of the step changes smoothly with $k$, the profiles are continuous in $\zeta$, see, e.g., the example reported in Fig.~\ref{fig:profiles}. 

\subsection{Quasistationary states after general inhomogeneous quenches}
\label{sec:inhomogeneous}

Let us now look at a more general inhomogeneous quench. We consider the case in which the system is prepared in the ground state of a Hamiltonian of the form \eqref{eq:FP_hamiltonian}, but with a chemical potential $h_x$ depending non-trivially on the position. At $t=0$ the chemical potential is then changed to a homogeneous value $h_x=h$ for all $x$ and the system is left to evolve unitarily.  

In this case there is no scaling limit in which the expectation value becomes exactly stationary. Intuitively, however, it is natural to expect that a form of quasi-stationarity emerges asymptotically in time. Namely, one expects that \eqref{eq:inhomoSSscaling} can be turned into a statement of the form 
\be
\limth\langle\Psi_0|\mathcal{O}_x(t)| \Psi_0\rangle \simeq \limth {\rm tr}[\mathcal{O}_0 \hat \rho_{s}(x,t)],
\label{eq:inhomoSS}
\ee 
where $\simeq$ denotes the leading contribution for large times, i.e., much larger than the time scale $\tau_{\rm th}$ of local relaxation. In writing Eq.~\eqref{eq:inhomoSS} one assumes that at large enough times, and at the leading order in time, the state becomes locally stationary and homogeneous. Therefore, it can be replaced by a space-time dependent stationary state $\hat \rho_{s}(x,t)$. In order for this assumption to be consistent, the state $\hat \rho_{s}(x,t)$ must be slowly varying, i.e., there must exist a volume element $\ell\times\tau$ of the space-time around $(x,t)$ such that 
\be \label{LDA_xt}
L_\tau \gg \tau \gg \tau_{\rm th},\qquad\qquad L_\ell \gg \ell \gg a,
\ee
where $a$ and $\tau_{\rm th}$ are respectively the lattice spacing of the chain \eqref{eq:FP_hamiltonian} (which we restored for the sake of clarity) and the local relaxation time, while $L_\ell$ and $L_\tau$ are the length and time scales for the variation of $\hat \rho_{s}(x,t)$, cf. Fig.~\ref{fig:HDapprox}. The condition on length scales in \eqref{LDA_xt} is known as \emph{local density approximation} (LDA) \cite{cazalilla2011one}. Note that, for systems on the continuum, the lattice spacing $a$ is replaced by the averaged interparticle distance $d$. Moreover, we remark that the above condition allows for the substitution \eqref{eq:inhomoSS} only for expectation values of local observables, i.e., observables with support of the order of $d$.

The assumption \eqref{eq:inhomoSS}  directly leads to an equation for the root density. Indeed, repeating the steps of the previous section, with \eqref{eq:inhomoSSscaling} replaced by \eqref{eq:inhomoSS}, we find 
\be
\partial_t \rho_{x,t}(k)+\varepsilon'(k)( \rho_{x,t}(k) - \rho_{x-1,t}(k) ) \simeq \partial_t \rho_{x,t}(k)+\varepsilon'(k) \partial_x \rho_{x,t}(k) \simeq 0\,,
\label{eq:continuity}
\ee
where $\rho_{x,t}(k)$ is the root density representing microcanonically the state $\hat \rho_{s}(x,t)$. In the second step we used that the equation is non-trivial only when $\partial_t \rho_{x,t}(k)$ and $\partial_x \rho_{x,t}(k)$ are of the same order in time and therefore higher derivatives go beyond the accuracy of \eqref{eq:inhomoSS}.  
\begin{figure}
\begin{centering}
\includegraphics[width=0.9\textwidth]{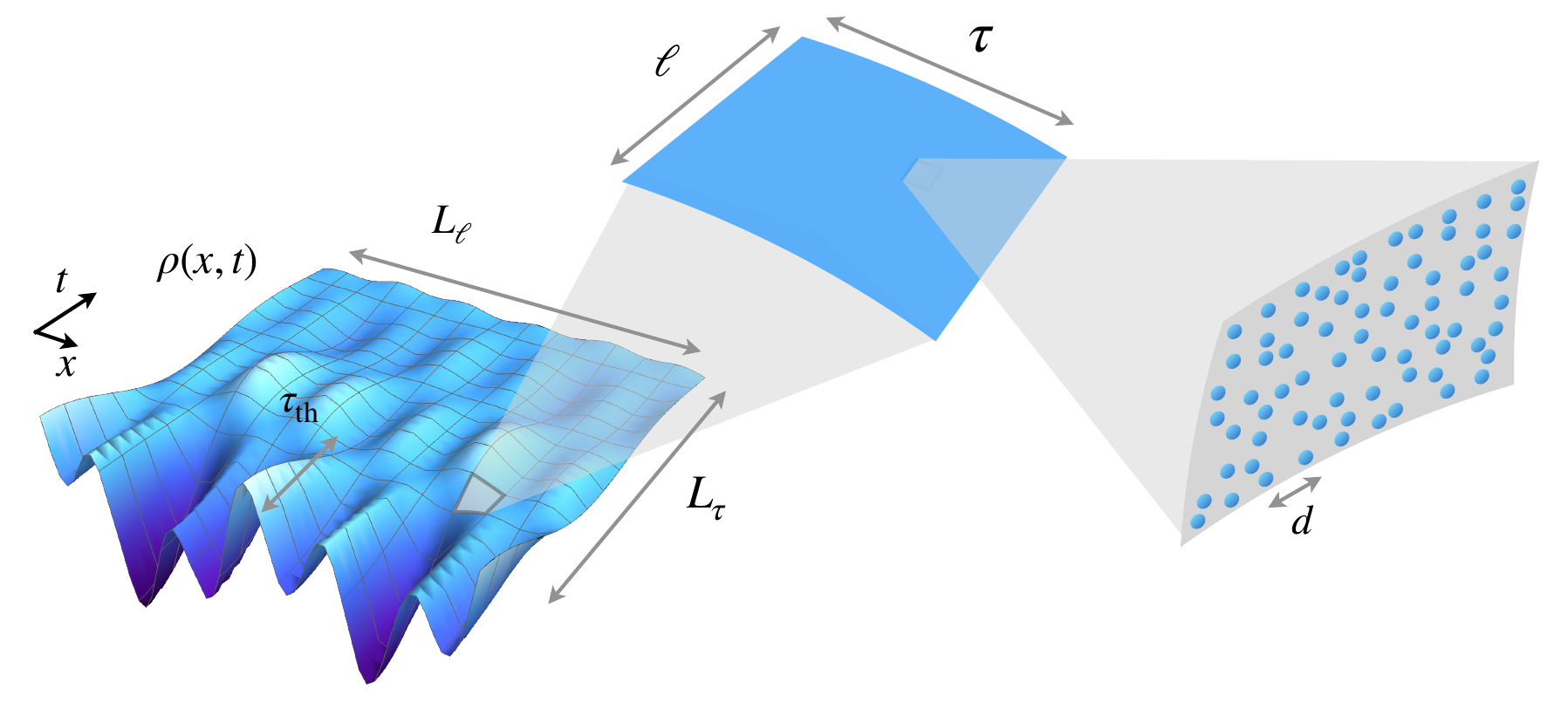}
\par\end{centering}
\begin{centering}
\caption{Pictorial representation of (space-time) scale separation. In the left side, $\rho (x,t)$ is the density profile of particles, as it looks like at large scales: the \emph{macroscopic} length- $L_{\ell}$ and time- $L_{\tau}$ scales are those at which inhomogeneities are important, and the global evolution takes place. The \emph{microscopic} scales, instead, are associated to the interparticle distance $d$ (right) and the local relaxation time $\tau_{\rm th}$ (left). The central part shows the \emph{mesoscopic} scale, i.e., the volume element $\ell \times \tau$ where the system can be considered at the same time homogeneous in space and locally relaxed to a quasi-stationary state $\hat{\rho}_s(x,t)$, while still containing a thermodynamically large number of particles. At this intermediate scale, the corresponding quasiparticle quasi-momentum distribution evolves according to a hydrodynamic equation (Eq.~\eqref{eq:continuity} for non-interacting systems).
}

\label{fig:HDapprox}
\par\end{centering}
\end{figure}

We see that the final result \eqref{eq:continuity} is a simple non-collisional Bolzmann equation for $\rho_{x,t}(k)$, interpreted as the distribution function of non-interacting classical particles moving with velocity $\varepsilon'(k)$. For a given initial condition imposed at time $t=t_0$ (large enough to be in the asymptotic regime) 
\be
\rho_{x,t_0}(k) = f(x;k),
\label{eq:initialcondition}
\ee
Eq.~\eqref{eq:continuity} determines the root density for all rescaled times $t\geq t_0$
\be
\rho_{x,t}(k) = f(x-\varepsilon'(k) t; k), 
\ee
giving a quantitative description of all the local properties of the system at large times after the quench.  

The discussion above is heuristic (for example, it assumes a finite relaxation time scale, $\tau_{\rm th}$, while this quantity typically diverges for integrable systems). However, it can be made more precise introducing an appropriate scaling limit. A convenient way to proceed is to introduce a length scale $\Lambda$ for the variation of $h_x$, namely
\be
h_x=g\left(\frac{x}{\Lambda}\right),
\ee
with $g(x)$ smooth function, and the rescaled variables
\be
\tau=v_{M}t/\Lambda,\qquad\qquad  \xi=x/\Lambda,
\ee
where $v_M=\max_k \varepsilon'(k)$ is the maximal velocity of the quasiparticles. In this language, the analogue of \eqref{eq:inhomoSSscaling} is obtained by taking the weak inhomogeneity limit $\Lambda\rightarrow\infty$ (see, e.g., Appendix B of Ref.~\cite{fagotti2020locally})
\be
\lim_{\Lambda\rightarrow\infty}\limth\langle\Psi_0|\mathcal{O}_x(t)| \Psi_0\rangle = \limth {\rm tr}[\mathcal{O}_0 \hat \rho_{s}(\xi,\tau)],
\label{eq:inhomoSSlambda}
\ee 
and leads to the following equation for the root density in rescaled variables  
\be
0=\lim_{\Lambda\rightarrow\infty}\Lambda[\partial_{\Lambda \tau} \rho_{\xi,\tau}(k)+\varepsilon'(k)( \rho_{\xi,\tau}(k) - \rho_{\xi-\frac{1}{\Lambda},\tau}(k) )] =\partial_\tau \rho_{\xi,\tau}(k)+\varepsilon'(k) \partial_{\xi} \rho_{\xi,\tau}(k)\,,
\label{eq:continuitylambda}
\ee
where, once again, $\rho_{\xi,\tau}(k)$ is the root density representing microcanonically the state $\hat \rho_{s}(\xi,\tau)$. This more rigorous point of view becomes necessary to properly account for subleading corrections to \eqref{eq:continuity}, which originate from higher orders in the gradient expansion, see Refs.~\cite{fagotti2020locally, fagotti2017higher}. For a more thorough discussion of this point we refer the reader to Sec.~\ref{sec:MF}.

\subsection{Generalization to interacting integrable models}
\label{sec:interactions}

In the previous sections, we have explained the main ideas underlying GHD in non-interacting theories. Here we discuss how the conceptual structure carries over to the interacting integrable case. In particular, here we considered the so-called Bethe ansatz integrable models~\cite{korepin1997quantum}. A key feature of these systems is that their spectral properties can always be understood in terms of stable quasiparticles, which display many analogies with the excitations of non-interacting theories. In particular, stable quasiparticles are parametrised by quasi-momenta, or rapidities, $\lambda_j$. However, due to the interactions, the latter cannot be quantized independently from one another: for an $N$-quasiparticle state, the simple relations~\eqref{eq:free_quantization} are typically replaced by~\cite{takahashi2005thermodynamics, korepin1997quantum}
\be\label{eq:bethe_eq}
e^{i L k(\lambda_r)}=\mathcal{F}_r(\lambda_1\,\ldots, \lambda_N)\,,\qquad r=1\,\ldots N\,,
\ee
where $\mathcal{F}_r$ depend on the specific model considered, while $k(\lambda_r)$ is the physical momentum associated with rapidity $\lambda_r$. The quantization conditions~\eqref{eq:bethe_eq} are customarily called ``Bethe equations". Although the structure of eigenstates in the presence of interactions becomes significantly more complicated, expectation values of conserved quantities are expressed as simple sums over $\{\lambda_j\}$. For instance, the momentum and energy associated with a given eigenstate are
\be
P\left[\{\lambda_j\}_j\right]=\sum_j k(\lambda_j)\,,\qquad E\left[\{\lambda_j\}_j\right]=\sum_j \varepsilon(\lambda_j)\,,
\ee
where the single-particle momentum $k(\lambda)$ and energy $\varepsilon(\lambda)$ are model-dependent functions.  

In general $\lambda_j$ need not to be real, and can take arbitrary values in the complex plane. In addition, there might also be different species of quasiparticles connected to different physical degrees of freedom. For instance, in a Bethe-ansatz integrable system of particles with spin (like the Hubbard model~\cite{essler2005one} or the Yang-Gaudin model~\cite{korepin1997quantum}) one has two distinct species of quasiparticles parametrised by disjoint sets $\{\lambda_j\}$ and $\{\mu_j\}$. Roughly speaking one is connected with spin and the other with charge degrees of freedom.  For simplicity, in this section we will restrict to the case of a single species.

For large volumes, the rapidities $\{\lambda_j\}$ arrange themselves in regular patterns in the complex plane which are obtained combining a number of ``basic" configurations where the rapidities stay at a fixed distance from one another. The latter can be specified by a real rapidity and are interpreted as bound states formed by the elementary quasiparticles. Moreover, in the thermodynamic limit the values that these real rapidities can take become dense and one can describe a solution of \eqref{eq:bethe_eq} in terms of sets of quasi-momenta distributions $\rho_n(\lambda)$. Here, $n$ takes discrete positive integer values, and $\rho_n(\lambda)$ is interpreted as the distribution of the quasi-momenta for a bound-state of $n$ quasiparticles. 

In summary, in all Bethe-ansatz integrable models eigenstates can be described using root densities as discussed in Sec.~\ref{sec:thermo} but, differently from the non-interacting case, for each eigenstate there is now a set of functions $\{\rho_n(\lambda)\}$ (rather than a single one) with $\lambda \in [-\Lambda,\Lambda]$ and $n=1,\ldots,N_b$. The maximal values $\Lambda$ and $N_b$ that $\lambda$ and $n$ can take depend on the details of the model and they both can be infinite.

Importantly, due to the non-trivial quantization conditions~\eqref{eq:bethe_eq}, the ``available'' values of $\lambda$ that could be occupied are not distributed uniformly. Accordingly, differently from the non-interacting case (cf.~\eqref{eq:freerhot}), the distribution $\rho_{t,n}(\lambda)$ of vacancies becomes non-trivial. The precise relation between $\rho_n(\lambda)$ and $\rho_{t,n}(\lambda)$ is found from~\eqref{eq:bethe_eq} (see e.g.~\cite{takahashi2005thermodynamics, korepin1997quantum}) and reads as  
\be
\rho_{t,n}(\lambda) = \frac{1}{2\pi} |\partial_\lambda k_n(\lambda)| - \sum_{m=1}^{N_b} \int_{-\Lambda}^{\Lambda} \!\!{\rm d}\mu\, T_{n,m}(\lambda-\mu)\rho_{m}(\mu)\,,
\label{eq:rhot}
\ee
where $k_n(\lambda)$ is the momentum of an $n$-particle bound-state, while $T_{n, m}(\lambda-\mu)$ encodes all information about the interactions (it is proportional to the logarithm of the scattering phase shift~\cite{takahashi2005thermodynamics, korepin1997quantum}). Note that Eq.~\eqref{eq:rhot} does not specify uniquely the  functions $\rho_n(\lambda)$, and stationary states must be determined by an independent equation, which is typically expressed in terms of the function
\be
\eta_n(\lambda)=\rho_{t,n}(\lambda)/\rho_{n}(\lambda)-1\,.
\label{eq:eta}
\ee
For example, in the case of thermal stationary states such additional equation takes the form
\be\label{eq:thermal_TBA}
\log \eta_n(\lambda)=-\beta \varepsilon_n(\lambda)+ \sum_{m=1}^{N_b} \int_{-\Lambda}^{\Lambda} \!\!{\rm d}\mu\, T_{n,m}(\lambda-\mu)\log \left[1+\eta^{-1}_m(\lambda)\right]\!,
\ee 
where $\varepsilon_n(\lambda)$ is the energy of a bound-state of $n$ quasiparticles. In analogy with the non-interacting case, the root densities completely specify the thermodynamic properties of the system. For instance, given a local charge $Q_n$, the corresponding expectation value on the state described by $\{\rho_{n}(\lambda)\}$ is simply
\be
\frac{1}{L}\braket{Q_n}_{\!\rho} = \braket{q_{n,x}}_{\!\rho} = \sum_{m=1}^{N_b}\int_{-\Lambda}^{\Lambda} \!\!{\rm d}\lambda\, q_{n,m}(\lambda) \rho_m(\lambda)\,.
\label{eq:chargesrhoint}
\ee
This formula has once again an intuitive kinetic theoretical interpretation, generalizing Eq.~\eqref{eq:chargesrho} to the interacting case. In fact, a generalization exists also for the expectation value of local currents, which reads
\be
\braket{j_{n}}_{\!\rho} = \sum_{m=1}^{N_s}\int_{-\Lambda}^{\Lambda} \!\!{\rm d}\lambda\, q_{n,m}(\lambda) v_m(\lambda) \rho_m(\lambda), 
\label{eq:currentsrhoint}
\ee 
where the velocities $\{v_n(\lambda)\}$ are ``dressed" by the interactions as described by the following integral equation
\be
v_n(\lambda) \rho_{t,n}(\lambda) = \frac{1}{2\pi} \partial_\lambda e_n(\lambda) - \sum_{m=1}^{N_s} \int_{-\Lambda}^{\Lambda} \!\!{\rm d}\mu\, T_{n, m}(\lambda-\mu)v_{m}(\mu)\rho_{m}(\mu),
\label{eq:vrhot}
\ee 
where $\{\rho_{t,n}(\lambda)\}$ are given in \eqref{eq:rhot}. Differently from Eq.~\eqref{eq:chargesrhoint}, which follows immediately from the definition of the root densities, formula \eqref{eq:currentsrhoint} is highly non-trivial. It was first conjectured in Ref.~\cite{bertini2016transport,castro-alvaredo2016emergent} but its rigorous proof has been accomplished only very recently~\cite{pozsgay2020algebraic} (see, however, Refs.~\cite{yoshimura2018equations, borsi2019current, yoshimura2020collision} for relevant partial results). For more detail on this aspect we refer the reader to the contributions by Cubero, Yoshimura, and Spohn; and Borsi, Pristy\'ak, and Pozsgay to this special issue. 

\subsubsection{Homogeneous quenches in interacting integrable models}

Despite the conceptual framework being completely analogous to the non-interacting case, the study of quantum quenches in interacting integrable systems is significantly more complicated on the technical level. In fact, explicit results are typically restricted to simple families of initial states. In essence, this is due to two main complications. 
\begin{itemize}
\item[(i)] Strictly local conservation laws are in general not enough to uniquely specify the root densities, and one also needs to consider \emph{quasi-local} conserved operators~\cite{ilievski2015complete,ilievski2016quasilocal}. These charges are again expressed as sums of densities (cf.~\eqref{eq:Charges2}) but the densities do not have finite support and exhibit exponentially decaying tails. It is customary to denote the combined set of local and quasi-local charges by $\{Q_{n,s}\}_{s=1/2,1,3/2,\ldots; n=0,1,\ldots}$ where $Q_{n,1/2}$ are the usual local charges. The expectation value of any charge $Q_{n,s}$ on a stationary state described by  $\{\rho_{m}(\lambda)\}$ can again be written in the form \eqref{eq:chargesrhoint} for some appropriate functions $\{q_{n,s,m}(\lambda)\}$. Importantly, the set of all local and quasi-local charges is complete, in the sense that   
\be
\sum_{m=1}^{N_b}\int_{-\Lambda}^{\Lambda} \!\!{\rm d}\lambda\, q_{n,s,m}(\lambda) f_m(\lambda)=0\quad \forall s,n \qquad \Leftrightarrow \qquad  f_m(\lambda)=0\,.
\ee
\item[(ii)] The analogue of \eqref{eq:initialdata}, i.e.\ an explicit expression for initial-state expectation values, is in general not available. Currently, this can be found only for low entangled states~\cite{piroli2016exact}. 
\end{itemize}
When these two complications can be overcome, i.e.\ a complete family of quasi-local conservation laws is known and the initial-state expectation values can be computed exactly, one can follow the ``string-charge duality" logic of Sec.~\ref{sec:homogeneous_quench} and arrive at a full description of the steady state in terms of the root densities $\rho_n(\lambda)$. Note that in the interacting integrable case there is typically no exact solution for the full dynamics to compare with and one should test the assumption \eqref{eq:homoSS} against numerical results. All the cases considered so far confirmed the validity of \eqref{eq:homoSS}, see e.g. Refs.~\cite{ilievski2015complete, wouters2014quenching, pozsgay2014correlations,piroli2016exact,piroli2017correlations}. 

Finally, we should stress that the string-charge duality  is not the only available approach to determine the post-quench stationary state. Indeed, two complementary methods are given by the so-called ``Quench Action"~\cite{caux2016quench,caux2013time} and ``Quantum-Transfer Matrix" approaches~\cite{piroli2017what,piroli2017from,piroli2019integrable,piroli2019integrable_II}. All these methods yield the same results but, depending on the specific model and initial state, some of them might be difficult (or even impossible) to implement. For further details on the latter methods we refer the reader to the relevant literature, see e.g.~\cite{ilievski2016string, caux2016quench, piroli2017what}.

\subsubsection{Inhomogeneous quenches in interacting integrable models}
\label{eq:inhomogeneousinteracting}

As for the case of homogeneous quenches, the GHD logic outlined in Secs.~\ref{sec:bipartitioning}--\ref{sec:inhomogeneous} can be directly extended to the interacting integrable case. Specifically, using Eqs.~\eqref{eq:chargesrhoint} and \eqref{eq:currentsrhoint} for the expectation values of quasi-local charge densities and related currents one finds the following continuity equation for the root densities $\{\rho_{n, x,t}(\lambda)\}$ of the locally quasi-stationary state   
\be
\partial_t \rho_{n, x,t}(\lambda) + \partial_x (v_{n,x,t}(\lambda)\rho_{n, x,t}(\lambda)) \simeq 0 \,.
\label{eq:continuityint}
\ee
In particular, in the case of bipartitioning protocols we can explicitly take the scaling limit $\limsc$ and obtain
\be
-\zeta \partial_\zeta \rho_{n, \zeta}(\lambda)+ \partial_\zeta( v_{n, \zeta}(\lambda) \rho_{n, \zeta}(\lambda)) = 0\,. 
\label{eq:continuityscalingint}
\ee
These equations differ from their non-interacting counterparts \eqref{eq:continuity} and \eqref{eq:continuityscaling} because of the presence of space-time dependent velocities. This is a direct consequence of the interactions (cf.~\eqref{eq:vrhot}) --- in fact, it is the only interaction effect in \eqref{eq:continuityint} and \eqref{eq:continuityscalingint} --- and has a very intuitive explanation. The motion of a given quasiparticle is perturbed by the scatterings with the others: this results in a change in its averaged velocity. Naturally, the change depends on $\{\rho_{n, x,t}(\lambda)\}$, the set of densities of the different species of quasiparticles at the space-time point $(x,t)$, yielding space-time dependent velocities. Assuming that, for any $\lambda$ and $n$, the equation $\zeta =v_{n, \zeta}(\lambda)$ has a unique solution, Eq.~\eqref{eq:continuityscalingint} can be immediately solved in terms of $\eta_n(\lambda)$ as follows~\cite{bertini2016transport} 
\begin{equation}
\label{eq:solution_hydro}
\eta_{n ,\zeta}(\lambda) = \eta_{n, R }(\lambda)\Theta({\zeta -v_{n, \zeta}(\lambda)})+\eta_{n, L } (\lambda)\Theta({v_{n, \zeta}(\lambda)-\zeta})\,,
\end{equation}
where $\Theta(x)$ is the step function, while the ``left" and ``right" functions $\eta_{n, L}(\lambda)$ and $\eta_{n, R}(\lambda)$ are those characterising the state at infinite distance from the junction on the right and on the left hand side, respectively. Note that \eqref{eq:solution_hydro} is still an implicit solution because $v_{n, \zeta}(\lambda)$ depends on $\eta_{n ,\zeta}(\lambda)$. The standard procedure to treat it is by iteration, see Sec.~\ref{sec:localphysics}.

Interestingly, the interaction-related complications outlined above do not arise when deriving Eqs.~\eqref{eq:continuityint} and \eqref{eq:continuityscalingint}. Indeed, in the derivation one only needs to assume that a complete set of quasi-local charges exists, without ever needing their explicit form. The aforementioned problems, however, emerge when trying to find appropriate boundary conditions. For example, let us consider Eq.~\eqref{eq:continuityscalingint}. Repeating the arguments of Sec.~\ref{sec:bipartitioning} we have that a unique solution is found by imposing the boundary conditions 
\eqref{eq:inhomoSSscalingboundaryL} and \eqref{eq:inhomoSSscalingboundaryR}. These are nothing but homogeneous quench problems and, as discussed before, in the interacting case they can be solved only for special classes of states. In particular, the simplest cases to treat are bipartitioning protocols where one joins two different stationary states with known root densities (in this case the homogeneous quench problems associated with boundary conditions are trivial). This includes the highly studied cases of two half chains prepared at different temperatures or at different filling. In the case of Eq.~\eqref{eq:continuityint} the problem of finding initial condition proved itself to be even harder. Up to now it has been solved only when the system is initialised in a slowly varying stationary state, see, e.g., Refs.~\cite{doyon2017large, bulchandani2017solvable, bulchandani2017bethe, caux2019hydrodynamics, ruggiero2020quantum, bastianello2020thermalization}. In this case $\rho_{n, x,0}(\lambda)$ is determined in terms of the equilibrium root densities within a local density approximation. 

\section{Local Physics of Inhomogeneous Quenches}
\label{sec:localphysics}

In  this section, we provide a survey of some recent applications of the GHD approach to the study of inhomogeneous quenches in interacting integrable lattice systems. We concentrate the discussion to the case of \emph{bipartitioning protocols} and only focus on the scaling limit \eqref{eq:scalinglimit} where observables become functions of the ray~$\zeta$. Moreover --- although we will also mention results in other models --- we will predominantly focus on the paradigmatic case of the XXZ Heisenberg chain (cf. \eqref{eq:heisenberg_XXZ}). We present the main physical results obtained in this regime and their qualitative interpretation, outlining connections with findings by alternative methods. Relevant results have also been obtained in other settings, models and limits but they are not covered in detail here. We refer the reader to the other contributions to this special issue (in particular see those by  Bastianello, De Luca, and Vasseur;  Bulchandani, Gopalakrishnan, and Ilievski; De Nardis, Doyon, Medenjak, and Panfil; Bouchoule and Dubail) for other examples of applications of GHD to the dynamics of local observables after inhomogeneous quenches. 

We begin by exemplifying the procedure discussed in Sec.~\ref{eq:inhomogeneousinteracting} for the case of a bipartitioning protocol in the paradigmatic case of the XXZ Heisenberg chain. The latter describes a system of spins on the lattice that interact as described by the following Hamiltonian 
\be\label{eq:heisenberg_XXZ}
H[\Delta, h]=\frac{1}{4} \sum_{x=-L/2+1}^{L/2} \left[\sigma_{1,x} \sigma_{1,x+1}+\sigma_{2,x} \sigma_{2,x+1}+\Delta \sigma_{3,x} \sigma_{3,x+1}\right]-h\sum_{x=-L/2+1}^{L/2} \sigma_{3,x}\,.
\ee
Here $\{\sigma_{\alpha,x}\}_{\alpha=1,\ldots,3}$ act as Pauli matrices on the local space $\mathbb{C}^2$ at site $x$ and like the identity elsewhere, while $h$ is a magnetic field. 

The Hamiltonian \eqref{eq:heisenberg_XXZ} is related to a chain of spinless fermions with a quartic interaction term via a Jordan-Wigner transformation (we again neglect issues arising from the boundary conditions), with $\Delta=0$ corresponding to the non-interacting point. The quasiparticle content of the model depends on the anisotropy parameter $\Delta$, with a particularly simple structure observed for $\Delta>1$. In that case, one has an infinite number of bound states, $N_b=\infty$, and $\lambda\in[-\Lambda,\Lambda]=[-\pi/2,\pi/2]$. The driving terms and kernels of Eqs~\eqref{eq:rhot}, \eqref{eq:thermal_TBA}, and \eqref{eq:vrhot} are given by 
\begin{align}
\partial_\lambda k_n(\lambda)&=2\pi a_n(\lambda),\\
\varepsilon_n(\lambda)&=-\pi \sinh (\eta) a_{n}(\lambda)+2 h n,\\
T_{n m}(\lambda)&=\left(1-\delta_{n m}\right) a_{|n-m|}(\lambda)+2 a_{|n-m|+2}(\lambda)+\ldots+2 a_{n+m-2}(\lambda)+a_{n+m}(\lambda),\\
a_{n}(\lambda)&=\frac{1}{\pi} \frac{\sinh (n \eta)}{\cosh (n \eta)-\cos (2 \lambda)}\,,
\end{align}
where we set $\eta=\operatorname{arcosh}\Delta$.

Let us consider a bipartitioning protocol, i.e., we initialise the system in the state 
\begin{equation}
\rho_0 =\rho_{\rm L} \otimes \rho_{\rm R}
\end{equation} 
and evolve it with Hamiltonian \eqref{eq:heisenberg_XXZ}. The GHD prescription to compute the profiles of local charges and currents can be summarised in the following steps:
\begin{itemize}
	\item[I.] Determine the left/right thermal stationary states (and hence the corresponding $\eta_{L,R}(\lambda)$) by solving the homogeneous quench problem. For example in the case where the left and right halves of the system are initialised in two different thermal states
	\be
	\hat \rho_{\rm L/R} = \frac{1}{Z_{\rm L/R}} e^{- \displaystyle \beta_{\rm L /R}H[\Delta, h_{\rm L/R}]},
	\ee
	one can obtain $\eta_{\rm L,R}(\lambda)$ from Eq.~\eqref{eq:thermal_TBA}.
	\item[II.] For each  ray $\zeta$, solve Eq.~\eqref{eq:solution_hydro}. Numerically, this can be done by a simple iterative scheme: one starts with an ansatz for $\eta^{(0)}_{n, \zeta}(\lambda)$, computes the corresponding velocities $v^{(0)}_{n, \zeta}(\lambda)$ using Eqs~\eqref{eq:rhot} and \eqref{eq:vrhot}, and obtains a new ansatz $\eta^{(1)}_{n, \zeta}(\lambda)$ using the right-hand side of \eqref{eq:solution_hydro}. One proceeds in this way to obtain subsequent approximations $\eta^{(k)}_{n, \zeta}(\lambda)$, until convergence is reached.
	\item[III.] For each ray $\zeta$, use the knowledge of $\eta_{n,\zeta}(\lambda)$ to compute $\rho_{n,\zeta}(\lambda)$ and $v_{n, \zeta}(\lambda)$ from Eqs~\eqref{eq:rhot} and \eqref{eq:vrhot}, and finally obtain the values of the charges and currents from Eqs~\eqref{eq:chargesrhoint} and \eqref{eq:currentsrhoint}. As discussed in Sec.~\ref{sec:thermo}, the knowledge of $\rho_{n,\zeta}(\lambda)$ allows for the computation of all local properties of the system beyond the density of conserved charges and their currents. In practice, however, explicit formulae expressing local observables in terms of root densities are scarce. Important examples have been found for simple few-point operators in the Heisenberg spin chain~\cite{mestyan2014short,pozsgay2017excited}, in the Lieb-Liniger model~\cite{kormos2009expectation,pozsgay2011local,piroli2016multiparticle,piroli2016quantum_II,bastianello2018exact,bastianello2018sinh}, and in the sinh-Gordon field theory~\cite{negro2013on, negro2014on, bertini2016quantum}. 

\end{itemize}
These steps are very simple to implement numerically and with straightforward modifications they can be applied to any integrable model treatable with the formalism of Sec.~\ref{sec:interactions}. This has been explicitly demonstrated in multiple studies of bipartitioning protocols in concrete models~\cite{bertini2016transport, bertini2018low, bertini2018universal, deluca2017nonequilibrium, mazza2018energy, piroli2017transport, bertini2019transport, castro-alvaredo2016emergent, mestyan2020molecular, alba2019entanglement, alba2018entanglement, mestyan2018spin, denardis2018hydrodynamic, nozawa2020generalized,nozawa2020generalized_2, mestyan2018spin,wang2020emergent}, see also Ref.~\cite{moller2020introducing} for a versatile, open-source numerical framework for solving typical equations appearing within GHD. Furthermore, while the above prescription typically allows one to access the values of the profiles numerically, there exist cases where fully analytic solutions can be obtained, as we discuss in the following.

\begin{figure}
\centering
\includegraphics[width=.45\textwidth]{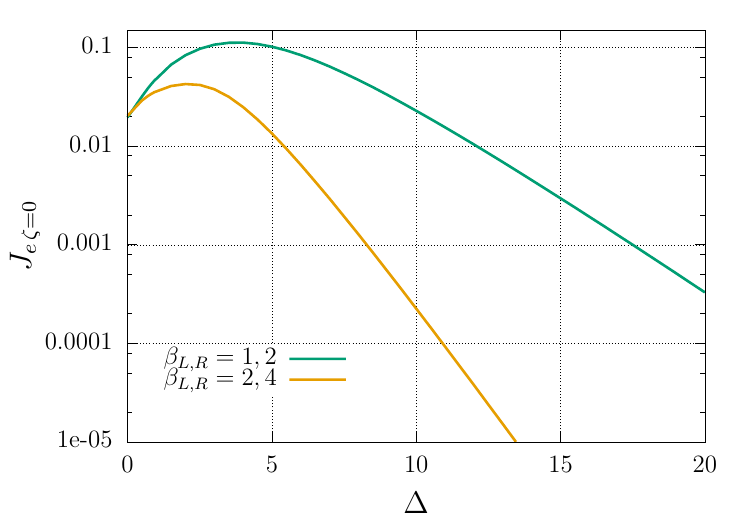}\qquad
 \includegraphics[width=.45\textwidth]{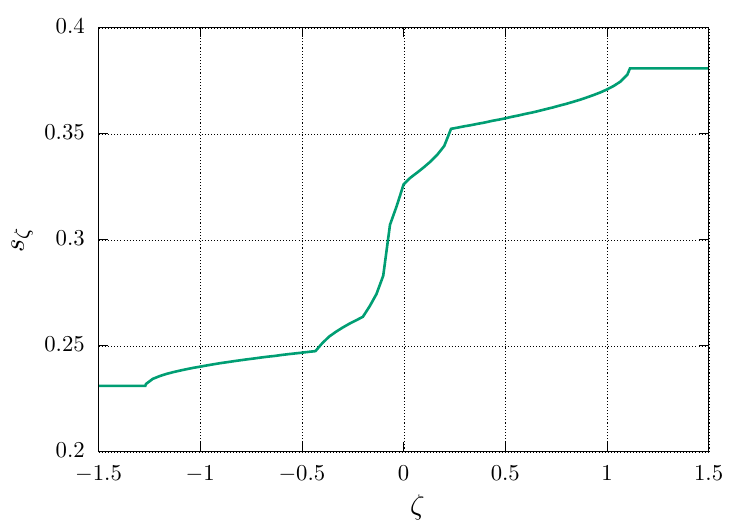}
 \caption{Left: energy current in the NESS of the Heisenberg chain~\eqref{eq:heisenberg_XXZ} as a function of the anisotropy  $\Delta$. The initial state is obtained by joining together two thermal states with inverse temperatures $\beta_L$, $\beta_R$ respectively. The current displays a maximum for $\Delta\sim \min \left(\beta_{R}^{-1}, \beta_{L}^{-1}\right)$, and vanishes exponentially as $\Delta\to\infty$. Right: profile of magnetization for a bipartitioning protocol in the Heisenberg chain, with $\Delta=2>1$. The initial state is obtained by joining together two Gibbs states with inverse temperatures $\beta_L=\beta_R=0$ and chemical potentials $(\beta h)_L=1$, $(\beta h)_R=2$.}\label{fig:profiles_interacting}
\end{figure}

\subsection{NESS}
\label{sec:NESS}

Arguably, the most interesting aspect of bipartitioning protocols in integrable systems is that they allow for the realisation of non-equilibrium steady states (NESS)s --- i.e. steady states supporting non-trivial currents --- in the context of isolated quantum lattice systems (i.e.\ without resorting to external driving). This contrasts with what happens in generic (non-integrable) lattice systems where the only local conservation law is the Hamiltonian. In the latter case assumption \eqref{eq:inhomoSSscaling}, together with some physical requirements on the form of $H$ (for example invariance under space inversion or time reversal), allows one to prove that the current vanishes at late times (see, e.g., Sec. IX B of Ref.~\cite{bertini2020finitetemperature}). The latter fact is in agreement with expectations coming from the Fourier law~\cite{fourier2009the, narasimhan1999fourier} --- which predicts a current proportional to the temperature gradient --- and the currently available numerical evidence~\cite{biella2016energy, biella2019ballistic, karrasch2013nonequilibrium}. The emergence of non-trivial NESSs after bipartitioning protocols in integrable systems was first pointed out in the non-interacting case~\cite{antal1999transport} and then in conformal field theories~\cite{bernard2012energy}. Proving that the NESS survives (integrable) interactions has been the first stark success of GHD~\cite{bertini2016transport, castro-alvaredo2016emergent}. In particular, in the context of lattice systems, the emergence of a NESS was first demonstrated for the XXZ Heisenberg model in Refs.~\cite{bertini2016transport,piroli2017transport}, where GHD allowed for a detailed study of the dependence of charges and currents on the interaction parameter. 

In the language of the previous section, the NESS is the LQSS associated with the ray $\zeta=0$ (cf.~\eqref{eq:inhomoSSscaling}). Namely, it is the state that captures the late-time properties of any finite region at infinite times after the quench. This state has been extensively investigated in non-interacting systems~\cite{antal1999transport,aschbacher2003non,aschbacher2006out,platini2007relaxation,lancaster2010quantum,eisler2013full,deluca2013nonequilibrium,collura2014quantum,eisler2014entanglement,collura2014non,deluca2015stationary,doyon2015non,viti2016inhomogeneous,allegra2016inhomogeneous,kormos2017temperature,perfetto2020dynamics,perfetto2017ballistic, mintchev2011non} and conformal field theories~\cite{sotiriadis2008inhomogeneous,doyon2014energy,bernard2012energy,bernard2015non,bernard2016hydrodynamic,bernard2016conformal,langmann2017steady,dubail2017conformal,dubail2017emergence} (see also the dedicated reviews \cite{bernard2016conformal} and \cite{vasseur2016nonequilibrium}). An important result of these studies is the determination of all higher cumulants of the NESS currents, which give access to the full counting statistics of the charges transferred through the junction (see also~\cite{doyon2019fluctuations, myers2020transport}). 

Within GHD, thermodynamic properties of the NESS can be studied directly from the general theory introduced in the previous section. For a given bipartitioning protocol, the value of the currents are generically found to be non-monotonic functions of the interaction parameter, see the example reported in Fig.~\ref{fig:profiles_interacting}. Note that it is not uncommon to see NESS currents growing with the interaction strength. Another interesting property of the NESS currents in integrable systems is that they cannot generically be written as sums of functions involving properties of a single lead only, i.e. 
\be
\braket{j_{n}}_{\!\rho_0}\neq f(\rho_L) + f(\rho_R). 
\ee
This is in contrast to what happens in free systems and CFTs, and can be viewed as a transparent signature of the interaction. The simplification discussed above happens only in some special cases, see e.g.\ the low-temperature regime for thermal reservoirs discussed in Sec.~\ref{sec:small_T}. 

While the GHD equations can be typically solved only numerically, closed analytic expressions for the profiles may be obtained in special cases. In the Heisenberg chain this happens trivially for $\Delta=0$, where the system becomes non-interacting. A more interesting example was found in Ref.~\cite{collura2018analytic}, which considered a quench from a ``domain-wall'' state, i.e., a bipartitioning protocol where the left and right halves of the system are initialised in completely polarized states, in opposite $z$-directions. While, in this case, transport at the hydrodynamic scale is trivial for $\Delta>1$, all local observables display non-trivial ballistic profiles in the regime $\Delta<1$, and fully analytic expressions may be obtained. Arguably, the most interesting result of Ref.~\cite{collura2018analytic} is that the NESS has a nowhere differentiable dependence on the strength of interactions. In particular, the magnetisation density and current profiles exhibit jumps in correspondence to values of the anisotropy $\Delta$ for which $\operatorname{arcos}(\Delta)/\pi$ is a rational number: this is in agreement with the nowhere differentiable spin Drude weight computed in the linear response regime~\cite{zotos1999finite,prosen2011open, prosen2013families, ilievski2017microscopic} (for further details see \cite{bertini2020finitetemperature} and the contribution by De Nardis, Doyon, Medenjak, and Panfil to this special issue). The analytic solution of Ref.~\cite{collura2018analytic} also allowed the authors to analyze the behaviour around the edge of the magnetisation profile, ruling out the presence of a Tracy-Widom scaling, typical of non-interacting behaviour (see Sec.~\ref{sec:lightcone}).

\begin{figure}
\centering
\includegraphics[width=.45\textwidth]{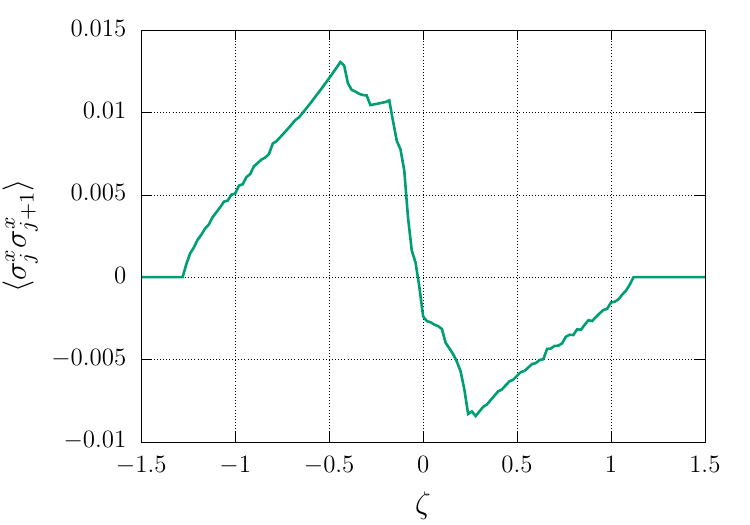}\qquad
 \includegraphics[width=.45\textwidth]{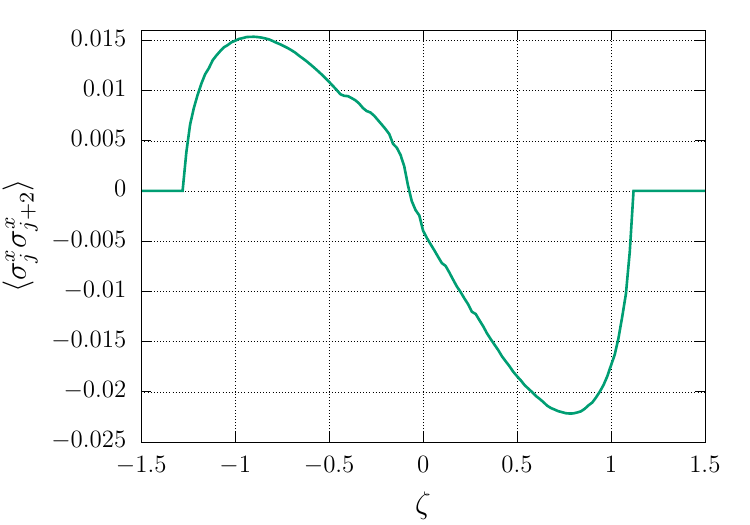}\qquad \caption{
Profiles of the expectation values for $\sigma_j^x \sigma_{j+1}^x$ (left) and $\sigma_j^x \sigma_{j+2}^x$ (right), for a bipartitioning protocol in the Heisenberg chain, with $\Delta=2>1$. The initial state is obtained by joining together two Gibbs states with inverse temperatures $\beta_L=\beta_R=0$ and chemical potentials $(\beta h)_L=1$, $(\beta h)_R=2$. The visible small ripples are numerical artefacts.}\label{fig:profiles_correlators_interacting}
\end{figure}

\subsection{Phenomenology of the profiles}
\label{sec:phenomenology}

Although the structure of the GHD equations is completely general, the details of the underlying microscopic model encoded in in the parameters $N_b$ and $\Lambda$ and functions $k_n(\lambda)$ and $T_{n, m}(\lambda)$ (cf. Sec.~\ref{sec:interactions}) give rise to a manifold phenomenology, with significant qualitative differences in distinct integrable systems. Examples of bipartitioning protocols have been studied, and often checked against independent numerical methods, in spin chains and lattices~\cite{bertini2016transport,piroli2017transport,bertini2018low,nozawa2020generalized,nozawa2020generalized_2, bertini2018universal, deluca2017nonequilibrium, mestyan2020molecular, alba2019entanglement, alba2018entanglement}, quantum gases~\cite{castro-alvaredo2016emergent,mestyan2018spin,wang2020emergent}, hard-rod systems~\cite{doyon2017dynamics}, classical~\cite{bastianello2018generalized,doyon2019generalized, bulchandani2019kinetic} and quantum~\cite{castro-alvaredo2016emergent, bertini2019transport} field theories.

In general, from the structure of the GHD equations, it is easy to see that different bound-states of quasiparticles give rise to non-analyticities $\{\zeta^\pm_n\}_n$ inside of the light cone and at its boundaries. These can be understood in terms of the quasiparticles' motion, as the non-analytic points $\zeta^\pm_n$ correspond to the maximum and minimum velocities of  the $n$-quasiparticle bound-states. The precise nature of such non-analyticities depends on the initial state and the model considered.

Once again, this could be already appreciated from GHD studies in the Heisenberg chain~\cite{bertini2016transport,piroli2017transport,collura2018analytic}. In the gapless regime of the model, $\Delta<1$, one has a finite number of bound states for rational values of $\operatorname{arcos}(\Delta)/\pi$, so that a finite number of non-analyticities appear~\cite{bertini2016transport}. On the contrary, the number of bound-states is always infinite for $\Delta\geq1$, giving rise to a series of non-analytic points, which accumulate inside the light cone~\cite{piroli2017transport}. Typically, $\zeta^\pm_n$ are easily visible for small values of $n$, see Figs.~\ref{fig:profiles_interacting}--\ref{fig:profiles_correlators_interacting} for concrete examples. As $n\to\infty$ the velocities converge to a $\lambda$-independent value, i.e. 
\be
\lim_{n \to \infty} v_{n,\zeta}(\lambda) = v_{\infty,\zeta}.
\ee
Accordingly, also the sequence $\zeta_n^+$ converges to a ray $\zeta_\infty$
\be
\lim_{n\rightarrow\infty}\zeta_n^+=\lim_{n\rightarrow\infty}\zeta_n^-=\zeta_\infty\,,
\ee
which corresponds to the velocity of the heaviest quasiparticle. 

Interestingly, it was shown in Ref.~\cite{piroli2017transport} that, depending on the initial state, the profiles of magnetisation and charges that are odd under spin-flip may exhibit abrupt jumps at the ray $\zeta_\infty$. This peculiar behaviour is ultimately related to the structure of the Bethe Ansatz for $\Delta>1$ and can be heuristically explained by saying that information about the overall sign of the magnetisation is carried by the heaviest quasiparticles~\cite{piroli2017transport}. An abrupt jump in $\braket{{\cal O}}_\zeta$ signals that the expectation value of ${\cal O}_x$ varies on length scales shorter than $t$ (i.e.\ proportional to  $t^a$ with $a<1$), implying that the transport of ${\cal O}_x$ is \emph{sub-ballistic}. This is in agreement with the numerical findings of Refs.~\cite{ljubotina2017spin, ljubotina2019kardar}, which identified diffusive spin-transport ($a=1/2$) for $\Delta>1$ and superdiffusive one ($a=2/3$) for $\Delta=1$. In particular, the former type of transport has later been described in GHD by introducing appropriate subleading corrections to \eqref{eq:continuityint}~\cite{denardis2018hydrodynamic, denardis2019diffusion}, while the latter is still subject to intensive research in relation to the observed Kardar-Parisi-Zhang scaling of the profiles~\cite{agrawal2020anomalous, bulchandani2020kardar, denardis2019anomalous, gopalakrishnan2019kinetic, weiner2020high, denardis2020superdiffusion}.

Non-analyticities also appear in the case of multiple quasiparticle species, which corresponds to integrable models with internal degrees of freedom describing for example particles with spin. Due to their physical relevance, inhomogeneous quenches in these systems have been widely investigated, resulting in applications of GHD to the Hubbard model~\cite{fava2020spin,nozawa2020generalized,nozawa2020generalized_2, ilievski2017ballistic}, spinful Fermi and Bose gases~\cite{mestyan2018spin,wang2020emergent}, sine-Gordon model~\cite{bertini2019transport}, non-linear sigma model~\cite{denardis2019anomalous}, and a special point of the two-component Bariev model~\cite{zadnik2020folded}. In these cases, non-analytic points correspond to either different species of quasiparticles or bound-states of quasiparticles of the same species. In fact, it is natural to wonder whether the presence of different species could be directly inferred from the profiles, or, in other words, if bipartitioning protocols could detect \emph{separation} effects. For instance, in the case of spinful fermions one could ask whether there exist some local observable whose light-cone profile only shows the effect of one of the two species of excitations. 

It turns out that, for bipartition protocols at finite energy densities, non-analyticities of all quasiparticle species are typically present in the profiles of arbitrary observables, so that no strict separation happens~\cite{mestyan2018spin}. However, some separation effects become manifest in special cases. One of them has been pointed out recently in the study of the Hubbard model, where an interesting phenomenon called \emph{clogging} emerges for some fine-tuned initial states ~\cite{nozawa2020generalized,nozawa2020generalized_2}. In essence, clogging consists in the fact that a vanishing charge current coexists with nonzero energy currents (or vice versa), within a finite region of the light cone. Its existence has been proven analytically in Ref.~\cite{nozawa2020generalized} in the case where half of the system is initially at half-filling and at infinite temperature, and it has been numerically observed in the high-temperature regime. In addition, different initial configurations resulting in clogging were studied in Ref.~\cite{nozawa2020generalized_2}, where it was confirmed that it could also take place in the NESS. In the next section, we will study a different case where some separation effects become visible, namely the low-temperature regime. 

\subsection{The low-energy limits}
\label{sec:small_T}

\begin{figure}
\centering
 \includegraphics[width=.45\textwidth]{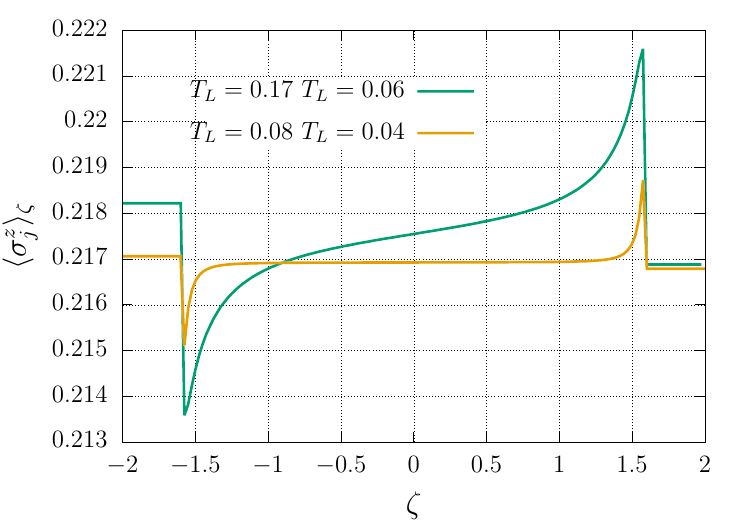}\qquad \includegraphics[width=.45\textwidth]{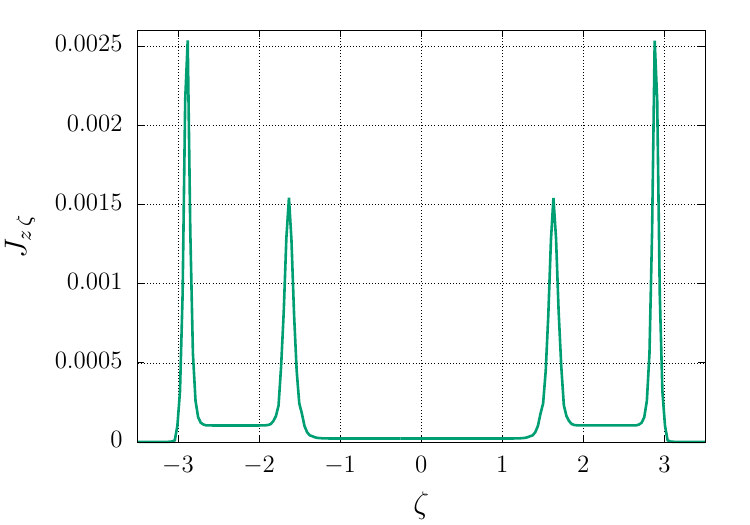}
 \caption{Left: low-temperature spin profile in the gapless phase of the XXZ Heisenberg chain. The plot corresponds to $\Delta=3$, $h=1.2$. Right: low-temperature profile of the particle current in the Yang-Gaudin model of spinful fermions~\cite{mestyan2018spin}. The initial state is obtained by joining together two thermal states at different (low) temperatures, and the same values of chemical potential and non-vanishing magnetic field.}\label{fig:lowT_profiles}
 \end{figure}

The low-temperature regime of the GHD equations turns out to be particularly interesting and has been subject to several investigations over the past few years~\cite{bertini2018universal,bertini2018low,mestyan2018spin,wang2020emergent,fava2020spin, denardis2019anomalous}.
\begin{figure}[b]
	\includegraphics[width=0.5\textwidth]{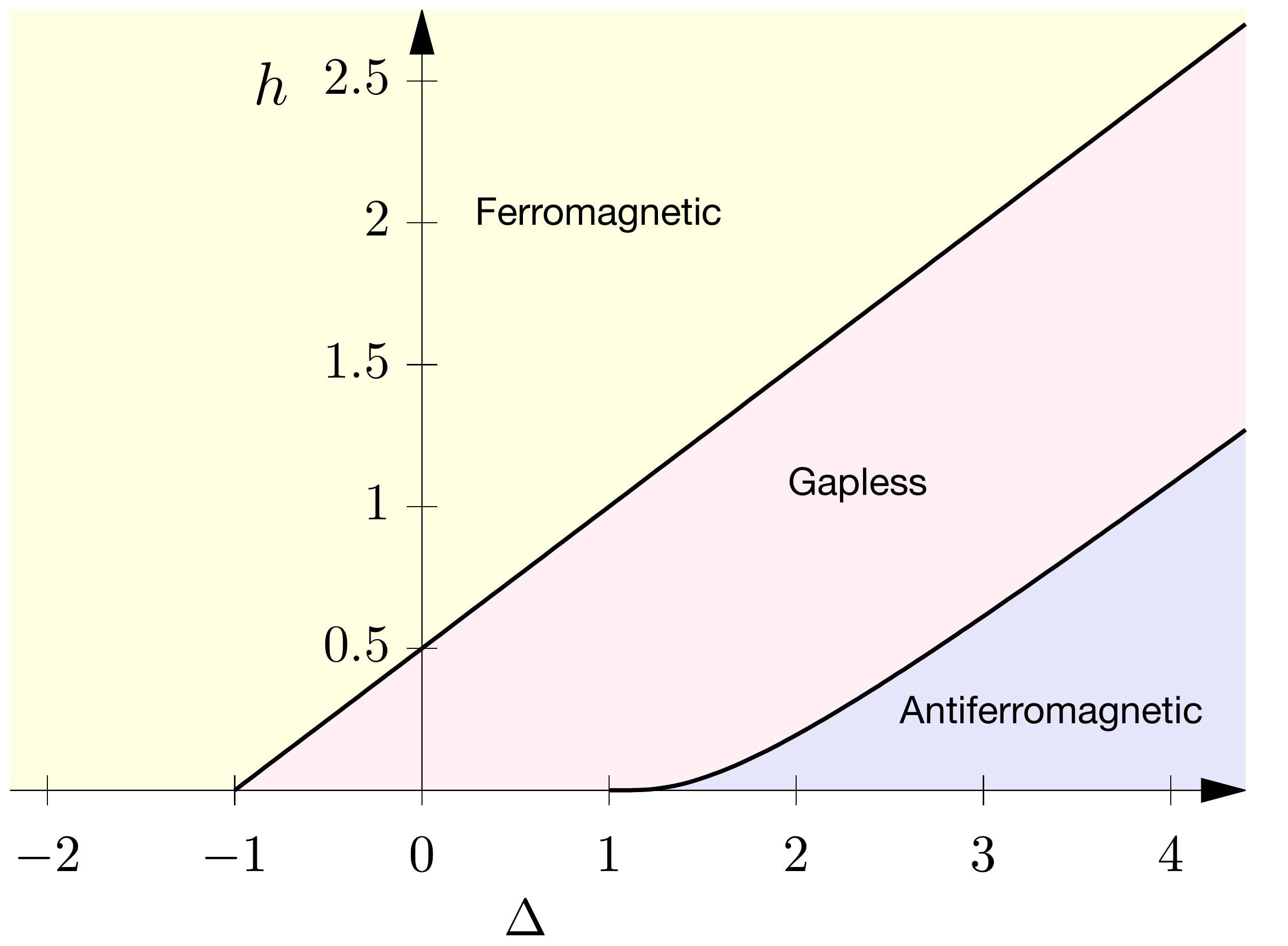}
	\caption{The phase diagram of the XXZ spin-$1/2$ chain in an external magnetic field $h$. The upper line is parametrised by $h_{\max}(\Delta)=J(1+\Delta)/2$ while the lower one by $h_{\min}(\Delta)= \sinh\eta \sum_{j=-\infty}^\infty  {e^{ i j \pi}}/{ (2\pi \cosh (j \eta))}$ for $\Delta={\rm arccosh}(\eta)>1$.}
	\label{fig:phasediagram}
\end{figure}

In particular, after a bipartitioning protocol from two thermal states at low (but different) temperatures, the GHD equations become analytically solvable, yielding qualitatively different results depending on whether or not the spectrum of the post-quench Hamiltonian has a gap. In the gapped phase, variations of the profiles are found to be exponentially small in the temperatures and are described by non-trivial functions of $\zeta$~\cite{bertini2018low}. In the gapless regime, instead, the leading order contributions for the profiles are polynomial functions of the temperature, which turn out to be universal. In fact, in this limit GHD allows one to recover the predictions of conformal field theory (CFT)~\cite{bernard2012energy,bernard2016conformal,bernard2016hydrodynamic}. In order to illustrate this, let us consider for concreteness the Hamiltonian~\eqref{eq:heisenberg_XXZ} in the gapless regime, which is realized for values $(\Delta,h)$ in a strip $h_{\rm min}(\Delta)\leq |h|\leq h_{\rm max}(\Delta)$ (see Fig.~\ref{fig:phasediagram}), and initialise the system by joining together two thermal states with inverse temperatures $\beta_L$, $\beta_R$. As $\beta_L,\beta_R\to \infty$ only low energy modes around the Fermi point are expected to contribute and the system is expected to be described by a CFT. This implies that the only relevant quasiparticles are those moving at the Fermi velocity and the profiles of local observables take the form of a ``three-step staircase''. In particular the NESS values (corresponding to the central step) can be computed exactly in the conformal limit, yielding the following results for the energy density and current~\cite{bernard2012energy,bernard2016conformal,bernard2016hydrodynamic}
\begin{subequations}\label{eq:cft_predictions}
\begin{align}
e_{\rm CFT}=\frac{c \pi}{12  v} \left(\beta_{L}^{-2}+\beta_{R}^{-2}\right)\,,\\
j_{\rm eCFT}=\frac{\pi c}{12} \left(\beta_{L}^{-2}-\beta_{R}^{-2}\right)\,,
\end{align}
\end{subequations}
where $c$ and $v$ are respectively the central charge and the speed of light in the CFT ($c=1$ for the critical Heisenberg chain, while $v$ is a non-trivial function of $\Delta$ and $h$). Note that the NESS current (and also the energy density) shows the ``non-interacting" structure discussed in Sec.~\ref{sec:NESS}: it is the sum of two terms, each one depending solely on the initial state of one lead. 

Predictions~\eqref{eq:cft_predictions} can be recovered from a low-temperature expansion of the GHD equations, which also allows one to access higher-order corrections in $\beta^{-1}_{L,R}$~\cite{bertini2018low}. In fact, GHD reveals modifications to the CFT picture for local observables different from the energy density and current. In particular, at the edges of the light cone of generic observables there appears a region of width $T_{L/R}= \beta^{-1}_{L,R}$ where the leading contribution is linear, rather than quadratic, in the temperature. An example is given in Fig.~\ref{fig:lowT_profiles}, where we report the magnetisation profile for an inhomogeneous quench in the critical Heisenberg chain. 

Surprisingly, it was found that such a broadening of light cone could also be described by a universal function, which can be derived via a non-linear Luttinger-Liquid approach~\cite{bertini2018universal}. Within this paradigm, one approximates the Hamiltonian spectrum near the Fermi points, using fermionic quasiparticles with dispersion relation
\be
\varepsilon_{r}(p)=v|p|+\frac{r}{2 m_{\ast}}|p| p+O\left(p^{3}\right),\qquad k_{r}(p)=r|p|,
\ee
where $r=\pm 1$, while $v$ and $m_\ast$ are phenomenological parameters respectively associated with the velocity and mass of the quasiparticles. From this approach, one finds that, in a neighbourhood of the light cone quantified by $\zeta-v \sim T_{L(R)} /\left(m_{*} v\right)$, the leading contribution for the profiles of a given observable $\mathcal{O}$ is
\be\label{eq:universal_function}
\delta\langle\mathcal{O}\rangle_{\zeta}=\frac{d }{2 \pi v^{2}}\left(T_{L}  \log \left(1+e^{m_{\ast} v (\zeta-v)/T_{L}}\right)-T_{R}  \log \left(1+e^{m_{\ast} v (\zeta-v)/T_{R}}\right) \right)+O(T_{L/R}^{2})\,,
\ee
while $d$ is a constant which depends on the observable of interest. Once again, Eq.~\eqref{eq:universal_function} can be exactly recovered based on low-temperature expansions of the GHD equations~\cite{bertini2018low}. Note that, also including this correction, the NESS currents are still of non-interacting type.

The technical steps involved with the low-temperature analysis of the GHD equations are largely independent of the details of the specific model considered. However, as we have mentioned, in this limit qualitative differences emerge in models with more than one quasiparticle species. This is best exemplified in the Yang-Gaudin model of spinful fermions~\cite{takahashi2005thermodynamics}, where the study of bipartitioning protocols at small temperature reveals spin-charge separation effects~\cite{mestyan2018spin} (similar features are also observed in Fermi-Bose mixtures~\cite{wang2020emergent}). In this case, the profiles of local observables display a five-step form, with two distinct light cones propagating from the junction, see Fig.~\ref{fig:lowT_profiles} for an example. Observing profiles with this structure, one can argue that they are produced by two decoupled nonlinear Luttinger Liquids, rather than a single one. It is important to stress, however, that for external magnetic field $h\neq 0$ local observables couple the two theories: this is due to the fact that the decoupled Luttinger Liquids do not describe individual spin and charge excitations, but a combination of the two \cite{giamarchi2004quantum,essler2005one}. For $h=0$, spin and charge completely decouple at the level of the Luttinger Liquid description~\cite{giamarchi2004quantum}. However, by setting $h=0$ in the post-quench Hamiltonian, and constructing the initial state by joining together two Gibbs states at different temperatures, it is not possible to create a magnetization imbalance. Therefore, in the ensuing dynamics the magnetization remains frozen (and equal to zero), without any light cone. In conclusion, within the bipartitioning protocol, it is not possible to observe a genuine separation of spin and charge in the form of two distinct light cones. On the other hand, it was recently shown that GHD is capable to predict such a separation for more general inhomogeneous initial states, where the gas is initially confined in a trap potential~\cite{scopa2021real}. In this case, for vanishing post-quench magnetic field and low temperatures, an initial spin-charge imbalance lead to the formation of two separate light cones for spin and charge, whose real-time dynamics can be quantitatively captured by the GHD equations~\cite{scopa2021real}.

Finally, we mention that low-temperature limits of the GHD equations have also been investigated in integrable quantum field theories~\cite{bertini2019transport, denardis2019anomalous}, where they allowed to clarify the connection between GHD and the semiclassical approach developed by Sachdev and collaborators~\cite{sachdev1997low,sachdev1997low_II,damle1998spin,damle2005universal}. Specifically, Ref.~\cite{bertini2019transport} considered bipartitioning protocols in the sine-Gordon model recovering the predictions of Refs.~\cite{moca2017hybrid,semiclassical2018}, based on the semiclassical approach, as a low-energy limit of the GHD equations. In this limit the transport of the topological charge was found to be sub-ballistic. Away from the low-energy limit, however, the numerical solution of the GHD equations showed that transport is always ballistic, in conflict with the semi-classical predictions~\cite{bertini2019transport}.

\section{Quantum Entanglement Generated by Inhomogeneous Quenches}
\label{sec:v}

The dynamics of quantum correlations are generically very hard to describe exactly, both in homogeneous and inhomogeneous settings. This is essentially due to the fact that they go beyond the purely hydrodynamic description that arises at large times. For example, although GHD gives us the exact asymptotic values of one-point functions after quenches from bipartite initial states, connected equal-time correlation functions between points located at different rays are subleading in the scaling limit \eqref{eq:inhomoSS}. There are, however, some exceptions to this empirical fact where non-trivial correlations can actually be accessed. For instance, dynamical correlation functions along ballistic light-cones are indeed accessible within GHD~\cite{doyon2018exact,moller2020euler} (see also the contribution by Doyon, De Nardis, Medenjak and Panfil in the present volume). Moreover, as discussed in Sec.~\ref{sec:vPR}, one can recover an effective description for time-dependent quantum correlations for quenches starting from a particular class of initial states.

In this section we consider another of such remarkable examples. In particular we show how, retaining some genuine quantum correlations generated at the time of the quench, one can describe the linear growth of several entanglement measures. The key for this to happen is the presence of well-defined quasiparticles protected by integrability. After a quantum quench, EPR (Einstein-Podolsky-Rosen) correlations are created between quasiparticles with opposite quasimomenta. The balistic propagation of these quasiparticles transports these correlations leading to the growth of entanglement. An important remark is that this ``quasiparticle picture'' for the entanglement spreading applies to quenches, both homogeneous and inhomogeneous,  in which the steady state is described by a statistical ensemble with finite density of thermodynamic entropy, i.e., entropy per volume. For stationary states with zero entropy density entanglement-related quantities exhibit a sublinear growth as function of time which is not captured by this approach (see Sec.~\ref{sec:vPR}).

More specifically, to quantify the entanglement we use the R\'enyi entropies~\cite{amico2008entanglement,calabrese2009entanglemententropy,eisert2010colloquium,laflorencie2016quantum}, defined as 
\begin{equation}
\label{eq:renyi}
S_A^{(n)}(t)=\frac{1}{1-n}\ln[\mathrm{tr}\rho^n_A(t)],
\end{equation}
where $n$ is a real parameter while $\rho_A(t)$ is the density matrix of the system at time $t$ reduced to a subsystem $A$, i.e.
\be
\rho_A(t)=   {\rm tr}_{\!\bar A} \ket{\Psi(t)}\!\bra{\Psi(t)},
\ee
with $\bar{A}$ the region complementary to $A$ and $\ket{\Psi(t)}$ the evolved state of the system. The R\'enyi entropies characterise the spectrum of $\rho_A(t)$, sometimes called entanglement spectrum~\cite{laflorencie2016quantum}, encoding information on how entanglement is shared between $A$ and $\bar A$. In particular, in the limit $n\to1$ one recovers the von Neumann entropy 
\be
S_A(t)=- {\rm tr}\rho_A(t)\ln\rho_A(t)\,.
\ee
Beside their theoretical interest, the quantities \eqref{eq:renyi} are also experimentally relevant. Indeed, in the last few years it has become possible to address the dynamics of entanglement-related quantities with cold-atom experiments~\cite{islam2015measuring,kaufman2016quantum,chiaro2019direct,elben2020mixed}, and Noisy Intermediate Scale Quantum (NISQ) computers~\cite{smith2019simulating}. 

In the remaining part of this section we show that, combining GHD with a simple  quasiparticle picture, one can describe exactly the asymptotic dynamics of the von Neumann entropy in a particular scaling limit. 
The structure of the section is as follows. In Sec.~\ref{sec:quasi} we introduce the quasiparticle picture. 
In Sec.~\ref{sec:homo} this is applied to describe the entanglement spreading after homogeneous quenches, both for interacting and non-interacting systems. In Sec.~\ref{sec:nonint} we discuss the entanglement dynamics after an inhomogeneous quench 
in free-fermion systems. Finally, in Sec.~\ref{sec:ent-inho} we consider inhomogeneous quenches 
in interacting integrable systems. 

\subsection{Quasiparticle picture: a semiclassical description of entanglement spreading}
\label{sec:quasi}

The quasiparticle picture was originally proposed in the context of 
CFT~\cite{calabrese2005evolution} to describe entanglement dynamics 
after global quenches from homogeneous initial states. In essence, the idea is that a homogeneous quench produces an extensive number of 
quasiparticle excitations, which are responsible for propagating the entanglement throughout 
the system. The quasiparticles that are 
produced far apart are assumed not to be entangled with one another, i.e.\ they do not contribute 
to the coherent quantum correlations. Only quasiparticles that are produced at the 
same point in space are entangled. As the entangled quasiparticles propagate through the system, quantum correlations spread accordingly. A further simplifying assumption is that only {\it pairs} of entangled quasiparticles are produced, the two members of the pair being emitted with velocities of opposite sign. Finally, entangled quasiparticles travel as 
free particles, i.e., they do not interact.

Let us now consider the case in which all the quasiparticles have velocity with the same magnitude $v$. This corresponds to models with perfect linear dispersion like CFTs. Within the quasiparticle picture the 
von Neumann entropy between a region $A$ and its complement at a generic time $t$ is 
proportional to the number of entangled pairs that are shared between them. Let us consider 
a finite interval $A$ of length $\ell$ embedded in an infinite chain. At short time $t\le \ell/(2v)$ the number of 
shared entangled pairs is proportional to the horizontal width of the shaded areas in Fig.~\ref{fig:quasi_free} (a) at that time, which is $4vt$. On the other hand, for $t>\ell/(2v)$ the number of entangled pairs is proportional to $\ell$. In conclusion, one has that
\begin{equation}
\label{eq:cft}
S_{\ell} (t)=4v t s\Theta(\ell-2v t)+\ell s\Theta(2v t-\ell).
\end{equation}
Here $s$ is the ``entanglement content" of each pair of entangled quasiparticles (it is the contribution of a single pair times the density of pairs). Note that, due to translational invariance, the only piece of information about $A$ we need to know is the length of the interval, while the  
position of $A$ within the chain is not important. For this reason we used the notation $S_{\ell}(t)$ in \eqref{eq:cft}. 

The interpretation of~\eqref{eq:cft} is straightforward, and it is shown in Fig.~\ref{fig:bip}. 
For $2vt<\ell$, $S_{\ell}(t)$ grows linearly. All the pairs that originated in the region 
of the $t=0$ axis shaded by the two light cones are shared between $A$ and $\bar A$. 
At any time $2vt>\ell$ the number of shared entangled pairs saturates to 
an extensive value ($S_{\ell}(t) \propto \ell$). 
Eq.~\eqref{eq:cft}, at this level, has to be regarded as a phenomenological 
description of the entanglement dynamics. In the next sections we will show, however, 
that with minimal modifications the quasiparticle picture can be made quantitatively accurate 
in specific integrable systems. 

\subsection{Entanglement dynamics after homogeneous quenches in integrable systems}
\label{sec:homo}

Let us now discuss how to apply the quasiparticle picture to microscopic integrable systems. 
We first focus on homogeneous quantum quenches in non-interacting models. To promote Eq.~\eqref{eq:cft} to a quantitative prediction, these have to be fixed from the microscopic data of the integrable model under consideration. As we will see, to do that one only needs ``thermodynamic information" about the system.

Let us begin by considering non-interacting systems. In this case it is quite natural to associate the entangling quasiparticles with the free modes that diagonalise the Hamiltonian. Unlike~\eqref{eq:cft}, such single-particle modes have a nontrivial 
dispersion, i.e. a mode $k$ has energy $\varepsilon(k)$ that depends on $k$ (see, e.g., Sec.~\ref{sec:toy-model}). Their velocities 
are then given by the group velocities of these modes 
\be
v(k)=\varepsilon'(k)=\frac{\rm d}{{\rm d}k}\varepsilon(k),
\ee
which is the same quantity appearing in~\eqref{eq:currentsrho}. Note that for non-interacting systems $v(k)$ depends only on the model's dispersion, and not on the pre-quench state. 

Adopting the quasiparticle picture, we restrict to the situation where only locally generated pairs are entangled, and we take them to have opposite momentum~\cite{Note10}.\footnotetext[10]{This can be justified by arguing that, together with being translational invariant, the initial state is typically also parity invariant.} This means that correlated pairs are specified by a single momentum $k$. Assuming an even dispersion relation $\varepsilon(k)$ we then find the following generalisation of \eqref{eq:cft}
\begin{equation}
\label{eq:quasi-1}
    S_{\ell} (t)=2t\int_{2|v(k)|t<\ell} {\rm d} k\,\, |v(k)|s(k)+
    \ell\int_{2|v(k)|t>\ell} {\rm d} k\,\, s(k). 
\end{equation}
Now, to completely fix \eqref{eq:quasi-1} we just have to determine the entanglement content of the pair $k$, which we denote by $s(k)$ to stress that it now depends of the quasimomentum. This can be fixed by imposing that the integrand of \eqref{eq:quasi-1} relaxes to the density of thermodynamic entropy of the GGE that describes the steady state after the quench~\cite{alba2017entanglement}. For free fermionic systems the latter is nothing but the integrand of the Yang-Yang entropy in Eq.~\eqref{eq:S-YY-free}, namely we have 
\be
\label{eq:sk}
s(k)=s_{\rm YY}[\rho(k)]\equiv\frac{1}{2\pi}\log\frac{1}{2\pi}-\rho(k)\log \rho(k)- \left(\frac{1}{2\pi}-\rho(k)\right)\log\left(\frac{1}{2\pi}-\rho(k)\right),
\ee
where $\rho(k)$ is the root density of the GGE. 

Crucially, for quenches in free fermionic systems the validity of \eqref{eq:quasi-1}--\eqref{eq:sk} can be proved~\cite{fagotti2008evolution}. Namely, by solving exactly the dynamics one can show that \eqref{eq:quasi-1} gives the leading order contribution for large $t$ and $\ell$, and, in particular, it becomes exact in the space-time scaling limit $t,\ell\to\infty$, with $\ell/t$ fixed.

As anticipated before, Eq.~\eqref{eq:quasi-1} contains only thermodynamic 
information about the system and the quench. Indeed, the group velocity of the 
quasiparticles is a property of the system's dispersion relation, and the entropy of the GGE is a thermodynamic quantity. For systems with a maximal velocity $v_M$ (for example those fulfilling the hypothesis of the Lieb Robinson bound~\cite{lieb1972finite}), Eq.~\eqref{eq:quasi-1} predicts a linear growth at short times $2v_Mt<\ell$, followed by a saturation to the volume law scaling $S_{\ell}(t)\propto\ell$ at asymptotically long times. In contrast with~\eqref{eq:cft} the saturation is not abrupt because the quasiparticles have a dispersion. Note that, while in the original quasiparticle picture 
only pairs of entangled quasiparticles are assumed to contribute to the entanglement (and such an assumption is verified in several quantum quenches in non-interacting models~\cite{fagotti2008evolution} 
and for some initial states in integrable interacting systems~\cite{alba2017entanglement,piroli2017what}, see also~\cite{delfino2014quantum,delfino2017theory}), this is not true in general. Specifically, for some initial states multiparticle entanglement, for instance involving triplets of 
quasiparticles, can be present. Some of these situations have been investigated in free-fermion systems, and the quasiparticle picture extended accordingly~\cite{bertini2018entanglementwith,bastianello2018spreading,bastianello2020entanglement_II}. 

%
\begin{figure}[t]
\centering
\includegraphics[width=.5\textwidth]{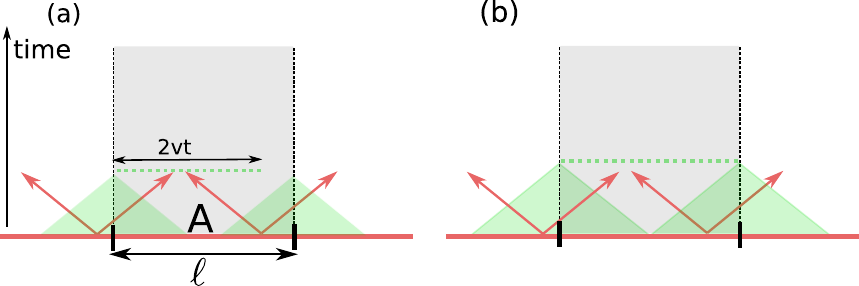}
\caption{Quasiparticle picture for the entanglement spreading 
 after a homogeneous quench in integrable systems. Pairs of entangled 
 quasiparticles are produced uniformly after the quench. We consider the 
 entanglement entropy $S_{\ell}(t)$ of an interval $A$ embedded in an infinite chain. 
 At a given time $t$, $S_{\ell}(t)$ is proportional to the total number of entangled pairs 
 shared between $A$ and its complement $\bar A$: One quasiparticle of the pair is inside 
 $A$, and the other one inside $\bar A$.  (a) Linear growth of the 
 entanglement entropy as $4 vt$ for $2vt<\ell$. The entanglement entropy is proportional 
 to the portion of the $t=0$ axis shaded by the two light cones. At time $t$ all the 
 right-moving members of the entangled pairs that are shared via the left edge of $A$ are 
 in the region denoted by the dotted line. (b) At $2v t>\ell$ the number of shared entangled 
 pairs is $\ell$ and the entanglement entropy exhibits a volume law $S_{\ell}(t)\propto\ell$. 
}
\label{fig:bip}
\end{figure}
%

Let us now discuss quenches from homogeneous initial states in {\it }interacting} systems. In particular, let us consider the case of Bethe Ansatz solvable models with a single species of quasiparticles (see Sec.~\ref{sec:interactions}). In this case the quasiparticle prediction~\eqref{eq:quasi-1} is modified as follows~\cite{alba2017entanglement}
\begin{equation}
\label{eq:conj}
S_{\ell}(t)= \sum_n\Big[ 2t\!\!\!\!\!\!\int\limits_{\!2|v_n|t<\ell}\!\!\!\!\!\!
{\rm d} \lambda\,\, |v_n(\lambda)|s_n(\lambda)+\ell\!\!\!\!\!\!\int\limits_{2|v_n|t>\ell}\!\!\!\!\!\!
{\rm d} \lambda\,\, s_n(\lambda)\Big]\,,
\end{equation}
where the sum runs over the quasiparticle bound states, labeled by $n$, while $\lambda$ denotes the rapidities. Apart from the sum over $n$, Eq.~\eqref{eq:conj} has the same 
structure as~\eqref{eq:quasi-1}.  Also the nature of the entangled quasiparticles is the same as compared to free systems: the quasiparticles in~\eqref{eq:conj} are constructed as the low-lying (particle-hole) excitations around the stationary thermodynamic macrostate, i.e., the GGE, that describes the steady state after the quench. However, the properties of the 
quasiparticles depend on the specific initial state, unlike the free-fermion case. 
For instance, the group velocities of the quasiparticles can be calculated by solving the system of integral equations in~\eqref{eq:vrhot}, and now depend on the pre-quench state (because the GGE depends on the initial state). Finally, in analogy with non-interacting systems, $s_n(\lambda)$ in~\eqref{eq:conj} is set to be equal to $s_{\rm YY}[\rho_{n}(\lambda)]$, the density of the Yang-Yang entropy of the GGE described by $\{\rho_n(\lambda)\}$. The latter is given by 
\begin{equation}
\label{eq:yy-int}
s_{\rm YY}[\rho_{n}(\lambda)] \equiv  \rho_{t,n}(\lambda)\ln\rho_{t,n}(\lambda)-\rho_n(\lambda)\ln(\rho_{n}(\lambda))
-(\rho_{t,n}(\lambda)-\rho_n(\lambda))\ln(\rho_{t,n}(\lambda)-\rho_n(\lambda)),
\end{equation}
where, in contrast to the free case (cf. Eq.~\eqref{eq:rhot}), $\rho_{t,n}(\lambda)$ is a nontrivial function of $\lambda$. We remark that Formula~\eqref{eq:conj} can also be generalized to the case of integrable models with multiple species of particles~\cite{mestyan2017exact,modak2019correlation}. 

Note that, differently from Eq.~\eqref{eq:quasi-1},  Eq.~\eqref{eq:conj} is a conjecture. Up to now it has been verified numerically in several quenches in the XXZ Heisenberg model~\cite{alba2017entanglement, alba2018entanglement} and in nested spin chains~\cite{modak2019correlation}, as well as analytically for some solvable quenches in the quantum cellular automaton ``Rule 54''~\cite{klobas2021exact,klobas2021exact_2,klobas2021entanglement}.

Note that, while the quasiparticle picture has been extended to study the R\'enyi entropies in the steady state~\cite{alba2017quench,alba2017renyi,mestyan2018renyi}, also for quenches from inhomogeneous initial states~\cite{alba2019towards}, the full dynamics of the R\'enyi entropies does not appear to be captured by a formula similar to \eqref{eq:conj}. We also remark that the quasiparticle picture can be easily adapted to describe the dynamics of the mutual information between two intervals~\cite{mestyan2017exact, alba2018entanglement,alba2019quantum,mestyan2020molecular}. Moreover,  it is possible to show that the so-called logarithmic negativity, which is a genuine entanglement measure between two disjoint intervals, becomes the same as the mutual information constructed from the R\'enyi entropy with R\'enyi index $n=1/2$, cf.~\eqref{eq:renyi}. This has been checked in free-fermion and free-boson models in Ref.~\cite{alba2019quantum_2}. Finally, we mention that, very recently, the quasiparticle picture has been applied to 
describe entanglement dynamics in simple free-fermion systems subject to gain and loss dissipation~\cite{alba2021spreading}. 

\subsection{Entanglement dynamics after inhomogeneous quenches: non-interacting systems}
\label{sec:nonint}

\begin{figure}[t]
\centering
\includegraphics[width=.7\textwidth]{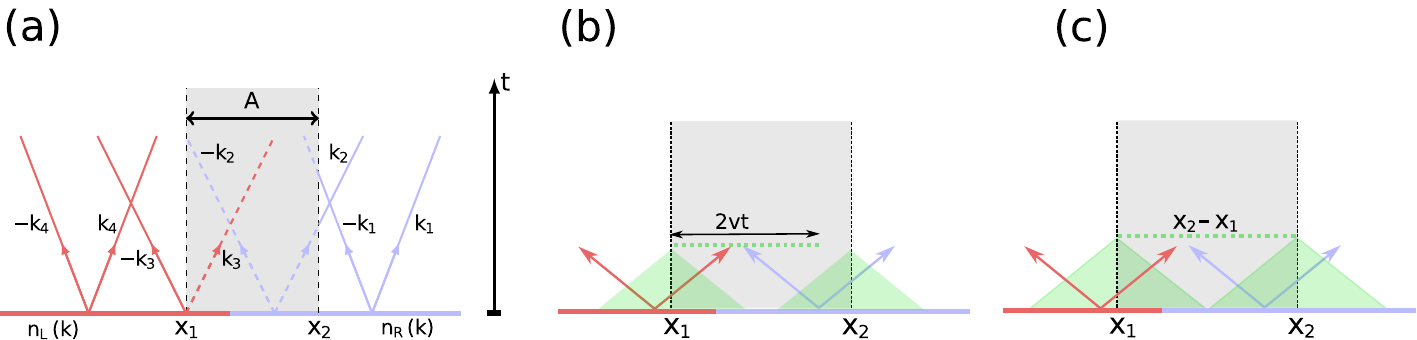}
\caption{ Hydrodynamic description of the entanglement spreading in free-fermion 
 models after a quench from an inhomogeneous initial state. (a) The 
 initial state obtained by joining two homogeneous but different 
 states (left (L) and right (R)) is left to evolve under a homogeneous 
 free-fermion Hamiltonian. 
 Both halves act as sources of entangled pairs of quasiparticles. The quasiparticles 
 are produced with opposite momenta and travel with opposite velocities. 
 Given an interval $A=[x_1,x_2]$, its entanglement entropy 
 is proportional to the number of entangled pairs shared 
 with its complement. The entanglement contribution of the pair 
 is the Yang-Yang entropy of the half where the pair is produced 
 at $t=0$. (b) Short-time entanglement dynamics.  
 The dotted line denotes the allowed position of the right-moving 
 member of an entangled pair that is shared via the left edge of $A$ at 
 $x_1$. The corresponding allowed position for the pairs shared via 
 the edge at $x_2$ is not shown. (c) Entanglement saturation 
 at $2 v t=x_2-x_1=\ell$. The number of shared pairs is $\ell=x_2-x_1$. 
}
\label{fig:quasi_free}
\end{figure}
%

Let us now see how the quasiparticle picture can be applied to inhomogeneous setups. We begin considering the case of non-interacting systems. Specifically, we focus on a bipartitioning protocol in the toy model introduced in Section~\ref{sec:toy-model}. The initial state is obtained by joining two semi-infinite chains (left and right) prepared in two homogeneous initial states. The setup is depicted in Fig.~\ref{fig:quasi_free}. The core of the original quasiparticle picture, namely the idea that entanglement propagates via quasiparticles initially produced in pairs in the bulk of the two chains, remains the same. In contrast to the homogeneous case, however, pairs produced in the left and right chains  have different entanglement content. On the other hand, since the system is evolved with a free homogeneous Hamiltonian the velocity of the quasiparticles does not depend on the region where they are produced. Putting all together, one arrives at the following quasiparticle prediction for the entanglement of an interval $[x_1,x_2]$~\cite{bertini2018entanglement}   
\begin{equation}
\label{eq:S-free}
	S_{[x_1,x_2]}(t)=\int {\rm d}k\,\, \Theta(-v(k))\int_{\max(x_2+2v(k)t,x_1)}^{x_2}
	 {\rm d}x\,\, s_{x-v(k)t}(k)+
\int {\rm d}k\,\, \Theta(v(k))\int^{\min(x_1+2v(k)t,x_2)}_{x_1}
	{\rm d}x\,\, s_{x-v(k)t}(k). 
\end{equation}
Note that, because of the absence of translational invariance, we have to keep track of both $x_1$ and $x_2$, defining the boundary of the subsystem under consideration. Accordingly, our notation for the entanglement entropy has changed to $S_{[x_1,x_2]}(t)$.
The integral over $k$ is over the Brillouin zone $[-\pi,\pi]$, $v(k)$ is the quasiparticles' group velocity, and $s_x(k)$ is the entanglement 
content of the quasiparticles, which is to be determined. 

Before doing that, let us comment on the physical interpretation of~\eqref{eq:S-free} (see Fig.~\ref{fig:quasi_free} (b)). 
Let us focus on the first term in~\eqref{eq:S-free}. 
This describes the entropy contribution of the quasiparticles (members of  
entangled pairs that are shared with the complement of $A$) with 
$v(k)<0$ that at time $t$ are within $A$. They were originated around 
the edge of the interval at $x_2$, and at a generic time 
$t$ are in the region $[\max(x_2+2v(k)t,x_1)]$. The second term 
takes into account the quasiparticles with $v(k)>0$ that were 
originated near the edge on the right of the interval at $x_1$ and are members of an entangled pair 
shared via that edge. At the generic time $t$ the allowed 
position of the quasiparticle is in the interval $[x_1,\min(x_1+2v(k)t,x_2)]$. 

We now discuss the entanglement content of the quasiparticles. Within the quasiparticle picture  there is no 
creation of entanglement during the evolution, but entanglement is simply 
created soon after the quench, and then ``transported'' by the quasiparticles.
Indeed, because 
of the argument $x-v t$ of $s_{x-v(k)t}$ in~\eqref{eq:S-free} the entanglement contribution is 
traced back to the sites where the entangled pairs were produced.  As it turns out~\cite{bertini2018entanglement} , the entanglement content $s_x$ of the quasiparticles is again fixed by the Yang-Yang entropy transported from the two 
initial chains
\begin{equation}
	s_x(k)=\Theta(-x)s_{\rm YY}[\rho_{\rm L}(k)]+\Theta(x) s_{\rm YY}[\rho_{\rm R}(k)]. 
\end{equation}
Here $\rho_{\rm L/R}$ are the root densities corresponding to the GGEs that describe 
the bulk of the two chains, i.e., for $x/t\to\pm\infty$, and $s_{\rm YY}$ is the associated 
Yang-Yang entropy density (cf. Eq.~\eqref{eq:sk}). The entanglement content of 
the quasiparticles can be rewritten in terms of the root density $\rho_{x/t}(k)$ 
associated with the GGE that describes the system at time $t$ and distance $x$ 
from the origin (cf.~\eqref{eq:solutionbipartitionfree}). To do so one uses the fact that the Yang-Yang entropy density satisfies the 
same GHD equation~\eqref{eq:continuityscaling} to write  
\begin{align}
s_{x-v(k)t}&=\Theta(x-v(k)t)s_{\rm YY}[\rho_{\rm R}(k)]+\Theta(v(k)t-x)s_{\rm YY}[\rho_{\rm L}(k)]\notag\\
                &=s_{\rm YY}[\Theta(x-v(k)t)\rho_{\rm R}(k)+\Theta(v(k)t-x)\rho_{\rm L}(k)]\notag\\
                &=s_{\rm YY}[\rho_{x/t}(k)]\,,
\end{align}
where in the last step we used \eqref{eq:solutionbipartitionfree}. Plugging this in~\eqref{eq:S-free}, one obtains 
\begin{equation}
\label{eq:S-free-a}
S_{[x_1,x_2]}(t)=\int {\rm d}k\, \Theta(-v(k))\int_{\max(x_2+2v(k)t,x_1)}^{x_2}
\!\!{\rm d}x\,\, s_{\rm YY}[\rho_{x/t}(k)]+
\int {\rm d}k \Theta(v(k))\int^{\max(x_1+2v(k)t,x_2)}_{x_1}
\!\!{\rm d}x\,\, s_{\rm YY}[\rho_{x/t}(k)]. 
\end{equation}
Importantly, despite the inhomogeneous setup, Eq.~\eqref{eq:S-free-a} still predicts a linear 
growth at short times, followed by a saturation to a volume-law entropy. 
The inhomogeneous initial condition is reflected in a non-monotonic behaviour at intermediate 
times~\cite{bertini2018entanglement}. For later comparison with the interacting case, we specify the above results to the entanglement 
production rate between two initial semi-infinite chains. Namely we fix $x_1=0$, $x_2\to\infty$ in~\eqref{eq:S-free-a} and neglect the contribution of the right boundary (which corresponds to choosing open boundary conditions). This yields 
\begin{equation}
\label{eq:s0}
S_{[0,\infty]}=t\int {\rm d}k\,\,  v(k)\Theta(v(k))(s_{\rm YY}[\rho_{\rm L}(k)]+s_{\rm YY}[\rho_{\rm R}(k)]).
\end{equation}
At this point it is straightforward to rewrite Eq.~\eqref{eq:s0} in terms of quantities depending only on the GGE that 
describes the interface between the two chains, namely  
\begin{equation}
\label{eq:s0-last}
S_{[0,\infty]}=t\int {\rm d}k\,\, |v(k)|s_{\rm YY}[\rho_{x/t=0}(k)].
\end{equation}
Several remarks are in order. First, the entanglement production rate (i.e., the rate at which the entanglement entropy grows) depends only on the physics at the 
interface between the two chains, which is described by $\rho_{x/t=0}$. 
This is expected because within the quasiparticle picture the entanglement 
growth reflects quasiparticles crossing the interface. Second, the quantity on r.h.s.\ of  Eq.~\eqref{eq:s0-last} is nothing but the rate at which the two chains exchange thermodynamic entropy. Implying that the latter coincides with the entanglement production rate. We anticipate that while the first property remains true 
for interacting integrable systems, the second one does not. 

%
\begin{figure}[t]
\centering
\includegraphics[width=.8\textwidth]{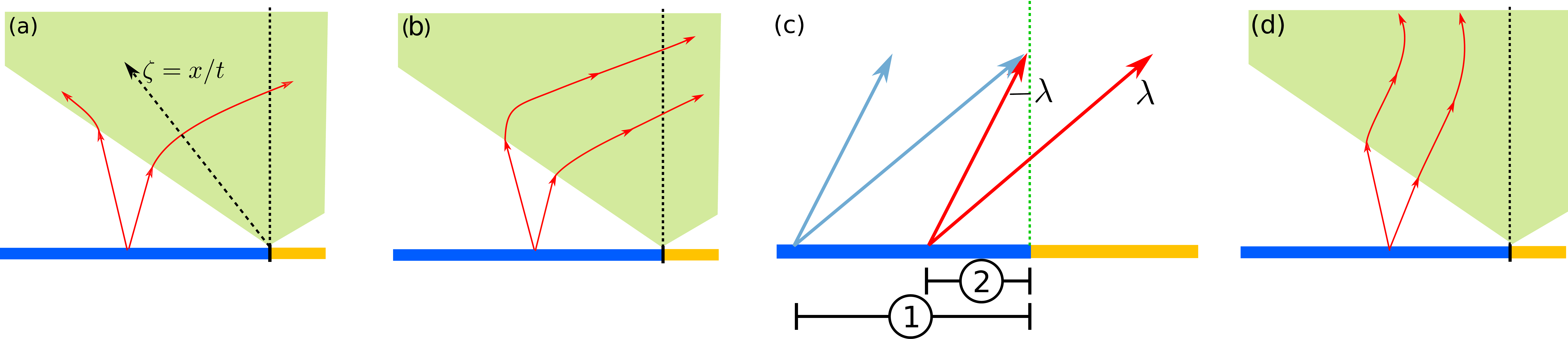}
\caption{ Dynamics of the entangled quasiparticles in interacting integrable systems. 
 Only entangled pairs of quasiparticles produced in one of the two subsystems are 
 considered. Before entering the light cone the quasiparticles follow straight 
 lines trajectories traveling with opposite velocities. Within the light cone 
 the velocities are nontrivial functions of the ray $\zeta=x/t$. This implies 
 that within the light cone the quasiparticles follow nonlinear trajectories. Three 
 different physical situations are possible: (a) at infinite time the two quasiparticles 
 forming the entangled pair are in different subsystems. (b)  Both members of the pair 
 are in the left subsystem. (c) Contribution to the entanglement of configurations  
 in (b). At time $t$ entangled pairs produced 
 in region $1$ are either shared or both members of the pair crossed the origin at $x=0$. Both members 
 of the pairs produced in $2$ crossed the origin 
 and are not shared anymore. 
 (d) Both entangled quasiparticles remain in the left chain and do not contribute to the 
 entanglement entropy. Note that both in (b) and (d) the velocities change sign before crossing 
 the interface between the two subsystems, due to the dressing. 
}
\label{fig:traj}
\end{figure}
%

\subsection{Entanglement dynamics after inhomogeneous quenches: interacting systems}
\label{sec:ent-inho}

Let us now discuss the entanglement dynamics after bipartitioning protocols in {\it interacting} integrable systems. Once again, the main idea of the quasiparticle picture is the same as for free models: entangled pairs of quasiparticles are produced in the bulk of the two 
chains, with entanglement content given by the density of thermodynamic Yang-Yang entropy in the GGE for $x/t\to\pm\infty$. Moreover, following the reasoning employed in the homogeneous interacting case (cf. Sec.~\ref{sec:homo}), here we assume that the quasiparticles are nothing but low-lying excitations on the (quasi) stationary state, which in this case is the LQSS $\hat \rho_{s}(\zeta)$. This allows us to identify the velocity of the excitations with $v_{n,x/t}(\lambda)$, obtained by solving \eqref{eq:rhot} and \eqref{eq:vrhot} with $\rho_{n}(\lambda)$ replaced by the root density $\rho_{n,x/t}(\lambda)$ of the LQSS (cf. Sec.~\ref{eq:inhomogeneousinteracting}). 

The main complication introduced by this identification is that the trajectories of the quasiparticles are not straight lines. This means that, given a pair of entangled quasiparticles with rapidities 
$\lambda$ and $-\lambda$ forming an entangled pair, it is a nontrivial task to trace back their trajectories and identify the subsystem in which they were produced. This step is, however, necessary to assign to the entangled pair the correct entanglement content. Still, as we are going to show, it is possible to provide compact analytical results for the entanglement dynamics. Moreover, these take a rather revealing form for the  entanglement production rate between the two chains. 

Let us start by defining as $X_{n,\lambda}(x,t)$ the position at time $t$ 
of a quasiparticle created at 
position $x$ with rapidity $\lambda$ and bound-state index $n$. 
The trajectory of the quasiparticle is determined by the 
following classical equation of motion~\cite{alba2019entanglement} 
\be
\frac{\mathrm d }{\mathrm d t}X_{n,\lambda}(x,t)=
v_{n,{X_{n,\lambda}(x,t)}/{t}}\left(\lambda\right).
\label{eq:eom}
\ee
Since $v_{n,x/t}(\lambda)$ can be determined for any value of $x,t$ 
by solving~\eqref{eq:vrhot}, the trajectory of a generic quasiparticle
can be obtained numerically from~\eqref{eq:eom} by knowing its initial position. 
Alternatively, on can use the the so-called ``flea gas'' method~\cite{doyon2018soliton}
to simulate numerically the motion of the quasiparticles and of the entangled pairs~\cite{mestyan2020molecular}.

From the analytical point of view, the integration of~\eqref{eq:eom} seems a daunting task. Yet, in some simple situations it is possible to obtain explicit expressions for the dynamics of the entanglement entropy~\cite{alba2019entanglement}. Here we do not provide the full derivation of the results but we illustrate the main steps of the reasoning and discuss the physical implications. Let us begin by noting that 
Eq.~\eqref{eq:eom} can be integrated as 
\be
\int_{x/t_0}^{X_{n,\lambda}(x,t)/t} \frac{d\zeta}{v_{n,\zeta}(\lambda)-\zeta}=\ln\frac{t}{t_0}, 
\label{eq:eom1}
\ee
with $x$ the initial position of the quasiparticle and $t_0$  a transient time sufficiently large 
to ensure the validity of the hydrodynamic description~\cite{alba2018entanglement}. 
In order to assign the correct entanglement content to the quasiparticles one has 
to trace back the quasiparticles' trajectory expressing their final position 
$X_{n,\lambda}(x,t)$ in terms of their initial one $x$. First, from~\eqref{eq:eom1} one 
can derive~\cite{alba2019entanglement}
\begin{align}
\label{eq:phi}
\frac{x}{t}=&\Theta\Big(\zeta_{n,\lambda}-\frac{X_{n,\lambda}}{t}\Big)[v^\mathrm{min}-v_{n,-\infty}(\lambda)]\exp\Big[\int_{v^\mathrm{min}}^{X_{n,\lambda}/t}\frac{dz}{z-v_{n,\lambda}(z)}\Big]\notag\\
&+\Theta\Big(\frac{X_{n,\lambda}}{t}-\zeta_{n,\lambda}\Big)[v^\mathrm{max}-v_{n,\infty}(\lambda)]\exp\Big[\int^{v^\mathrm{max}}_{X_{n,\lambda}/t}\frac{dz}{v_{n,z}(\lambda)-z}\Big],
\end{align}
where $\zeta_{n,\lambda}$ is the solution of $\zeta=v_{n,\zeta}(\lambda)$, while $v^\mathrm{min}=\min_n\min_\lambda v_{n,-\infty}(\lambda)$, and $v^\mathrm{max}=\max_n\max_\lambda v_{n,\infty}(\lambda)$. Eq.~\eqref{eq:phi} gives the initial position $x$ of a generic 
quasiparticle in terms of its final one $X_{n,\lambda}(x,t)$ at time $t$. 
The method used above is similar to the so-called method of characteristics to solve 
first order partial differential equations~\cite{courant1989methods}. 

The crucial step to derive the quasiparticle picture is to establish a relation between the trajectory of the two quasiparticles forming an entangled pair, i.e., having rapidity $\lambda$ and $-\lambda$. To this end, one defines a function $J_{n,\lambda}(\zeta)$ such that $J_{n,\lambda}(x/t)t$ is the position at time $t$ of the quasiparticle labelled by $(n,\lambda)$ that started at the same point in space with the quasiparticle labelled by $(n,-\lambda)$, which at time $t$ is at position $x$. We do not provide the explicit expression of $J_{n,\lambda}(x/t)$, because is rather unwieldy, and we refer the interested reader to the original reference~\cite{alba2019entanglement}. 

In terms of $J_{n,\lambda}(x/t)$ one can express the dynamics of the von Neumann entropy of a given subsystem $A=[x_1,x_2]$ as follows~\cite{alba2019entanglement} 
\begin{align}
{S}_{[x_1,x_2]}(t)= t\sum_n\int\mathrm d 
\lambda\, &\left
\{\mathrm{sgn}(J_{n,-\lambda}(\zeta_1)
-\zeta_1)\mathrm{sgn}(\zeta_2-J_{n,-\lambda}(\zeta_1
))
(\zeta_1-v_{n,\zeta_1}(\lambda))s_{n,
\zeta_1}(\lambda)\right. 
\notag\\
&\left.
-\mathrm{sgn}(J_{n,
-\lambda}(\zeta_2)-\zeta_1)\mathrm{sgn}(\zeta_2
-J_{n,-\lambda}(\zeta_2))(\zeta_2-v_{n,
\zeta_2}(\lambda))s_{n,\zeta_2}(\lambda)\right\},
\label{eq:finalresult}
\end{align}
where we set $\zeta_i=x_i/t$. The entropies $s_{n,\zeta_{1/2}}(\lambda)$ are obtained by using~\eqref{eq:rhot}, \eqref{eq:solution_hydro} and the definition of the Yang-Yang entropy density~\eqref{eq:yy-int}. Eq.~\eqref{eq:finalresult} is the analog of~\eqref{eq:S-free-a} for interacting integrable systems. The first term in~\eqref{eq:finalresult} takes into account contributions to the entanglement entropy due to quasiparticles at the boundary at $x_1$, whereas the second one accounts for the boundary at $x_2$. Eq~\eqref{eq:finalresult} becomes much simpler for the entanglement production rate between two semi-infinite chains. By fixing $x_1=0$, $x_2\to\infty$ in~\eqref{eq:finalresult} and neglecting the contribution of the right boundary one finds 
\begin{equation}
\label{eq:eprate}
{S_{[0,\infty]}}(t)= t\sum_n\int\mathrm d \lambda\, 
\mathrm{sgn}(\lambda) v_{n,0}(\lambda)
s_{n,0}(\lambda),
\end{equation}
where we assumed $\mathrm{sgn}( v_{n,\pm\infty}(\lambda))=\mathrm{sgn}(\lambda)$, which is the typical situation in bipartitioning quenches~\cite{alba2019entanglement}. Eq.~\eqref{eq:eprate} predicts a linear growth of the entanglement 
entropy, and depends only on the macrostate with $x/t\to0$, similar to 
free fermions (cf.\ Eq.~\eqref{eq:s0-last}). 
An interesting result is that in the presence of interactions 
the growth rate of the entanglement entropy is different from the exchange rate 
of thermodynamic entropy, given by 
\begin{equation}
\label{eq:s-exch}
    S_{\mathrm{exch}}=t\sum_n\int d\lambda |v_{n,0}(\lambda)|s_{n,0}(\lambda). 
\end{equation}
As noted in the previous subsection this does not happen in the non-interacting case.
The origin of this difference is depicted in Fig.~\ref{fig:traj}, and it can be traced back 
to the dressing of the quasiparticles velocities due to the interactions (see~\eqref{eq:vrhot}). In particular,
Fig.~\ref{fig:traj} shows the possible trajectories of the quasiparticles forming 
entangled pairs, which are 
straight lines outside the light cone, while they are curved inside of it. 

Three different situations are possible: In the first one (see Fig.~\ref{fig:traj} (a)) the quasiparticles forming the entangled pair are in different subsystems at $t\to\infty$. Apart from the dressing of the velocities, this is the same situation observed in quenches from inhomogeneous initial states in free-fermion models, and in quenches from homogeneous initial states in interacting integrable ones. However, as shown in Figs.~\ref{fig:traj} (b) and (d), interactions can be so strong that the quasiparticles forming an entangled pair are in the same subsystem at $t\to\infty$. In particular, in the example depicted in Fig.~\ref{fig:traj} (b) both members of the entangled pair generated on the left chain are on the right one at $t\to\infty$, while the pair in Fig.~\ref{fig:traj} (d) is confined within the left chain where it originated. The latter configuration does not contribute to the entanglement. This happens because only the pairs that are shared between $A$ and $\bar A$  contribute to their mutual entanglement, and the pairs in Fig.~\ref{fig:traj} (d) are never shared. On the other hand the configuration in Figs.~\ref{fig:traj} (b) does contribute to~\eqref{eq:eprate}. Specifically, an entangled pair as in  Fig.~\ref{fig:traj} (b) contributes to the entanglement until both members of the pair are in the right subsystem. This leads to the factor 
$\mathrm{sign}(\lambda)$ in~\eqref{eq:eprate}. An important remark is that the situation in Fig.~\ref{fig:traj} (b) occurs for ``slow'' quasiparticles, for which the dressing of the velocities is stronger~\cite{alba2019entanglement}. In particular, this is true for bound states. Indeed, typically, the larger the bound-state size $n$ is, the smaller is its group velocity. An interesting fact is that for some inhomogeneous quench protocols by tuning the interaction strength, and hence changing the quasiparticles velocities, it is possible to completely suppress the contribution of the bound states to the entanglement dynamics and to transport in general. 
In this case  the scenario in Fig.~\ref{fig:traj} (d) 
holds for bound states with arbitrary $\lambda$.  Moreover, there is a ``critical'' value 
of interactions strength below which bound-state transport is permitted. Interestingly, the 
critical interaction  depends on the bound-state size $n$, and it decreases 
on increasing $n$. This means that 
by lowering the interaction strength it is possible to progressively allow 
for the transport of bound states with larger and larger $n$. 
These effects have  been investigated in Ref.~\cite{alba2019towards} (see also Ref.~\cite{alba2018entanglement}).

\begin{figure}
\centering
\includegraphics[width=.5\textwidth]{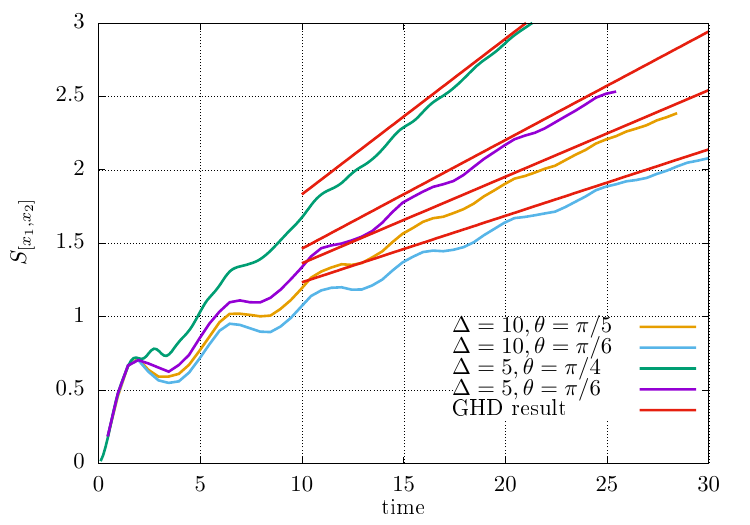}
\caption{Entanglement dynamics from inhomogeneous initial states in the 
 XXZ spin chain. Here the initial state is obtained by joining two semi-infinite 
 chains prepared in the N\'eel state and in the tilted ferromagnet. In the legend, $\theta$ 
 is the tilting angle and $\Delta$ the chain anisotropy. The black lines in the 
 figure are tDMRG data. The dashed lines are the analytic predictions obtained 
 by using \eqref{eq:eprate}. 
}
\label{fig:dmrg}
\end{figure}

Let us finally discuss some numerical checks of~\eqref{eq:eprate}. We consider once again the XXZ Heisenberg spin 
chain, and focus on the 
initial state obtained by joining together the N\'eel and the tilted ferromagnetic states, which are defined 
as 
\begin{align}
\label{eq:neel}
& |N\rangle=\left|\uparrow\downarrow\uparrow\cdots\right\rangle\,,\\
\label{eq:tferro}
& |F,\theta\rangle=e^{i\theta\sum_j \sigma_j^y/2}\left|\uparrow\uparrow\uparrow\cdots\right\rangle\,,
\end{align}
where we denoted by $\ket{\uparrow}$, $\ket{\downarrow}$ the two local basis states. We focus on the entanglement entropy between two semi-infinite chains prepared in~\eqref{eq:neel} and~\eqref{eq:tferro}. In Fig.~\ref{fig:dmrg} we report exact numerical data for the dynamics of the entanglement entropy between the two chains obtained by using the time-dependent Density Matrix Renormalization Group (tDMRG)~\cite{schollwock2011density}. The figure shows $S_{[x_1,x_2]}(t)$ plotted versus time. The different lines correspond to different values of the anisotropy $\Delta$ of the chain (cf.~\eqref{eq:heisenberg_XXZ}) and different tilting angles $\theta$ (cf.~\eqref{eq:tferro}). Clearly, the numerical data exhibit a nontrivial transient dynamics. Note in particular the presence of oscillating corrections. Still, at moderately long times $t\approx 10$ a clear linear behaviour sets in. In all the cases, this appears to be accurately captured by Eq.~\eqref{eq:eprate}, which is reported as red straight lines in Fig.~\ref{fig:dmrg}. 


\section{Quantum Fluctuations around Inhomogeneous Backgrounds}
\label{sec:vPR}

While in the previous section we have assumed a quasiparticle picture to gain insight into the entanglement dynamics, in this section we consider another approach to complements GHD and access \emph{exact} leading-order quantum correlations out of equilibrium. This approach, first introduced in Ref.~\cite{ruggiero2020quantum}, is based on the combination of GHD and a recently-developed inhomogeneous Luttinger-Liquid description~\cite{dubail2017conformal, brun2018inhomogeneous, ruggiero2019conformal} (see also Ref.~\cite{allegra2016inhomogeneous}), and allows one to compute exactly generic time-dependent correlation functions when evolving from low-energy states. 

The initial idea was that, at low energy, inhomogeneous systems can still be treated using CFT methods at the price of working in a curved background. While the first works were restricted to equilibrium situations~\cite{dubail2017conformal}, the method was later extended to time-dependent problems~\cite{ruggiero2019conformal}.
Meanwhile, it was realized that, giving up conformal invariance, the same framework allows us to treat much more general situations both in~\cite{brun2018inhomogeneous} and out-of-equilibrium~\cite{ruggiero2020quantum}. These ideas are better illustrated considering a specific model. In particular, following most of the relevant literature, here we focus on the case of the 1D Bose gas. This choice is also experimentally motivated as 1D Bose gases in external potentials are now routinely realised in cold-atom experiments, see e.g.~\cite{olshanii1998atomic,vanamerongen2008yang,vogler2013thermodynamics,schemmer2019generalized,moller2021extension} and the contribution by Bouchoule and Dubail to this special issue.

We organize the discussion as follows. In Sec.~\ref{sec:BoseGases_equilibrium} we focus on trapped 1D Bose gases at equilibrium, while in Sec.~\ref{sec:BoseGases_dynamics} we move to the out-of-equilibrium situation.

\subsection{Luttinger-Liquid treatment of trapped 1D Bose gases at equilibrium} \label{sec:BoseGases_equilibrium}

A 1D Bose gas with point-wise interactions --- also known as Lieb-Lininger model~\cite{lieb1963exact,korepin1997quantum} --- in the presence of an external potential $V(x, t)$, is described by the following Hamiltonian
\begin{equation} \label{model}
	H  \, = \, \int {\rm d}x  \left(  \frac{1}{2m}  ( \partial_x \Psi^\dagger ) (\partial_x \Psi) + (V(x, t)-\mu) \Psi^\dagger \Psi  + g \Psi^{\dagger 2} \Psi^2 \right), 
\end{equation}
where $\mu$ is a chemical potential and $\Psi^\dagger=\Psi^\dagger(x) $, $\Psi=\Psi(x)$ are operators that create/annihilate a boson at position $x$, satisfying the canonical commutation relations
\begin{equation}
[\Psi(x), \Psi^\dagger(x')]= \delta(x-x').
\end{equation}
There are several differences with respect to the model considered in the previous sections: besides its bosonic nature, this model is defined in the \emph{continuum} and, for a generic value of $g$, it is interacting. Moreover, while in the absence of confining potential the model is Bethe-Ansatz solvable~\cite{takahashi2005thermodynamics}, $V(x,t)$ breaks integrability.

\subsubsection{Homogeneous Luttinger Liquid}

In the homogeneous case (i.e.\ $V(x,t)=0$) and at equilibrium, the low energy physics (i.e.\ the physics on length and time scales that are large compared to microscopic ones) of the system \eqref{model} is captured by the Luttinger-liquid (LL) theory~\cite{giamarchi2004quantum} or, equivalenty, the CFT of a free compactified boson $\phi$~\cite{di1997conformal}. 
For later convenience, let us briefly review how such a bosonic field, sometime referred to as ``height field''~\cite{Nienhuis1987}, emerges from the microscopic degrees of freedom. 
In general, any local operator $\mathcal O(x)$ in the microscopic model can be written as a sum of local operators in the CFT
\begin{equation}
\mathcal O(x)= \sum_j A_{\mathcal O,\phi_j} \phi_j (x) \, ,
\end{equation}
with $A_{\mathcal O,\phi_j}$ non-universal constants, and $\{ \phi_j \}_j$ CFT operators, which can be ordered from the most to the least relevant one, according to their scaling dimension.
If we denote the microscopic density of the physical bosons by $\hat{\rho} (x)=\Psi (x)^{\dagger} \Psi(x)$, then its leading order in the CFT operators is given by 
\be
\hat{\rho} (x) =\rho_0 + \frac{1}{2\pi} \partial_x \phi (x) + \cdots,
\ee
with $\rho_0$ the average (constant) density. The boson correlations are fixed (in imaginary time) by the following action
\begin{equation}
S= \frac{1}{2\pi K} \int {\rm d}z\, {\rm d}\bar{z}\, \partial_z \phi (z,\bar{z}) \, \partial_{\bar{z}} \phi (z,\bar{z}) \, .
\end{equation}
Here $z= x + i v \tau$, $\bar{z}=x- i v \tau$, and $\tau$ is the imaginary time. The parameters of the model are $K$, known as Luttinger parameter and related to the interactions, and $v$, the sound velocity~\cite{haldane1981effective,giamarchi2004quantum}.  

\subsubsection{Inhomogeneous Conformal Field Theory in the Tonks-Girardeau limit}

Let us now turn to the case of non-zero external potential. We begin by considering the limit of infinite repulsion, $g \rightarrow + \infty$, also known as Tonks-Girardeau (TG) limit. This limit describes hard-core bosons or, equivalently (after a Jordan-Wigner trasformation) free fermions.
Specifically, by introducing the fermionic operators 
\be
\Psi_F^{\dagger} (x)= e^{i \pi \int_{y<x} \hat{\rho} (y) dy} \Psi^{\dagger} (x)\,,
\label{eq:JW}
\ee 
the Hamiltonian \eqref{model} becomes quadratic
\begin{equation} \label{f-model}
H = \int {\rm d}x \left( \frac{1}{2m} \partial_x \Psi_F^{\dagger} (x) \partial_x \Psi_F (x) + V(x,t) \Psi_F^{\dagger} (x) \Psi_F (x) \right).
\end{equation}
Note that bosonic and fermionic densities are the same under the mapping \eqref{eq:JW}, i.e.
\be
\hat{\rho} (y)=\Psi (y)^{\dagger} \Psi (y)=\Psi_F (y)^{\dagger} \Psi_F (y).
\ee 

Following the historical development (cf.\ Ref.~\cite{dubail2017conformal}) we begin by considering a gas at equilibrium at zero-temperature in a time-independent potential $V(x)$. We also assume that the inhomogeneity changes ``slowly enough'', namely we are in the regime of Eq.~\eqref{LDA_xt}: at this stage, since we are at equilibrium, it is enough to assume it in the ``space'' direction, which is nothing but LDA~\cite{cazalilla2011one}.

In this regime, the system can be considered homogeneous in a small patch (i.e.\ a small spatial region) and one can use results from the homogeneous theory. Matching these patches together, however, imposes constraints on the global low energy theory. In particular, Ref.~\cite{dubail2017conformal} showed that the only consistent field theory defined globally on the entire domain such that its propagator has the required local behaviour everywhere is a CFT of a Dirac fermion or, equivalently, a free boson, in a curved space-time. The associated action reads
\begin{equation} \label{actionCurvedCFT}
S [\phi]= \frac{1}{2 \pi} \int {\rm d}z\, {\rm d}\bar{z} \, g^{ab} \partial_a \phi (z,\bar{z}) \, \partial_b \phi(z, \bar{z}) \, \sqrt{-\det g},
\end{equation}
where $g^{ab}$ is the metric field ($a,b \in \{z,\bar{z} \}$) and we used that at the free Fermi point $K=1$. Importantly, the latter fact is not modified by the presence of the trapping potential. 

As in the homogeneous case, the bosonic field $\phi$ is related to the density fluctuations, but now around an inhomogeneous average. Namely, at the leading order in the scaling dimensions, the density is now written as 
\be
\hat{\rho}(x) =\rho_{0} (x) + \frac{1}{2 \pi} \partial \phi (x)+ \cdots,
\ee
where $\rho_0(x) $ is fixed by the LDA. In our case, where the inhomogeneity comes from a generic external potential $V(x)$, it is given by $\rho_{0} (x) =\sqrt{\mu -V(x)}$.
The curved metric in \eqref{actionCurvedCFT} is fixed by such (inhomogeneous) classical limit of the density. In particular, there exists a set of coordinates (known as \emph{isothermal} coordinates) for which the metric takes the simple form 
\begin{equation} \label{metricCurvedEq}
{\rm d}s^2= \pi^2 \rho_0 (x)^{2} \, {\rm d} {z'}\, {\rm d}\bar{z}'\,,
\end{equation}
with ($z',\bar{z}'$) given explicitly by
\begin{equation}
z'= \tilde{x} + i \tau,\qquad\qquad \tilde{x} = \int^x \frac{{\rm d}x'}{\pi \rho_0(x')}\,. 
\end{equation}
In such ``stretched" coordinates the system appears again uniform, except for the (overall) conformal factor $\pi^2 \rho_0 (x)^2$ in Eq.~\eqref{metricCurvedEq} (which can be eliminated by a Weyl transformation). Namely, the action is still of the form~\eqref{actionCurvedCFT}, with $g^{a b}$  replaced by the flat (euclidean) metric. The main difference with respect to the homogeneous case, is that now the sound velocity characterizing the CFT is not constant, but depends explicitly on the position, i.e.\ $v(x)=\pi \rho_0(x)$. This feature leads to the emergence of \emph{curved light cones}, as pointed out in~\cite{dubail2017emergence}.

In principle this approach gives access to all the correlation functions. Specifically, in the case of the Bose gas it leads to explicit predictions for the entanglement entropy (both in the ground state~\cite{dubail2017conformal} and in the low-lying excited states~\cite{murciano2018inhomogeneous}) and the one-particle density matrix $\langle\Psi^\dagger (x) \Psi(0) \rangle$~\cite{brun2017one} for a \emph{generic} form of the external potential. 

Other systems which can be described in terms of a CFT in a curved space are those where the Hamiltonian is  deformed via an envelope function $f(x)$, i.e.
\be
\mathcal{H}= \int {\rm d}x\, f(x)\, H(x)\,.
\label{eq:Hdeformed}
\ee 
Examples studied in the literature include sine-square and Moebius deformations~\cite{maccormack2019holographic}, and the so-called rainbow model~\cite{rodriguez2017more}. In the latter case, the underlying conformal invariance has been exploited to get analytic predictions and explicit results for entanglement-related quantities (i.e. different entanglement measures, entanglement Hamiltonian, and contour~\cite{tonni2018entanglement}).

\subsubsection{Inhomogeneous Luttinger Liquid for generic interaction strength}

Let us move away from the TG limit, and look at generic values of the interaction strength, $g$ (cf.~\eqref{model}). Also in this case we expect that the low energy physics is described by a quadratic action. This is ultimately related to the universality of the Luttinger-Liquid description and can be understood using the following argument (cf.~Ref.~\cite{brun2018inhomogeneous}).

First we assume that  the local degrees of freedom of the system of interest are captured by a single bosonic field $\phi$ associated to a local Hamiltonian. Next, we also assume that physical observables, and hence the action, are invariant under constant shifts $\phi \to \phi +\textrm{const}$. This automatically leads to an action of the form 
\begin{equation}
\mathcal{S}[\phi] = \int {\rm d}x\, {\rm d}\tau\,\,\mathcal{L} (\partial_x \phi, \partial_{\tau} \phi, \cdots),
\end{equation}
where $\mathcal{L}$ is the Lagrangian density. At this point, assuming that the action is minimised by a unique classical configuration $\phi_{cl}$, we expand at second order around the minimum and define
\begin{equation} \label{igff}
S[\phi] \equiv \mathcal{S}[\phi + \phi_{cl}] - \mathcal{S}[\phi_{cl}] = \frac{1}{8 \pi} \int {\rm d}x\, {\rm d}\tau\, \, \frac{\sqrt{-\det g}}{K(x)} g^{ab} \partial_a \phi (x,\tau) \, \partial_b \phi(x,\tau)\,, 
\end{equation}
with metric tensor
\begin{equation}
g = 
\begin{pmatrix} 
\displaystyle \frac{\partial^2 \mathcal{L}}{\partial (\partial_x \phi)^2} & \displaystyle \frac{\partial^2 \mathcal{L}}{\partial (\partial_x \phi) \partial (\partial_{\tau} \phi)}\\
\\
\displaystyle \frac{\partial^2 \mathcal{L}}{\partial (\partial_x \phi) \partial (\partial_{\tau} \phi)} &\displaystyle \frac{\partial^2 \mathcal{L}}{\partial (\partial_{\tau} \phi)^2} \\ 
\end{pmatrix},
\end{equation}
written in the coordinates $(x,\tau)$, and $K(x) \equiv [ 4\pi \sqrt{\det \nabla^2 \mathcal{L}}]^{-1}$ is interpreted as a space-dependent Luttinger parameter. 
All higher order terms in the expansion of the same action have scaling dimension larger than 2, and therefore are irrelevant in two dimensions in an RG sense.
We refer to this model as \emph{inhomogeneous Luttinger Liquid} (or iLL in short): literally, the inhomogeneous (and in general also time-dependent) generalization of the standard Luttinger-Liquid model~\cite{luttinger1963exactly,haldane1981effective}.
Note, however, that also other names appeared in literature (e.g., it was dubbed \emph{inhomogeneous Gaussian free field} in~\cite{brun2018inhomogeneous}).

The action in \eqref{actionCurvedCFT} is a special example of iLL which is also conformally invariant. This is essentially due to the fact that only the sound velocity of the Luttinger Liquid becomes position dependent, $v \to v(x)$, while the Luttinger parameter remains constant. The latter, however, turned out to be a special property of the infinite interaction limit (free fermionic point): for finite interaction also the Luttinger parameter generically acquires a spatial dependence, $K \to K(x)$, breaking conformal symmetry~\cite{brun2018inhomogeneous}. Nonetheless, since the resulting iLL theory is still quadratic, it can be used to numerically evaluate all correlation functions in terms of the two-point function, while the latter is found by solving an appropriate generalised Poisson equation~\cite{brun2018inhomogeneous}. All the inhomogeneous parameters are again obtained from the microscopic model relying on separation of scales, and using the exact Bethe Ansatz solution of the homogeneous case~\cite{shashi20122nonuniversal,shashi2012exact,kitanine2012form}. This approach has been used to compute the density profile (including density ripples)~\cite{brun2018inhomogeneous}, the one-particle density matrix~\cite{brun2018inhomogeneous}, and the entanglement entropy~\cite{bastianello2020entanglement} in the ground state of the trapped Lieb-Liniger model (and its anyonic generalization~\cite{scopa2020one}) for generic values of the interactions. Finally, we remark that there are special situations where also at finite values of the interactions $K$ remains constant~\cite{dubail2017emergence}, and therefore the model retains conformal invariance. An example is given by the stationary state resulting from a bipartitioning protocol, where two XXZ chains prepared in two ferromagnetic states with opposite magnetizations are joined together. There CFT methods were exploited in Ref.~\cite{collura2020domain} to analytically characterise the leading behaviour of the entanglement entropy. Moreover, in some cases, $K$ has been observed to vary very slowly~\cite{eisler2017front} and therefore it is still possible to rely on conformal symmetry to get analytic results.

\subsection{Luttinger-Liquid treatment of quenches in 1D trapped Bose gases} \label{sec:BoseGases_dynamics}

The inhomogeneous and time-dependent LL point of view outlined above can be extended to non-equilibrium situations. Every time one has a well-defined (semi)classical limit (which at equilibrium is given by the LDA), one can include quantum fluctuations by exhibiting an action, whose minimum reproduces that classical limit, and whose quadratic expansion gives quantum fluctuating corrections. This action, in turn, defines a path integral, and, therefore, a standard ``quantization" procedure which has been applied in many different context, from superfluidity~\cite{khalatnikov2018introduction} to Hall liquids~\cite{wiegmann2014anomalous}. The problem, then, becomes how to write explicitly an action satisfying the above requirements (whose existence and uniqueness is anyway not guaranteed). Below we are going to illustrate the solution to this problem, as found in \cite{ruggiero2020quantum}.

Before proceeding, however, it is worth recalling the main logic behind this quantization.
Indeed, our starting point is a fully quantum model (in our case the one described by the Hamiltonian~\eqref{model}), where correlations are all contained in the time evolved quantum state so that no further quantization is needed at this level. However, in typical situations the full state cannot be constructed explicitly and its correlations are practically inaccessible. The idea is then to circumvent the problem by first finding the (semi)classical limit of the theory (the hydrodynamic solution in our case) and interpreting small fluctuations around such solution as classically propagating linear sound waves. The latter are eventually ``re-quantised'' by constructing the aforementioned action. Note, in particular, that this is \emph{not} the action associated to the initial model. In fact, the relation between the two is not generically understood.

As before we will illustrate the method focusing on 1D Bose gases in traps, now subject to a quantum quench to induce the dynamics.

\subsubsection{Quantum Hydrodynamics} \label{sub:QCH}

We begin once again by considering the model in \eqref{model}  in the TG limit, and focus on a \emph{harmonic trap quench. Namely, we prepare the system in the ground state of \eqref{f-model} with
\be
V(x,0)=\frac{1}{2}m \omega_0^2 x^2, 
\ee
and} suddenly change the frequency to a different value  $\omega \neq \omega_0$ (note, however, that for harmonic traps the quench problem can be solved for generic smooth functions $\omega(t)$~\cite{ruggiero2019conformal}). Since in the TG limit the theory is quadratic and the initial state is Gaussian, all observables can be computed in terms of two-point functions. In particular, a very convenient quantity to consider is the so-called Wigner function~\cite{wigner1932on}
\be \label{fourier}
n_{x,t} (k) = \int {\rm d}y\, e^{i ky} \braket{\Psi_F^\dag (x+y/t,t)\Psi_F(x-y/t,t)},
\ee
which fulfils the following evolution equation  
\begin{equation}
\partial_t n_{x,t} (k) + \frac{ k}{m} \partial_x n_{x,t} (k)  =  \partial_x V(x, t) \partial_k n_{x,t} (k) .
\label{eq:Wfevolution}
\end{equation}
While this equation is \emph{exact} for harmonically trapped non-relativistic particles, the same is not true when considering a relativistic system or free fermions on a lattice (see Sec.~\ref{sec:MF}).  In addition, from \eqref{fourier} it is clear that $n_{x,t} (k)$ contains the same information as the two-point function $\langle \Psi_F^\dag (x,t)\Psi_F(y,t) \rangle $, while it does not capture ``off-diagonal'' contributions (cf. Sec.~\ref{sec:MF}) coming from $\langle \Psi_F (x,t)\Psi_F(y,t) \rangle$ (when it is non-zero) or higher point correlation functions (present for non-Gaussian initial states).

Here, however, we are not interested in the exact microscopic dynamics of $n_{x,t} (k)$. Instead, we interpret $n_{x,t} (k)$ as a coarse grained (or semiclassical) slowly varying distribution function in position and momentum, very much like $\rho_{x,t}(k)$ in Eq.~\eqref{eq:continuity}. In fact, the evolution equation \eqref{eq:Wfevolution} is nothing but the GHD equation in the presence of a slowly varying harmonic potential~\cite{doyon2017anote}. We remark that such a coarse grained interpretation of the Wigner function has been extensively used to characterise the semiclassical regime of Fermi gases in 1D~\cite{bettelheim2006orthogonality,bettelheim2011universal,bettelheim2012quantum}.

\begin{figure}
\centering
 \includegraphics[width=0.7\textwidth]{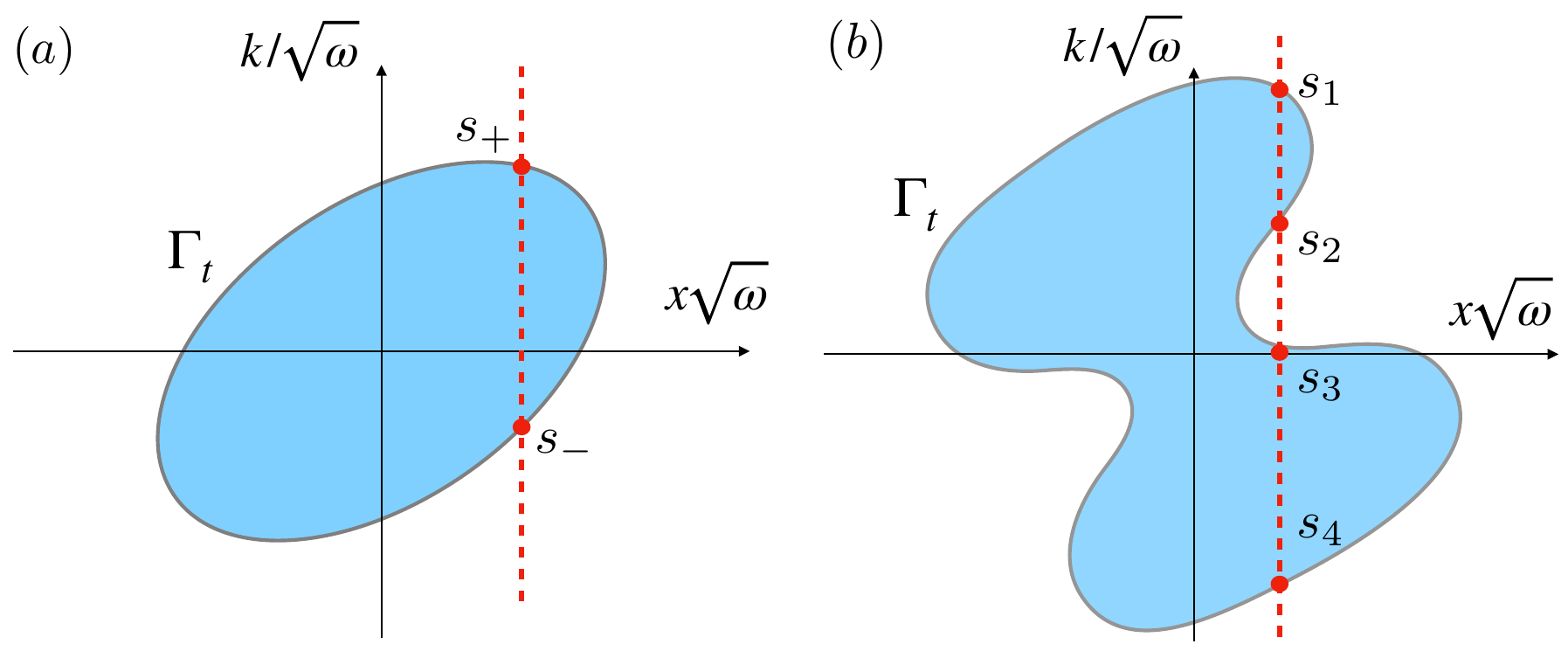}
 \caption{The curve $\Gamma_t$ encircling the points $(x,k)$ at which the Wigner function $n_{x,t}(k)$ is equal to one. Panel (a): Simple situation occurring in the quench in frequency of the single well potential ($\omega_0 \to \omega$), where at any given position $x$ there are only two Fermi points on the contour $\Gamma_t$. Panel (b): More general situation with more than two Fermi points, occurring in the quench from the double to the single well (at frequency $\omega$).
 \label{fig:wigner} }
	\end{figure}

Our goal here is to simplify \eqref{eq:Wfevolution} by using the structure of the initial state. To this end we observe that our initial state (the ground state in the harmonic trap with $\omega(t)=\omega_0$), is a particular example of \emph{zero-entropy} state, namely, it has zero entropy density (such expression was first introduced in \cite{doyon2017large}).
 Those are states whose Wigner function (in the semiclassical limit) reduces to a characteristic function, parametrised by a curve $\Gamma_t$ (the \emph{Fermi contour})
\begin{equation} \label{wigner}
n_{x,t} (k) = 
\begin{cases}
1 & (x,k) \; \textrm{is inside} \; \Gamma_t\\
0 & (x,k) \; \textrm{is outside} \; \Gamma_t
\end{cases}.
\end{equation}
Locally, they look like \emph{split Fermi seas}~\cite{fokkema2014split,eliens2016general,vlijm2016correlations},  labelled by a finite number of Fermi points $\{ k_a \}_{a=1,\cdots, 2Q}$ (see Fig.~\ref{fig:wigner}). Importantly, since entropy is conserved at the hydrodynamic level, these states remain zero-entropy states under GHD evolution. Note also, in connection to Sec.~\ref{sec:v}, that evolution from such states would give zero entanglement entropy within the quasi-particle picture (which captures \emph{linear} growth only). However, they still display a sublinear growth of entanglement entropy that can be exactly accessed with the method reviewed in the present section~\cite{ruggiero2021quantum}.

When evolving from zero entropy states the dynamics of the Wigner function is encoded in that of the contour $\Gamma_t$, or, equivalently, in that of the Fermi points.
Therefore the evolution of $n_{x,t} (k)$ can be described in terms of the $2Q$ evolution equations of the Fermi points $\{ k_a\}_{a=1,\cdots, 2Q}$, which take the form of Burgers equations~\cite{abanov2005hydrodynamics}
\begin{equation} \label{burgers}
\partial_t k_a (x,t) + \frac{ k_{a} (x,t)}{m} \partial_x k_a (x,t) = -\partial_x V(x, t)\,.
\end{equation}
In particular, since in our quench both initial and trapping potentials are harmonic, the Wigner function is just an ellipse that rotates in time \cite{ruggiero2021quantum} (see Figure \ref{fig:wigner} (a)). This means that for any given position $x$, at any time $t$ there are always no more than two Fermi points. As a result, the two associated Burgers equations can be re-expressed in terms of the local density $\rho (x,t)$ (continuity equation) and the hydrodynamic velocity $u (x,t)$ (Euler equation), thus recovering the conventional hydrodynamic equations 
\begin{equation}
\begin{cases}
\partial_t \rho +\partial_x ( u \rho)=0\\
\partial_t u +u\partial_x u = \frac{1}{m\rho} \partial_x P - \frac{1}{m} \partial_x V
\end{cases}.
\label{eq:eulerhydro}
\end{equation}
Here $P= {\pi^2 \rho^3}/{(3m)}$ is the quantum pressure (which explicitly depends on $\hbar$, here set to $1$) of the Tonks-Girardeau gas at zero temperature, the only ``quantum" input at this stage. We stress that the reduction of GHD to standard hydrodynamics is not related to the free nature of the TG gas, and in fact holds for zero-entropy states of generic interacting integrable models, as long as the number of Fermi points is two. In general, however, during the evolution the number of Fermi points might increase, leading to \emph{shocks} in the solution of the hydrodynamic equations (note that shocks do not occur, instead, in zero temperature GHD, as pointed out in~\cite{doyon2017large}).

An action reproducing the Equations \eqref{eq:eulerhydro} has been constructed in Ref.~\cite{ruggiero2019conformal} (see also Ref.~\cite{abanov2005hydrodynamics}). The result is again an iLL of the form~\eqref{igff}. Interestingly, in the TG limit the action is still of the form \eqref{actionCurvedCFT}. Namely, it is a CFT (only the sound velocity is inhomogeneous), with metric explicitly given in terms of the hydrodynamic parameters as follows
\begin{equation}
ds^2= \pi^2 \rho(x, t)^2 \left( (dx - u(x, t) dt)^2 -\pi^2 \rho(x,t)^2 dt^2 \right).
\end{equation}
This action was used to compute several correlation functions, including a closed-form expression for the $n$-particle density matrix~\cite{ruggiero2019conformal}. The same technique has also been used to engineer non-equilibrium scenarios where entanglement entropy shows a non-standard (namely, non-linear) growth~\cite{kosior2020nonlinear}. A similar approach has also been used in~\cite{gawkedzki2017finite,moosavi2019inhomogeneous} to study transport and correlation functions starting from an inhomogeneous temperature profile. Finally, Ref.~\cite{langmann2018diffusive} proposed a generalisation of the approach to the case of a CFT with random sound velocity: this can be used to elucidate how purely ballistic waves in standard CFT acquire normal and anomalous diffusive contributions.

\subsubsection{Generalized Quantum Hydrodynamics}

Even in the TG limit, and for evolutions from zero-entropy states, it is possible to find cases where the simplified description \eqref{eq:eulerhydro} does not hold. Indeed, it is sufficient to look at situations where more than two Fermi points occur~\cite{doyon2017large,ruggiero2021quantum}. This is the case, for instance, of a quench from a double- to a single-well potential, see Figure~\ref{fig:wigner} (b) (note that this setup is also relevant from the experimental point of view, see, e.g., Ref.~\cite{wilson2020observation,malvania2020generalized,schemmer2019generalized}; for instance, it can model the celebrated quantum Newton cradle experiment~\cite{kinoshita2006quantum}). In this case the hydrodynamic regime is still described by the Wigner evolution equation, Eq.~\eqref{eq:Wfevolution}, and, for zero-entropy states, it is still equivalent to a set of Burgers equations for the Fermi points, Eq.~\eqref{burgers} (now four of them, i.e.\ $Q=2$). However, the latter cannot be recast in terms of the two equations for $\rho (x,t)$ and $u(x,t)$ anymore.

To re-quantize the theory we then have to write an action whose equations of motion reproduce the Burgers equations (namely the GHD equations) and whose second order expansion describes (classical) fluctuations around the GHD solution. We refer to this framework as ``Generalized Quantum Hydrodynamics'' (note that the theory was  originally dubbed ``Quantum Generalized Hydrodynamics''~\cite{ruggiero2020quantum}; the rearrangement of terms is here aimed at reducing some potential confusion in relation to the results reviewed in the next section). In this more general situation, the assumption for the local degrees of freedom to be captured by a single bosonic field $\phi$ becomes questionable. Indeed, from the point of view of bosonization~\cite{giamarchi2004quantum}, it is necessary to associate a chiral excitation (field) to each Fermi point. Therefore, if there are more than two components it is not possible to end up in a field theory of a single bosonic field.

The crucial observation is that, in phase-space, the problem can be recast in that of quantising fluctuations of incompressible regions (see~\cite{ruggiero2020quantum} for details), which is well known in the literature on the quantum Hall effect~\cite{wen1990chiral,wen1992theory,iso1992fermions,cappelli1993infinite}. Borrowing from those results, it is possible to exhibit an action with the desired property, which turns out to be the one of a \emph{chiral} boson $\varphi$ living on the Fermi contour (equivalently, around a given point $x$, it appears as the sum of $2Q$ chiral fields $\{ \varphi_a \}$). Parametrising this contour as $(x(s), k(s))$, and introducing a density operator which measures the excess number of occupied states around $(x(s), k(s))$, 
\be
\delta \rho (s) = \partial \varphi (s),
\ee
one finds indeed the following commutation relations
\begin{equation} \label{symplectic}
[\partial \varphi (s), \partial \varphi (s')] = \frac{1}{2 \pi i} \delta' (s-s').
\end{equation}
Passing to the usual coordinates $(x,t)$, and to the Hamiltonian formalism, the Hamiltonian $H_{\rm GqHD}$ -- where the subscript stands for Generalized Quantum Hydrodynamics -- generating \eqref{burgers} can be taken as
\begin{equation} \label{H_QGHD}
H_{\rm GqHD}= \frac{1}{4\pi} \sum_{a, b} \int dx\,  \partial \varphi_a (x) A^{ab} \partial \varphi_b (x)\,,
\end{equation}
where $A^{ab} \equiv \partial \varepsilon_a /\partial k_b$  is known as \emph{flux Jacobian} (with $\varepsilon_a, k_b$ energy and momentum, respectively, of a small excitation around a given Fermi point) and is diagonal in the TG limit.

The program described above can be directly generalised to fully interacting integrable models. In fact, it was first carried out for a 1D Bose gas \eqref{model} with generic interaction $g$~\cite{ruggiero2020quantum}. In this case the role of $n_{x,t} (k)$ is played by the ``filling function"
\be
\vartheta_{x,t} (\lambda)= \frac{1}{1+\eta_{x,t}(\lambda)} = \frac{\rho_{x,t}(\lambda)}{\rho_{t,x,t}(\lambda)},
\ee
where $\rho_{t,x,t}(\lambda)$ and $\eta_{x,t}(\lambda)$ are defined in terms of $\rho_{x,t}(\lambda)$ in \eqref{eq:rhot} and \eqref{eq:eta} respectively (we suppressed the index $n$ assuming no bound states). To proceed one replaces \eqref{eq:Wfevolution} with the general GHD equation for $\vartheta_{x,t} (\lambda)$ in the presence of external potentials~\cite{doyon2017anote}, and considers the class of \emph{zero-entropy} states of the interacting system~\cite{fokkema2014split,eliens2016general,vlijm2016correlations}, now labelled by a finite number of ``Fermi rapidities'' $\{ \lambda_a \}_{a=1,\cdots, 2Q}$. The associated filling function $\vartheta_{x,t} (\lambda)$ takes again the form of a characteristic function, and small fluctuations can be seen as deformations of the Fermi rapidities $\{ \lambda_a \}$. It has been shown in Ref.~\cite{ruggiero2020quantum} that the momenta $\{k_a \}$ of quasiparticles satisfy an exact continuity equation (generalizing the Burger equation valid in the TG limit). Therefore they are the right quantum variables to be quantised also in the interacting case. The final result is exactly Eq.~\eqref{H_QGHD}, where the flux Jacobian $A^{ab}$ can generically couple different chiral excitations.

The Hamiltonian \eqref{H_QGHD}, together with the symplectic structure \eqref{symplectic}, defines the theory of Generalized Quantum Hydrodynamics, which has been applied in both the free~\cite{ruggiero2021quantum} and in the interacting~\cite{ruggiero2020quantum} case to get results for correlations functions and entanglement entropy. Note that it takes the form of a \emph{multi-component}, spatially inhomogeneous, time-dependent Luttinger Liquid, but now, importantly, constructed on top of the \emph{exact profile} (at the Euler scale) given by GHD, as derived from the full quantum initial model. Note, finally, that once again, \eqref{H_QGHD} acquires conformal invariance in the TG limit.

\section{Dynamical generation of quantum correlations} 
\label{sec:MF}

If we try to reintroduce $\hbar$ in the GHD equation~\eqref{eq:continuityint} describing time evolution in the low inhomogeneity limit, we realise that it does not appear explicitly: the equation is essentially classical. Therefore, that equation can only describe how quantum correlations are transported over time, rather than generated. We have seen in the previous section that, for a certain class of initial states, one can describe the propagation of initial-state correlations also in the presence of interactions by means of the theory of generalised quantum hydrodynamics. Instead, we  discuss  here  how  quantum  correlations  can  be  generated  dynamically, even when they are not present in the initial state (for example, when the system is initialised in a product state). To do that we will assigning a precise, quantitative meaning to the space-time dependent root density that we have qualitatively introduced in the previous sections. Consequently, the forthcoming discussion is less elementary than the rest of the review.

We identify two mechanisms for the generation of quantum correlations: 
\begin{enumerate}
\item \label{p:2} There are quantum corrections to the asymptotic generalised hydrodynamic equation~\eqref{eq:continuityint}.
\item \label{p:3} The state is not completely characterised by root densities.
\end{enumerate} 

Let us start with clarifying Point~\ref{p:2}. Despite the asymptotic GHD equation being classical, the full quantum evolution is doubtlessly not classical. In fact, the dependence on $\hbar$ in the von Neumann equation  disappears practically only if the density matrix is stationary. This indicates that there must be quantum corrections to the GHD equation itself. 

Point~\ref{p:3} is clear already in the homogeneous limit, as it is equivalent to the statement that not every state is stationary (see Section~\ref{sec:homogeneous_quench}). Trivial as it sounds, this point can nevertheless be a source of confusion: in relevant scaling limits, some observables are described by expressions that depend only on the root densities, a remarkable example being the growth of the entanglement entropy after a quantum quench from a product state --- cf. Sec.~\ref{sec:v}. In this case the initial state is not stationary but the asymptotic expression in Eq.~\eqref{eq:conj} depends on the state only through the root density. This paradox is resolved after understanding two points: (i) quantities that time evolve independently are coupled by the condition that the state is pure; (ii) the growth of entropy is essentially due to dephasing, which gets rid of the quantities that cannot be interpreted as root densities (in the non-interacting case, they are parametrised by the field $\Psi$ defined below) (see, \emph{e.g.}, Refs~\cite{fagotti2008evolution,alba2017entanglement}). 

Points \ref{p:2} and \ref{p:3} are intimately connected and can be better understood by adopting the point of view described in the second part of Sec.~\ref{sec:inhomogeneous}, i.e., interpreting \eqref{eq:continuityint} as the leading order contribution in a low inhomogeneity limit. In this way the terms generating quantum corrections will be identified with the subleading contributions in the limit. Let us describe how this can be done in the case of free fermion systems, like the one described by the Hamiltonian $H(h)$ in  Eq.~\eqref{eq:FP_hamiltonian}.

To access the subleading contributions in \eqref{eq:continuityint} one has to first understand what is the meaning of a space-time dependent root density beyond the leading order. To this end, there are two possibilities to envisage
\begin{enumerate}[(a)]
\item 
Lifting the root density to a function of space and time is just an effective way to capture the asymptotic evolution.
\item \label{s:2} There is a way to define a space-time dependent root density so that it \emph{exactly} describes the time evolution of a class of states. 
\end{enumerate}
In the first scenario it is usually sufficient to define the root density so as to approach the standard root density in the homogeneous limit. An example of such a line of attack is provided by Ref.~\cite{fagotti2017higher} (and, to some extent, by Ref.~\cite{denardis2018hydrodynamic} in an interacting system), which attempted to go beyond the low inhomogeneity limit by fixing  the definition of root density a priori. The main drawback of a similar description is the impossibility to describe time evolution exclusively in terms of the root density. For example, if we enforce definition~\eqref{eq:rhoop} for a given choice of functions $g_\mu(k)$, the evolution of the root density will not be closed (see the discussion around \eqref{eq:Gamma}). 

On the other hand, at least for non-interacting fermions or equivalent spin chain models, Approach \eqref{s:2} can be successfully applied and one can provide a nonperturbative definition of a special class of states described entirely by the root density~\cite{fagotti2020locally}. Specifically, the root density can be associated with a Wigner function~\cite{wigner1932on} of the Bogoliubov fermions diagonalising the system and the special states are such that no correlations other than those captured by the root density are present. The latter condition is necessary because the relation between root density and correlation matrix is not one-to-one --- cf. Point~\eqref{p:3}. These states extend the notion of locally quasistationary states beyond the low-inhomogeneity limit~\eqref{eq:inhomoSSlambda}. 

In this section we will assume that the Hamiltonian is homogeneous, like that in Eq.~\eqref{eq:FP_hamiltonian}. A similar approach can however be applied also to fermions described by inhomogeneous quadratic Hamiltonians~\cite{fagotti2020locally}, and, in particular, to free fermions in a trap~\cite{dean2019nonequilibrium, eisler2013universality,eisler2009entanglement,ledoussal2018multicritical}.

Let us consider the 2-by-2 block of the correlation matrix associated to sites $(\ell,n)$ 
\be
\Gamma^{\ell n}=\begin{bmatrix}
\delta_{\ell n}-\braket{(c_\ell^{\phantom{\dag}}+c_\ell^\dag)(c_n^{\phantom{\dag}}+c_n^\dag)}&-i \braket{(c_\ell^{\phantom{\dag}}+c_\ell^\dag)(c_n^{\phantom{\dag}}-c_n^\dag)}\\
-i \braket{(c_\ell^{\phantom{\dag}}-c_\ell^\dag)(c_n^{\phantom{\dag}}+c_n^\dag)}&\delta_{\ell n}+ \braket{(c_\ell^{\phantom{\dag}}-c_\ell^\dag)(c_n^{\phantom{\dag}}-c_n^\dag)}\, ,
\end{bmatrix}
\ee
where $c_n$ can be the spinless fermions of our example with Hamiltonian~\eqref{eq:FP_hamiltonian}.
For the LQSSs of the Hamiltonian~\eqref{eq:FP_hamiltonian}, this can be parametrised as follows
\begin{align}\label{eq:GammaLQSS}
\Gamma^{\ell n}=&\hbar\sum_{m\in\mathbb Z}\quad \iint\limits_{[-\pi,\pi]^2}\frac{\mathrm d^2 q}{\pi}e^{i (\ell-n) q_2+i(\ell+n-m) q_1} e^{-\frac{i}{2}\Theta_h(\hbar(q_2+q_1))\sigma^z} e^{\frac{i}{2}\Theta_h(\hbar(q_2-q_1))\sigma^z}\notag\\
&\qquad\qquad\qquad\qquad\times\Bigl\{\1 \rho_{\frac{m}{2},\mathrm{o}}(\hbar q_2)+\sigma^y \Bigl[\rho_{\frac{m}{2},\mathrm{e}}(\hbar q_2)-\frac{1}{4\pi\hbar }\Bigr]\Bigr\}\,,
\end{align}
where $\rho_{x, e/o}(p)$ can be interpreted as the even and the odd part (w.r.t. $p$) of the root density, respectively. This  matrix has indeed two important properties: first, in the homogeneous limit it describes a generic stationary state, and, second, it is closed under time evolution~\cite{fagotti2020locally}.

We point out that the space dependence in the root density (highlighted by the subscript $m/2$ in \eqref{eq:GammaLQSS}) is the result of a particular convention, explained below, in assigning a position to a product of operators acting on different sites (e.g., $c^\dag_\ell c_n$), which is however irrelevant in the low-inhomogeneity limit. As long as quadratic operators like $c_\ell c_n$ are concerned, arguably, the most intuitive definition of position is the average $x=(\ell+n)/2$. This convention has for example the advantage that the position changes in a simple way under chain inversion. There is, however, a possibly unexpected consequence: quadratic operators with odd  range have integer positions, whereas those with even range lie on an effective lattice with  half-integer positions. 
To avoid the complication of specifying which lattice the operator belongs to, we allow $x$ to run over all positions, independently of whether they are physical or not. 
Given that, the root density $\rho_x(p)$ does not need to capture the expectation value of the operators that cannot have position $x$. And this holds true even for values of $x$ that are neither integers nor half-integers, in which case $\rho_x(p)$ could in principle assume any value. Ref.~\cite{fagotti2020locally} proposed to use these degrees of freedom to promote $\rho_x(p)$ to an entire function of $x\in\mathbb C$. This simplifies the asymptotic expansion in the limit of low inhomogeneity and makes it more transparent the correspondence with the kinetic interpretation of the root density. 

Keeping this in mind, we can invert \eqref{eq:GammaLQSS} and find 
\be\label{eq:rhoK}
\rho^{({\rm K})}_x(p)=\frac{1}{4\pi\hbar}+\sum_{y\in\frac{1}{2}\mathbb Z}{\rm K}(x-y)\int_{-\pi}^\pi\frac{\mathrm d q}{8\pi^2\hbar}\sum_{\ell,n\in\mathbb Z}e^{i(n-\ell)\frac{p}{\hbar}+i (2y-n-\ell) q}  \mathrm{tr}\left[ \Gamma^{\ell,n} e^{-i\frac{1}{2}\Theta_h(p-\hbar q)\sigma^z}\frac{\mathrm{I}+ \sigma^y}{2}e^{i \frac{1}{2}\Theta_h(p+\hbar q)\sigma^z}\right]\!,
\ee
where ${\rm K}(x)$ is any entire function that equals $0$ at nonzero integers and $1$ at $x=0$ (in fact, one could also consider more generic interpolations, but we aim at simplicity).

The choice of ${\rm K}(x)$ is arbitrary because the root density is defined so as to capture only the expectation values of the operators at either integer or half-integer positions. This degree of freedom becomes superfluous in the homogeneous limit, where we would like to drop any dependence on $x$. It is then convenient to choose ${\rm K}(x)$ satisfying two auxiliary properties: 
\be
\sum_{z\in\frac{1}{2}\mathbb Z}{\rm K}(x+z)=2
\ee
and 
\be
\sum_{z\in\frac{1}{2}\mathbb Z}{\rm K}(x+z)(-1)^{2z}=0
\ee 
for any $x\in \mathbb R$. A kernel satisfying all the aforementioned constraints is the sinc interpolation 
\be
{\rm K}(x)=\frac{\sin(\pi x)}{(\pi x)}.
\ee 
It is then easy to check that, in a stationary state, $\rho^{({\rm K})}_x(p)$ returns the standard root density. Note that $p$ in \eqref{eq:rhoK} has the dimensions of a momentum (for the sake of simplicity we have set the lattice spacing equal to $1$) and the root density is a density in phase space, which, in turn, has the dimensions of the inverse of an action. 

Apart from being consistent with the thermodynamic description of expectation values in homogeneous states, Definition~\eqref{eq:rhoK} is convenient because $\rho_{x,t}^{({\rm K})}(p)$ turns out to depend on the state only through $\rho_{x,0}^{({\rm K})}(p)$ and vice versa. Specifically,  the root density as defined in \eqref{eq:rhoK} evolves in time according to the Moyal dynamical equation
\be\label{eq:GHDcomplete}
\partial_t\rho^{({\rm K})}_{x,t}(p)=\{\{E_h(p),\rho^{({\rm K})}_{x,t}(p)\}\},
\ee
where $E_h(k)=2J \varepsilon(p/\hbar)$, the Moyal bracket is defined as~\cite{moyal1949quantum}
\be
\{\{f_x(p),g_x(p)\}\}=-i/\hbar(f_x(p)\star g_x(p)-g_x(p)\star f_x(p)),
\ee
and, for  $f_x(p)$ and $g_x(p)$ $2\pi$-periodic functions of $p$, the star product $\star$ reads as
\be
f_x(p)\star g_x(p)=\iint\limits_{[-\pi,\pi]^2}\frac{\mathrm d^2 q}{(2\pi)^2}\sum_{n,m\in\mathbb Z} e^{-i(n-2x) q_1} f_{-\frac{m}{2}}(p+\hbar q_1)e^{-i (m+2x) q_2} g_{\frac{n}{2}}(p+\hbar q_2)\, .
\ee
Ref.~\cite{fagotti2020locally} has interpreted \eqref{eq:GHDcomplete} as a non-perturbative version of generalised hydrodynamics. From that perspective, GHD is the phase-space formulation~\cite{moyal1949quantum} of quantum mechanics in a special sector of the Hilbert space spanned by LQSSs. Since we are considering a homogeneous Hamiltonian, Eq.~\eqref{eq:GHDcomplete} can be readily inverted
\be\label{eq:solGHD}
\rho^{({\rm K})}_{x,t}(p)=
\sum_{z\in\frac{1}{2}\mathbb Z}\int_{-\pi}^{\pi}\frac{\mathrm d q}{2\pi}  e^{2i q z}e^{-\frac{i}{\hbar}(E_h(p+\hbar q)-E_h(p-\hbar q))t} \rho^{({\rm K})}_{x-z,0}(p)\, .
\ee
Note that the sum over $z$ could also be replaced by an integral, as done in Ref.~\cite{fagotti2020locally}, but then the interpolation would explicitly depend on time (instead, \eqref{eq:solGHD} is such that the interpolation of the time evolution is the time evolution of the interpolation).  
Just like phase-space quantum mechanics makes the semiclassical expansion in the limit $\hbar\rightarrow 0$ transparent, the GHD equation \eqref{eq:GHDcomplete} can be readily expanded in the limit of low inhomogeneity. To show this point, it is again convenient to choose the sinc kernel. Indeed, the latter has the property 
\be
\sum_{z\in\frac{1}{2}\mathbb Z}e^{2i q z}\frac{\sin[\pi(x-z)]}{\pi(x-z)}=\int_{-\infty}^\infty \mathrm d z e^{i q z}\frac{\sin[\pi(x-\frac{z}{2})]}{\pi(x-\frac{z}{2})},
\ee
which implies that the sum over $z$ can be exactly replaced by an integral over $z$. Therefore, we can write 
\be\label{eq:solGHD1}
\rho^{(\mathrm{sinc})}_{x,t}(p)=
\int_{-\infty}^\infty\mathrm d z\int_{-\pi}^{\pi}\frac{\mathrm d q}{2\pi}  e^{i q z}e^{-\frac{i}{\hbar}(E_h(p+\hbar q)-E_h(p-\hbar q))t} \rho^{(\mathrm{sinc})}_{x-\frac{z}{2},0}(p)\, .
\ee
As anticipated in Section~\ref{sec:inhomogeneous}, the low-inhomogeneity expansion can be carried out by introducing an auxiliary parameter $\bar b\equiv \Lambda/a$ that quantifies the typical scale of the inhomogeneity. Indeed, let us consider a family of smooth initial conditions 
\be
f_{b,0}(x,p)=f_{1,0}\left(\frac{x }{b},p\right),
\ee
such that $f_{\bar b ,0}(x,p)=\rho^{(\mathrm{sinc})}_{x,0}(p)$, and calculate $f_{b,t}(x,p)$. If $\bar b$ is large, we can approximate \eqref{eq:solGHD1} by its asymptotic expansion  in the limit of large $b$
\begin{align}\label{eq:GHDasympt}
f_{b,t}(x,p)&=
\int_{-\infty}^\infty\mathrm d z\int_{-\pi b}^{\pi b}\frac{\mathrm d q}{2\pi}  e^{i q  z}e^{-\frac{i}{\hbar}(E_h(p+\hbar \frac{q}{b})-E_h(p-\hbar \frac{q}{b}))t} f_{1,0}\left(\frac{x}{b}-\frac{z}{2},p\right)\notag\\
&\sim\int_{-\infty}^\infty\mathrm d z\int_{-\infty}^{\infty}\frac{\mathrm d q}{2\pi}  e^{i q  z}\exp\left[{-2i\sum_{n=0}^N\frac{\hbar^{2n}q^{2n+1}}{(2n+1)!b^{2n+1}}E_h^{(2n+1)}(p)t+O(b^{-2N-3})}\right] f_{1,0}\left(\frac{x}{b}-\frac{z}{2},p\right)\, .
\end{align}
We call this  $(2N+1)$-th order generalised hydrodynamics.
In particular, truncating the expansion at the third order $(N=1)$, setting back $b$ to $\bar b$, and integrating over $q$, we find the solution to third-order generalised hydrodynamics 
\be\label{eq:GHD3_root}
\rho^{(3)}_{x,t}(p)=\int\mathrm d y\mathrm{Ai}[y] \rho^{(3)}_{x-E'_h(p)t+\frac{y}{2}(E'''_h(p)\hbar^2 t)^{1/3},0}(p)\, ,
\ee
where the physically irrelevant dependency on ${\rm K}$ is understood. From \eqref{eq:GHDasympt}, we can also easily derive the $(2N+1)$-th order hydrodynamic equation: on the one hand, the time derivative generates a polynomial in ${q}/{b}$; on the other hand, the space derivative acts as 
\be
\partial_xf_{1,0}\left(\frac{x}{b}-\frac{z}{2}\right)=-\frac{2}{b}\partial_z f_{1,0}\left(\frac{x}{b}-\frac{z}{2}\right),
\ee
which is equivalent to 
\be
\frac{2 iq }{b} f_{1,0}\left(\frac{x}{b}-\frac{z}{2}\right)
\ee
after integrating \eqref{eq:GHDasympt} by parts in $z$. We then find 
\be\label{eq:GHD3}
\partial_t\rho_{x,t}(p)+E'_h(p)\partial_x \rho_{x,t}(p)=\hbar^2\frac{E'''_h(p)}{24}\partial_x^3 \rho_{x,t}(p)+O(\hbar^4 \partial_x^5)\, .
\ee 
From this equation we can extract a sufficient criterion for the irrelevance of the corrections
\be
\left |\frac{\varepsilon'''(p/\hbar)}{24\varepsilon'(p/\hbar)}\right |\ll \left |\frac{\partial_x \rho_{x,t}(p)}{\partial_x^3 \rho_{x,t}(p)} \right |\sim O(b^2),
\ee
where we assumed that the root density  with $b=1$ has $O(1)$ derivatives. In our specific case with dispersion relation \eqref{eq:disp_rel}, there could be problems only when $h$ is close to $1$, in which case one should have $b\gg {|h-1|}^{-1}$ otherwise observables with range $O(|h-1|^{-1})$ and larger could be affected by the correction.

In bipartitioning protocols, the most visible effect of the corrections in Es.~\eqref{eq:GHD3} is arguably at the rays where profiles are not smooth in the ballistic limit (cf. Sec.~\ref{sec:localphysics}). For example, we will see that \eqref{eq:GHD3} correctly captures the  Airy scaling behaviour around the light-cone edge~\cite{eisler2013full}. Remarkably, the more naive definition of space-time dependent root density considered in Ref.~\cite{fagotti2017higher} fails in the same task. In order to understand why we recall that the relation between $\rho_x^{({\rm K})}(p)$ and the correlation matrix --- Eq.~\eqref{eq:rhoK} --- is not one-to-one: if we try to express the correlation matrix in terms of the root density, we realise that there is more information contained in the former. Refs~\cite{fagotti2017higher, fagotti2020locally} proposed to introduce an auxiliary odd complex field, called $\Psi_x^{(\tilde {\rm K})}(p)$, to complete the information. In terms of $\rho^{({\rm K})}_{x}(p)$ and $\Psi_x^{(\tilde {\rm K})}(p)$, the correlation matrix can be written as 
\begin{multline}\label{eq:Gamma}
\Gamma^{\ell n}=\hbar\sum_{m\in\mathbb Z}\quad \iint\limits_{[-\pi,\pi]^2}\frac{\mathrm d^2 q}{\pi}e^{i (\ell-n) q_2+i(\ell+n-m) q_1} e^{-\frac{i}{2}\Theta_h(\hbar(q_2+q_1))\sigma^z}e^{\frac{i}{2}\Theta_h(\hbar(q_2-q_1))\sigma^z} \\
 \Bigl\{\1 \rho^{({\rm K})}_{\frac{m}{2},\mathrm{o}}(\hbar q_2)+\sigma^y \Bigl[\rho^{({\rm K})}_{\frac{m}{2},\mathrm{e}}(\hbar q_2)-\frac{1}{4\pi\hbar }\Bigr]+\sigma^z \Psi^{(\tilde {\rm K})}_{\frac{m}{2},\mathrm{R}}(\hbar q_2)-\sigma^x\Psi^{(\tilde {\rm K})}_{\frac{m}{2},\mathrm{I}}(\hbar q_2)\Bigr\}\, ,
\end{multline}
where 
\be
\Psi^{(\tilde {\rm K})}_{x,\rm{R}}(p)=\mathrm{Re}[\Psi^{(\tilde {\rm K})}_{x}(p)],\qquad\qquad \Psi^{(\tilde {\rm K})}_{x,\rm{I}}(p)=\mathrm{Im}[\Psi^{(\tilde {\rm K})}_{x}(p)].
\ee
The function $\tilde {\rm K}(x)$, appearing as a superscript of $\Psi^{(\tilde {\rm K})}_x(p)$, represents a degree of freedom in its definition, as ${\rm K}(x)$ does for $\rho^{({\rm K})}_x(p)$, and satisfies the same properties of ${\rm K}(x)$. In particular, the auxiliary field can be expressed in terms of the correlation matrix as follows
\be
\Psi_x^{(\tilde {\rm K})}(p)=\sum_{y\in\frac{1}{2}\mathbb Z}\tilde {\rm K}(x-y)\int_{-\pi}^\pi\frac{\mathrm d q}{8\pi^2\hbar}\sum_{\ell,n\in\mathbb Z}e^{i(n-\ell)\frac{p}{\hbar}+i (2y-n-\ell) q}  \mathrm{tr}\left[ \Gamma^{\ell,n} e^{-i\frac{1}{2}\Theta_h(p-\hbar q)\sigma^z}\frac{\sigma^z-i\sigma^x}{2}e^{i \frac{1}{2}\Theta_h(p+\hbar q)\sigma^z}\right]\, .
\ee
Importantly, the auxiliary field time evolves independently of the root density and satisfies a different dynamical equation
\be\label{eq:GHDPsi}
i\hbar\partial_t \Psi_{x,t}(p)=E_h(p)\star \Psi_{x,t}(p)+\Psi_{x,t}(p)\star E_h(-p)\sim\\
2E_h(p)\Psi_{x,t}(p)-\hbar^2\frac{E''_h(p)}{4}\partial_x^2\Psi_{x,t}(p)+O(\hbar^4\partial_x^4)
\ee
where  the physically irrelevant dependence on $\tilde {\rm K}$ is understood and we have also shown the first orders of the low-inhomogeneity expansion using that the dispersion relation is even. 
In particular, in the low-inhomogeneity limit
\be
e^{2i E_h(p)t/\hbar}\Psi_{x,t}(p)
\ee
satisfies the Schr\"odinger equation of a free particle with effective mass $2/E''_h(p)$. As for the complete GHD equation, also the dynamical equation of the auxiliary field, \eqref{eq:GHDPsi}, can be readily inverted
\be
\Psi^{(\tilde {\rm K})}_{x,t}(p)=
\sum_{z\in\frac{1}{2}\mathbb Z}\int_{-\pi}^{\pi}\frac{\mathrm d q}{2\pi}  e^{2i q z}e^{-\frac{i}{\hbar}{(E_h(p+\hbar q)+E_h(-p+\hbar q))t}} \Psi^{(\tilde {\rm K})}_{x-z,0}(p)\, .
\ee
The solution to \eqref{eq:GHDPsi} truncated at a given order can be obtained in the same way as the solution to $n$-th order GHD; in particular, expanding \eqref{eq:GHDPsi} up to the third order (which vanishes for even dispersion relations like $E_h(p)$) we have
\be
\Psi_{x,t}^{(3)}(p)=e^{-2 \frac{i}{\hbar} {E_h(p)t}}
\int_{-\infty}^\infty\mathrm d y\frac{e^{i\mathrm{sgn}(\hbar E''_h(p)t) y^2}}{ \sqrt{i \pi 
\mathrm{sgn}(\hbar E''_h(p)t)
}} \Psi^{(3)}_{x-y\sqrt{|\hbar E_h''(p)t|},0}(p)\, .
\ee
In contrast to what happens for the root density~\eqref{eq:GHD3_root}, the leading correction to the auxiliary field is captured by a Gaussian kernel. 

We are now in a position to understand that a different definition of root density, $\tilde\rho_x(p)$, mixing the fields $\rho_{x}(p)$ and $\Psi_{x}(p)$ and approaching $\rho_{x}(p)$ in the low-inhomogeneity limit, which can be formally expressed as 
\be
\tilde\rho_x(p)=\rho_{x}(p)+ \Upsilon[\partial_x\rho_{x}(p),\partial_x^2\rho_{x}(p),\dots;\partial_x\Psi_{x}(p),\partial_x^2\Psi_{x}(p),\dots]\, ,
\ee
would satisfy a different dynamical equation, irremediably coupled with independent degrees of freedom. In addition, even neglecting the contributions independent of $\tilde \rho_x(p)$, the dynamical equation would have a different low-inhomogeneity expansion, which is exactly the problem affecting this definition of inhomogeneous root density.

\subsection{Behaviour at the light-cone edges}
\label{sec:lightcone}

In this section we consider the physics around the edges of the light cone when the root density describing the initial state is piece-wise continuous. Note that this does not exactly correspond to the bipartitioning protocols considered so far, where the initial density matrix takes the form $\hat \rho(0)=\hat\rho_L\otimes\hat\rho_R$: correlations between  left and right operators are still present around the discontinuities of the root density (and of the auxiliary field). Note however that a state with a piece-wise constant root density and the corresponding partitioned state are almost equivalent everywhere except in regions, localised around the junctions, of extent proportional to the correlation length. The bipartitioning protocol will be considered at the end of this section. We will follow the approach of Ref.~\cite{fagotti2017higher} but we will define the root density as in the previous section. Specifically, using the sinc kernel, the correlation matrix is given by Eq.~\eqref{eq:GammaLQSS} with the root density  
\be\label{eq:varrho}
\rho_{x,t}(p)=\frac{1}{4\pi\hbar}+\lim_{\epsilon\rightarrow 0^+}
 \sum_{s=\pm 1}\int_{-\frac{\pi}{2}}^{\frac{\pi}{2}}\frac{\mathrm d q}{4\pi i}e^{2is q x}e^{-\frac{i}{\hbar}{(E_h(p+\hbar sq)-E_h(p-\hbar sq))t}}  \left[\cot(\frac{q}{2}-i\epsilon)\varrho_{s}(p)-\tan(\frac{q}{2})\varrho_{s}(p+\pi)\right]\, ,
\ee
where  we called $\varrho_{\pm}(p)=\rho_{R/L}(p)-{1}/{(4\pi\hbar)}$ the shifted root density at time zero at the right and at the left of the discontinuity, respectively. 
Expectation values around a light-cone edge are completely characterised by the correlation matrix restricted to a moving subsystem that contains the edge. We consider the scaling limit in which the time is large and the displacement from the edge scales as $t^{1/3}$.
For pedagogical purposes, we start by assuming first-order generalised hydrodynamics (which, we remind the reader, corresponds to expanding the argument of the exponential in \eqref{eq:varrho} at the first order in $q$). By carrying out the scaling limit, we find that the correlation matrix behaves as
\be\label{eq:DeltaGamma1}
\Gamma_{1-\mathrm{GHD}}^{\ell,n}-\Gamma_R^{\ell,n}\sim
2^{7/3}\pi \hbar  [\rho_L(p_M)-\rho_R(p_M) ]
\frac{i\sin[\frac{(\ell-n) p_M}{\hbar}]\1+\sigma^y\cos[\frac{(\ell-n) p_M}{\hbar}+
\Theta_h(p_M)\sigma^z]}{({-t E_h'''( p_M)\hbar^2})^{\frac{1}{3}}}K_{0}(x_\ell,x_n)\, ,
\ee
where $\Gamma_R$ is the correlation matrix associated with $\rho_R(p)$ (hence, at the right of the light-cone edge) and we defined the dimensionless displacement as
\be
x_\ell=\frac{2^{1/3}(\ell-v_{\rm M}t)}{({-t E_h'''(p_M)\hbar^2})^{1/3}}\, ,
\ee
the kernel $K_{0}(x_\ell,x_n)$ as
\be
K_{0}(x_\ell,x_n)=\theta\Bigl(-\frac{x_\ell+x_n}{2}\Bigr)\frac{\sin[\sqrt{-\frac{x_\ell+x_n}{2}}(x_\ell-x_n)]}{\pi(x_\ell-x_n)}\, ,
\ee
and $p_M$ is the momentum associated with the maximal velocity, i.e.
\be
E_h'(p)\leq E'_h(p_M)=: v_M,
\ee
for any $p$. Eq.~\eqref{eq:DeltaGamma1} is the result of transporting the quantum correlations in the initial state at time $t$ through first-order GHD: no additional quantum correlations are created.
Direct comparison with numerical data shows that \eqref{eq:DeltaGamma1} gives us the correct  qualitative behaviour but does not exactly capture the scaling limit of the correlations. This gap is filled by third-order generalised hydrodynamics, which, as we have seen, carries the leading quantum correction to the asymptotic hydrodynamic equation. Note however that the third order is not expected to be sufficient in situations where the second derivative of the velocity at its maximum/minimum vanishes (see also Ref.~\cite{ledoussal2018multicritical}). In order to distinguish the quantum correlations generated dynamically from the ones present in the initial state, we attach an auxiliary variable $\chi$ to the reduced  Planck constant  $\hbar$ explicitly appearing in Eq.~\eqref{eq:GHD3}: $\hbar \rightarrow\chi\hbar$. The correct result is then recovered by setting $\chi$ to $1$. 
We obtain
\be\label{eq:GammaGHD3}
\Gamma^{\ell, n}-\Gamma_R^{\ell, n}=
2^{7/3}\pi \hbar  [\rho_L(p_M)-\rho_R(p_M) ]
\frac{i\sin[\frac{(\ell-n)  p_M}{\hbar}]\1+\sigma^y\cos[\frac{(\ell-n)  p_M}{\hbar}+
\Theta_h(p_M)\sigma^z]}{({-t E_h'''(p_M)\hbar^2})^{\frac{1}{3}}}K_{\chi}(x_\ell,x_n)+o(t^{-\frac{1}{3}})\, .
\ee
where the kernel $K_\chi[x_\ell,x_n]$ is given by
\be
K_\chi[x_\ell,x_n]=\chi^{\frac{1}{3}}K_{1}\left[\chi^{-\frac{2}{3}}\frac{x_\ell+x_n}{2}+\chi^{\frac{1}{3}}\frac{x_\ell-x_n}{2},\chi^{-\frac{2}{3}}\frac{x_\ell+x_n}{2}-\chi^{\frac{1}{3}}\frac{x_\ell-x_n}{2}\right]\, ,
\ee
and $K_1[x_\ell,x_n]$ is the Airy kernel
\be
K_1[x_\ell,x_n]=\frac{\mathrm{Ai}(x_\ell)\mathrm{Ai}'(x_n)-\mathrm{Ai}'(x_\ell)\mathrm{Ai}(x_n)}{x_\ell-x_n}\, .
\ee
Note that the kernel associated with first-order GHD, namely $K_0[x_\ell,x_n]$, can be obtained by taking the limit 
\be
\lim_{\chi\rightarrow 0}K_\chi[x_\ell,x_n].
\ee

Let us use \eqref{eq:GammaGHD3} to compute the connected two-point function of the spin in the $z$ direction in the vicinity of the light-cone edge. If we assume that the root density describing the right state is smooth, the two-point function in the right state will decay exponentially in the distance 
\be
|\ell-n|=|\delta\ell-\delta n|= (x_\ell-x_n)\left(\frac{-t E_h'''(\hbar \bar q_M)\hbar^2}{2}\right)^{1/3}.
\ee
Therefore, in the scaling limit under exam we can neglect $\Gamma_R^{v_{M}t+\delta\ell, v_{M}t+\delta n}$ with respect to $\Gamma^{v_{M}t+\delta\ell, v_{M}t+\delta n}$. Proceeding in this way we then find
\begin{align}\label{eq:szszGHD3}
\braket{s_{v_{M}t+\delta\ell}^z s_{v_{M}t+\delta n}^z}_c &=\frac{\hbar^2}{4}(\braket{\sigma_{v_{M}t+\delta\ell}^z\sigma_{v_{M}t+\delta n}^z}-\braket{\sigma_{v_{M}t+\delta\ell}^z}\braket{\sigma_{v_{M}t+\delta n}^z})=\frac{\hbar^2}{4}\det\Gamma^{v_{M}t+\delta\ell,v_{M}t+\delta n}\notag\\
&\sim \frac{\hbar^2}{4}\det[\Gamma^{v_{M}t+\delta\ell, v_{M}t+\delta n}-\Gamma_R^{v_{M}t+\delta\ell, v_{M}t+\delta n}]\notag\\
&\sim -\frac{2^{8/3}\pi^2 \hbar^4  [\rho_L(p_M)-\rho_R(p_M) ]^2\cos^2[\Theta_h(p_M)]}{({-t E_h'''(p_M)\hbar^2})^{\frac{2}{3}}}[K_{\chi}(x_\ell,x_n)]^2\, .
\end{align}
The only difference between this and the result obtained within first-order GHD is in the kernel, which appears as a multiplicative factor. 
This shows that the (quantum) corrections to first-order GHD do not necessarily translate into corrections in connected correlation functions: they can have leading effects!

If the state is not an LQSS (for which $\Psi_{x,t}(p)=0$), we should also consider the contribution from the auxiliary field. One can however show that the latter is subleading: specifically, the contribution from $\Psi$ to the fermionic two-point function  asymptotically behaves as $O(t^{-2})$. This cancellation is convenient, as it allows us to use \eqref{eq:szszGHD3} even in quench settings, where the state evolves in time also in the bulk. 

Finally, let us clarify the relation between the results presented in this section and the bipartitioning protocols $\hat\rho(0)=\hat\rho_L\otimes\hat\rho_R$ discussed in the previous sections. To this end, we assume that in the two states joined together connected correlations decay exponentially. Focusing on the part of \eqref{eq:Gamma} involving the root density, the correlation matrices of the two settings differ in a term of the form 
\begin{multline}\label{eq:junction}
\hbar\int\mathrm d y\mathrm{Ai}[y] \sum_{m\in\mathbb Z}\quad \iint\limits_{[-\pi,\pi]^2}\frac{\mathrm d^2 q}{\pi}e^{i(\ell+n-m) q_1}\delta \rho^{(K;3)}_{\frac{m}{2}-E'_h(\hbar q_2)t+\frac{y}{2}(E'''_h(\hbar q_2)\hbar^2 t)^{1/3},0}(\hbar q_2)\\
\left\{e^{i (\ell-n) q_2}
e^{-\frac{i}{2}{\Theta_h(q_2+q_1)}\sigma^z} \frac{\1+\sigma^y}{2}
 e^{\frac{i}{2}{\Theta_h(q_2-q_1)}\sigma^z}-
e^{-i (\ell-n) q_2}
e^{\frac{i}{2}{\Theta_h(q_2-q_1)}\sigma^z} \frac{\1-\sigma^y}{2}
 e^{-\frac{i}{2}{\Theta_h(q_2+q_1)}\sigma^z}\right\}
\end{multline}
where $\delta \rho_{x,0}(p)$ decays exponentially in $|x|$. We can express the latter property as a smoothness condition on its Fourier transform:
\be
\delta \rho_{x,0}(\hbar q_2)=\int_{-\pi}^{\pi}\frac{\mathrm d \omega}{2\pi}e^{2i\omega x}\tilde \rho_{\omega}(\hbar q_2)\, ,
\ee
where $\tilde \rho_{\omega}(\hbar q_2)$ is a smooth function of $\omega$. By summing over $m$ in \eqref{eq:junction} we then get 
\begin{multline}
\hbar\int\mathrm d y\mathrm{Ai}[y] \iint\limits_{[-\pi,\pi]^2}\frac{\mathrm d^2 q}{\pi}e^{i(x_\ell+x_n)(\frac{-t E_h'''(p_M)\hbar^2}{2})^{1/3} q_1}e^{iq_1[2(E_h'(p_M)-E'_h(\hbar q_2))t+y(E'''_h(\hbar q_2)\hbar^2 t)^{1/3}]}\tilde \rho_{q_1}(\hbar q_2)\\
\left\{e^{i (\ell-n) q_2}
e^{-\frac{i}{2}{\Theta_h(q_2+q_1)}\sigma^z} \frac{\1+\sigma^y}{2}
 e^{\frac{i}{2}{\Theta_h(q_2-q_1)}\sigma^z}-
e^{-i (\ell-n) q_2}
e^{\frac{i}{2}{\Theta_h(q_2-q_1)}\sigma^z} \frac{\1-\sigma^y}{2}
 e^{-\frac{i}{2}{\Theta_h(q_2+q_1)}\sigma^z}\right\}\, .
\end{multline}
Since all functions are smooth, a change of variables to 
\be
k=\left (\frac{-t E_h'''(p_M)\hbar^2}{2}\right )^{1/3}q_1\qquad \text{and}\qquad q=\left (\frac{-t E_h'''(p_M)\hbar^2}{2}\right )^{1/3}\left (q_2-\frac{p_M}{\hbar}\right)
\ee
is sufficient to show that the expression is $O(t^{-\frac{2}{3}})$, and hence subleading with respect to \eqref{eq:GammaGHD3}.
In conclusion, \eqref{eq:GammaGHD3} and \eqref{eq:szszGHD3} describe the light-cone edge even in the standard bipartitioning protocols where there are no initial correlations between operators acting on opposite sides of the junction.

\section{Conclusions and Outlook}
\label{sec:conclusions}

We have given a pedagogical account of GHD seen as a theory to describe the large-time behaviour of quantum many-body systems after inhomogeneous quenches. We have described the basic ideas of the theory --- paralleling the now understood case of homogeneous quantum quenches --- and its simplest predictions concerning the expectation values of local observables at infinite times. We have discussed certain extensions of the theory to keep track of the evolution of quantum correlations. Specifically, we have showed how GHD can be combined with the quasiparticle picture of Ref.~\cite{calabrese2005evolution} to describe the entanglement growth after inhomogeneous quenches, and how it can be combined with the inhomogeneous Luttinger-Liquid framework of Ref.~\cite{brun2018inhomogeneous} to account for the evolution of quantum correlations for a certain class of small inhomogeneous quenches. Finally we have discussed a systematic approach to determine higher order corrections to the GHD equations in non-interacting systems. In spite of the extremely rapid development of GHD (the original papers appeared less than five years ago and the theory is already the subject of a special issue) there are still many open questions. Here we discuss some of them, following their order of appearance in the review.    
\begin{itemize}
\item Arguably one of the most pressing open questions concerning the development of the theory is the ``initial condition" problem for Eq.~\eqref{eq:continuityint}. Namely, there is currently no standard procedure that, given a slowly varying inhomogeneous initial state $\ket{\Psi_0}$, produces a suitable initial condition for Eq.~\eqref{eq:continuityint}, enabling a quantitative description of the late-time dynamics. As discussed in Sec.~\ref{eq:inhomogeneousinteracting}, for interacting systems this can be done only for a very limited class of initial states.   

\item 
As discussed in Sec.~\ref{sec:v} the quasiparticle picture for the entanglement spreading 
gives us the leading-order description in the space-time scaling limit. An interesting question is to understand 
subleading corrections. These are expected due to the fact that the quasiparticles follow classical trajectories only on average. In fact, due to the interactions,
the motion of a given quasiparticle performs a random walk around the classical trajectory. This effect is responsible for the diffusive correction to the GHD equation~\cite{denardis2018hydrodynamic, gopalakrishnan2018hydrodynamics, denardis2019diffusion} but its role in the entanglement dynamics is not yet understood. A related question is whether and how exotic transport properties of integrable systems at some special points, such as superdiffusion, are reflected in the entanglement dynamics. Other open issues pertaining to the study of entanglement are $(i)$ describe the full-time dynamics of the R\'enyi entropies based on a quasiparticle structure, $(ii)$ clarify how integrability breaking affects the entanglement spreading~\cite{bertini2020prethermalization}, and $(iii)$ incorporate the effects of dissipation in the quasiparticle 
picture~\cite{alba2021spreading,alba2021noninteracting,alba2021unbounded,carollo2021emergent}. Finally, another interesting direction is to investigate the entanglement spreading 
during the out-of-equilibrium dynamics with a time-dependent Hamiltonian, for instance, 
by using the approach developed in Ref.~\cite{bastianello2019generalized}. 

\item  
As discussed in Sec.~\ref{sec:vPR}, the current construction of Generalized Quantum Hydrodynamics is only valid for zero-entropy states. It would be interesting to extend the theory to include other initial states, such as, e.g., states at finite (low) temperature. In analogy with the standard Luttinger-Liquid theory, we expect it to be feasible and to potentially reveal an interesting competition between thermal and quantum fluctuations. Moreover, as a theory of \emph{linear} fluctuations, Generalized Quantum Hydrodynamics cannot give access to \emph{all} quantum corrections. There are different ways to take them into account, corresponding to different kinds of (subleading) contributions. A natural next step, would be to investigate corrections coming from non-linear quantum fluctuations (namely, taking into account the ``curvature'' of the dispersion relation): this is expected to lead to a generalization of standard \emph{non-linear Luttinger-Liquid theory}~\cite{imambekov2012one}. This is related to the problem of studying the effects of irrelevant and marginal perturbations of the LL theory, which are far from understood out of equilibrium (see, e.g., \cite{sotiriadis2017equilibration} and references therein). Finally another interesting direction pertains to the experimental validation of such extension of GHD in the context of cold-atom experiments, which would be a natural next step to pursue after the recent experimental confirmations of GHD itself~\cite{schemmer2019generalized,malvania2020generalized}. 

\item The non-perturbative formulation of generalised hydrodynamics described in Section~\ref{sec:MF} strongly relies on the applicability of Wick's theorem. The possibility of generalising it to interacting integrable systems is an open question. In particular, in the presence of interactions (in the Hamiltonian), it is unclear whether a class of inhomogeneous states described by higher-order GHD can be at all defined. We also mention that the non-perturbative formulation of Section~\ref{sec:MF} has not yet been extended to the non-interacting time evolution of states that do not satisfy the Wick's theorem. Another aspect that deserves more investigation is the study of connected correlation functions, which, as shown in Section~\ref{sec:MF}, could depend strongly on contributions that are generally subleading when considering the expectation value of local operators.
\end{itemize}

\acknowledgements
We are grateful to many colleagues for collaborations on topics connected with this review and for inspiring discussions over the years. Even though it is impossible to mention all of them we would like to give a special thank to Pasquale Calabrese, Mario Collura, Jacopo De Nardis, Benjamin Doyon, Jer\^ome Dubail, Fabian Essler, Tony Jin, Toma\v{z} Prosen, and Takato Yoshimura. We thank especially Fabian Essler for valuable comments on the manuscript. BB thanks the Galileo Galilei Institute in Florence, where part of the material covered in this review was presented in the lecture course ``Transport in closed one-dimensional systems" as part of the PhD school SFT 2019. VA thanks the organisers of the summer school on ``Clean  and  disordered  systems  out  of  equilibrium'', held  in  Carg\`ese in September  2020, where part of the material covered in this paper was presented. This work has been supported by the Royal Society through the University Research Fellowship No.\ 201101 (BB), the Swiss National Science Foundation under Division II (PR), the DFG (German Research Foundation) under Germany's Excellence Strategy --- EXC-2111 --- 390814868 (LP), the European Research Council under ERC Advanced grant No.\ 743032 DYNAMINT (VA) and under the Starting Grant No.\ 805252 LoCoMacro (MF).

\bibliography{./bibliography}

\end{document}